\documentclass[12pt]{article}
\usepackage{graphicx}
\usepackage{latexsym,amsmath,amsfonts,amssymb}
\usepackage{bbm}
\usepackage{slashed}
\usepackage{cite}
\usepackage[linktoc=page]{hyperref}

\renewenvironment{thebibliography}[1]{%
\begin{oldthebibliography}{#1}%
\setlength\itemsep{-1mm}%
}%
{%
\end{oldthebibliography}%
}

\newcommand{\dvol}{d\mathrm{vol}}
\newcommand{\vol}{\mathrm{vol}}

\newcommand{\parfrac}[2]{\frac{\partial #1}{\partial #2}}
\newcommand{\delfrac}[2]{\frac{\delta #1}{\delta #2}}
\newcommand{\ud}[2]{^{#1}_{\phantom{#1}#2}}
\newcommand{\du}[2]{_{#1}^{\phantom{#1}#2}}
\newcommand{\rep}[1]{\ensuremath{\mathbf{#1}}}

\newcommand{\eg}{\textit{e.g.}}

\newcommand{\ie}{\textit{i.e.}}

\numberwithin{equation}{section}

\newcommand{\pslash}{p \!\!\! \slash\,}
\newcommand{\nn}{\nonumber}
\newcommand{\mat}[1]{\begin{pmatrix} #1 \end{pmatrix}}
\newcommand{\smat}[1]{\big( \begin{smallmatrix} #1 \end{smallmatrix} \big)}
\newcommand{\be}{\begin{equation}} \newcommand{\ee}{\end{equation}}
\newcommand{\bea}{\begin{equation} \begin{aligned}} \newcommand{\eea}{\end{aligned} \end{equation}}
\newcommand{\bmu}{\begin{multline}} \newcommand{\emu}{\end{multline}}
\newcommand{\tabs}{\rule[-1ex]{0pt}{3.5ex}}

\newcommand{\cA}{\mathcal{A}}

\newcommand{\cF}{\mathcal{F}}

\newcommand{\cJ}{\mathcal{J}}

\newcommand{\cL}{\mathcal{L}}

\newcommand{\cN}{\mathcal{N}}
\newcommand{\cO}{\mathcal{O}}
\newcommand{\cP}{\mathcal{P}}
\newcommand{\cQ}{\mathcal{Q}}
\newcommand{\cR}{\mathcal{R}}

\newcommand{\bC}{\mathbb{C}}

\newcommand{\bH}{\mathbb{H}}

\newcommand{\bP}{\mathbb{P}}

\newcommand{\bR}{\mathbb{R}}
\newcommand{\bZ}{\mathbb{Z}}
\newcommand{\fg}{\mathfrak{g}}

\DeclareMathOperator{\Tr}{Tr}

\newtheorem{teo}{Theorem}

\begin{document}


\begin{titlepage}

\vspace*{2.3cm}

\begin{center}
{\LARGE \bf Two-dimensional SCFTs from wrapped branes \\[10pt] and $c$-extremization} \\

\vspace*{2cm}

{\bf Francesco Benini and Nikolay Bobev }
\bigskip
\bigskip

Simons Center for Geometry and Physics\\
Stony Brook University \\
Stony Brook, NY 11794-3636, USA  \\
\bigskip
fbenini@scgp.stonybrook.edu, nbobev@scgp.stonybrook.edu  \\
\end{center}

\vspace*{1.8cm}

\begin{abstract}

\noindent  We apply $c$-extremization \cite{Benini:2012cz}, whose proof we review in full detail, to study two-dimensional $\cN=(0,2)$ superconformal field theories arising from the low-energy dynamics of D3-branes wrapped on Riemann surfaces and M5-branes wrapped on four-manifolds. We compute the exact central charges of these theories using anomalies and $c$-extremization. In all cases we also construct AdS$_3$ supergravity solutions of type IIB and eleven-dimensional supergravity, which are holographic duals to the field theories at large $N$, and exactly reproduce the central charges computed via $c$-extremization.

\end{abstract}

\end{titlepage}


{\small
\setlength\parskip{-0.5mm}
\tableofcontents
}

\section{Introduction}

Conformal field theories (CFTs) in two spacetime dimensions play a central r\^ole in describing critical phenomena and appear prominently in string theory. In two dimensions the conformal group is infinite-dimensional, and similarly other symmetries usually form infinite-dimensional algebras. As a consequence such theories are tightly constrained and sometimes exactly solvable. In this paper we will be concerned with theories with at least $\cN=(0,2)$ superconformal symmetry:
they are interesting in their own right, play a pivotal r\^ole in type II and heterotic superstring compactifications \cite{Candelas:1985en, Gepner:1987qi, Banks:1987cy}, and are of some significance in mathematics because of their connection to holomorphic vector bundles on Calabi-Yau manifolds \cite{Witten:1993yc}.

The $\cN=(0,2)$ superconformal algebra contains an Abelian right-moving (\ie{} holomorphic) Kac-Moody current $\omega(z)$, called the R-symmetry current, under which the complex supercharge $\cQ$ is charged (we conventionally fix the charge to 1). This current is important because, for instance, it determines the exact dimension of chiral primary operators and the Virasoro right-moving central charge $c_R$ of the theory. In a non-conformal $\cN=(0,2)$ supersymmetric theory with an R-symmetry $U(1)_R$ and other Abelian flavor symmetries (under which the supercharges are not charged), the R-symmetry is not uniquely defined. Any mixing of $U(1)_R$ with the flavor symmetries, \ie{} any linear combination of those symmetries under which $\cQ$ has charge 1, produces an equally good R-symmetry. In fact the R-current is in the same supermultiplet as the stress tensor $T_{\mu\nu}$ and the supercurrent, and mixing corresponds to linear improvement transformations of the multiplet \cite{Dumitrescu:2011iu}. On the contrary if the theory flows to an IR fixed point, the superconformal R-symmetry \emph{is} singled out because improvement transformations would not preserve the tracelessness of $T_{\mu\nu}$ or current conservation equations, and it is a non-trivial task to determine it.

In \cite{Benini:2012cz} we found%
\footnote{Similar ideas were discussed in unpublished work by A.~Adams, D.~Tong and B.~Wecht \cite{ATW}.}
a general principle, dubbed $c$-extremization, which allows to identify the exact R-symmetry in a unitary superconformal field theory (SCFT) with normalizable vacuum. One considers all possible Abelian currents that assign charge 1 to $\cQ$:
$$
\Omega_\mu^\text{tr}(t) = J^\text{r}_\mu + \sum_{I \, (\neq\, \text{r})} t_I J_\mu^I \;,
$$
where $J^\text{r}_\mu$ is a choice of R-symmetry current and $J_\mu^I$ (with $I \neq \text{r}$) are all Abelian flavor symmetry currents. We will call $\Omega_\mu^\text{tr}$ the ``trial R-current''. Out of $\Omega_\mu^\text{tr}(t)$, one constructs the quadratic function $c_R^\text{tr}(t)$ which is proportional to the 't~Hooft anomaly of $\Omega_\mu^\text{tr}(t)$:
$$
c_R^\text{tr}(t) = 3 \Big( k^\text{rr} + 2 \sum_{I \,(\neq\, \text{r})} t_I k^{\text{r}I} + \sum_{I,J \, (\neq\,\text{r})} t_I t_J k^{IJ} \Big) \;,
$$
where $k^{IJ}$ are the 't~Hooft anomaly coefficients defined in (\ref{anomaly equations}), (\ref{anomaly coefficients weak coupling}) or (\ref{anomalies CFT}). The exact superconformal R-symmetry current is equal to  $\Omega_\mu^\text{tr}(t^*)$, where $t_I^*$ are the values of $t_I$ such that the function $c_R^\text{tr}(t)$ is extremized. Moreover the value $c_R^\text{tr}(t^*)$ at the critical point is equal to the right-moving central charge $c_R$ at the IR fixed point. The left-moving central charge is simply $c_L = c_R - k$, where $k$ is the gravitational anomaly coefficient in (\ref{anomaly equations}), (\ref{anomaly coefficients weak coupling}), (\ref{anomalies CFT}).

Two-dimensional $c$-extremization is similar in spirit to four-dimensional $a$-maximization \cite{Intriligator:2003jj}. In both cases the relevant function to be extremized is a polynomial constructed out of 't~Hooft anomalies. These anomalies are RG invariant and can be easily computed in the UV if the theory is weakly coupled. Given their topological nature they can often be computed even in theories without a Lagrangian description.%
\footnote{Two-dimensional $c$-extremization is also similar to three-dimensional $F$-maximization \cite{Jafferis:2010un, Closset:2012vg}, although in three dimensions there are no conformal and 't~Hooft anomalies and the computation of $F$ is harder.}

In this paper we consider various applications and checks of the $c$-extremization principle. In particular we consider examples of two-dimensional $\cN=(0,2)$ theories obtained by twisted compactifications of higher-dimensional field theories on compact manifolds. We study two classes of examples. First, we consider twisted compactifications of four-dimensional $\cN=4$ super-Yang-Mills (SYM) on Riemann surfaces \cite{Bershadsky:1995vm, Maldacena:2000mw}. The $c$-extremization principle allows for the computation of the IR R-symmetry and the central charges $c_R$ and $c_L$. For gauge group $U(N)$ such theories describe the low-energy dynamics of $N$ D3-branes wrapping a holomorphic two-cycle in a Calabi-Yau fourfold. At large $N$ they are described by dual type IIB supergravity solutions that we present in detail. The central charges can be computed holographically and exactly match with the result obtained by $c$-extremization.
As a by-product of this analysis, we answer a question raised by Almuhairi and Polchinski in \cite{Almuhairi:2011ws} about how to compute the central charges in the theories described above arising from genus-zero Riemann surfaces.

Another large class of examples of interesting two-dimensional SCFTs, for which we can apply and test $c$-extremization, is provided by twisted compactifications of the six-dimensional $\cN=(2,0)$ theory on four-manifolds. In this case a Lagrangian description of the six-dimensional theory is not known (nor do we find a Lagrangian description of the two-dimensional theories), but anomalies can nevertheless be extracted with the methods of \cite{Harvey:1998bx, Benini:2009mz, Alday:2009qq, Bah:2011vv, Bah:2012dg} and this provides enough information to apply $c$-extremization. For the six-dimensional theory of type $A_{N-1}$, the two-dimensional theories describe the low-energy dynamics of $N$ M5-branes on four-cycles inside non-compact special holonomy manifolds. We pay particular attention to four-cycles that are a product of closed Riemann surfaces inside Calabi-Yau fourfolds. We also discuss K\"ahler four-cycles in Calabi-Yau fourfolds, complex special Lagrangian four-cycles in hyper-K\"ahler fourfolds, as well as co-associative four-cycles in $G_2$-holonomy manifolds. At large $N$ these wrapped M5-branes are described by AdS$_3$ solutions of eleven-dimensional supergravity which are dual to the SCFTs in questions. Most of the solutions are new and we present them explicitly. We then use holography to extract the central charges of the dual SCFTs and the result again matches the one obtained by $c$-extremization.

We should notice that there is another class of interesting examples, \ie{} $\cN=(0,2)$ gauged linear sigma-models and Landau-Ginzburg models \cite{Witten:1993yc, Distler:1993mk, Silverstein:1994ih}, where the IR R-symmetry has been determined. In these cases $c$-extremization produces the same answer, and indeed an extremization principle was recognized in \cite{Melnikov:2009nh}.

The paper is organized as follows. In the next section we review the $c$-extremization principle of \cite{Benini:2012cz} and its proof in full detail. In Section \ref{sec: D3-branes} we study four-dimensional $\cN=4$ SYM on Riemann surfaces. We also present the supergravity solutions that are holographic duals to these theories. Section \ref{sec: vectors} is devoted to a discussion of some aspects of vector fields in AdS$_3$/CFT$_2$ that pertain to our examples. In Section \ref{sec: M5-branes} we study the six-dimensional $\cN=(2,0)$ theory on a product of closed Riemann surfaces and construct new supergravity solutions dual to the IR SCFTs. We discuss twisted compactifications on other four-manifolds in Section \ref{sec: other 4manifolds}. We briefly conclude with some open problems in the last section and present the details of various computations in the appendices.

\section{The superconformal R-symmetry in two dimensions}
\label{sec: c-ext}

In this section we study two-dimensional $\cN=(0,2)$ unitary SCFTs with normalizable vacuum and review the $c$-extremization principle that allows to identify the exact IR superconformal R-symmetry \cite{Benini:2012cz}. In Appendix \ref{app: free theories} we show how the method applies to free fields, and to what extent one needs to assume normalizability of the vacuum.

\subsection{Anomalies in two dimensions}
\label{subsec: 2danom}

Local quantum field theories in two dimensions suffer from gauge and gravitational anomalies, but not mixed gauge-gravitational anomalies \cite{AlvarezGaume:1983ig}. Consider a theory where $U(1)^N$ is the Abelian part of the internal continuous global symmetry group. This means that the theory has conserved current operators $J^I_\mu(x)$, with $I=1,\dots,N$, and a conserved stress tensor $T_{\mu\nu}(x)$. When the theory is coupled to non-dynamical (external) vector fields $A_\mu^I$ and to a curved background with metric $g_{\mu\nu}$, the anomalous violations of current conservations are
\be
\label{anomaly equations}
\nabla^\mu J^I_\mu = \sum_M \frac{k^{IM}}{8\pi} \, F_{\mu\nu}^M \varepsilon^{\mu\nu} \;,\qquad\qquad \nabla_\mu T^{\mu\nu} = \frac k{96\pi} \, g^{\nu\alpha} \varepsilon^{\mu\rho} \partial_\mu \partial_\beta \Gamma^\beta_{\alpha\rho} \;,
\ee
where $F_{\mu\nu}^I$ are the field strengths associated to $A_\mu^I$, $\Gamma^\beta_{\alpha\rho}$
is the Levi-Civita connection for $g_{\mu\nu}$ and $\varepsilon^{\mu\nu}$ is the covariant antisymmetric tensor in two dimensions. The coefficients $k^{IM}$ and $k$ are the 't~Hooft anomaly coefficients.%
\footnote{We stress that despite calling them ``gauge anomalies'', we always consider anomalies of global currents.}
We have chosen a renormalization scheme in which $k^{IM}$ is a symmetric matrix, the stress tensor is symmetric and local Lorentz rotations are non-anomalous.
The information in equations (\ref{anomaly equations}) can be encoded in the anomaly polynomial
\be
I_4 = \frac12 \sum_{I,M} k^{IM} c_1(F^I) \wedge c_1(F^M) - \frac{k}{24} \, p_1(R) \;,
\ee
where $c_1(F^I)$ are first Chern classes and $p_1(R)$ is the first Pontryagin class of the two-dimensional spacetime, constructed out of the field strengths $F_{\mu\nu}^I$ and the curvature two-form $R_{\mu\nu}$. $I_4$ is a formal four-form from which anomalies can be extracted through the descent formalism \cite{Wess:1971yu, AlvarezGaume:1983ig, AlvarezGaume:1983cs, AlvarezGaume:1984dr}. These standard results are reviewed in Appendix \ref{app: anomaly polynomial}.%
\footnote{A different, and perhaps more physical, way of understanding the r\^ole of the anomaly polynomial is that the partition function $Z(g_{\mu\nu}, A_\mu^I)$ of the theory in external background is a section of a line bundle, whose first Chern class is the integral of the anomaly polynomial on spacetime \cite{AlvarezGaume:1983ig, Witten:1985xe}. We thank Y. Tachikawa for discussions on this point.}

If the theory has a weakly coupled Lagrangian description, the coefficients $k^{IM}$ and $k$ receive contributions from chiral fermions and bosons and are computed exactly by one-loop diagrams with two current insertions.
Spin-$\frac12$ (complex) Weyl fermions contribute as
\be
\label{anomaly coefficients weak coupling}
k^{IM} = \Tr\limits_\text{Weyl fermions} \gamma^3 Q^I Q^M \;,\qquad\qquad k = \Tr\limits_\text{Weyl fermions} \gamma^3 \;,
\ee
where $\gamma^3$ is the two-dimensional chirality matrix which we take positive on right-movers, and $Q^I$ are the charge operators. Majorana-Weyl fermions contribute to $k$ as half of a Weyl fermion. Real chiral bosons contribute to $k$ as Weyl fermions, and, if linearly coupled to the vector fields as $\frac{Q^I}{\sqrt\pi} A^{I\mu} \partial_\mu \phi$, they also contribute to $k^{IM}$ as Weyl fermions. These standard results are presented in more detail in Appendix \ref{app: anomalies}.

Regardless of the existence of a weakly coupled description, the anomaly coefficients $k^{IM}$ and $k$ are well defined by the operator equations (\ref{anomaly equations}) and---as long as the symmetries are not broken---are invariant under RG flow \cite{tHooft:1980}. In fact there is no need of turning on external backgrounds: the anomaly coefficients can be reconstructed from the poles at zero momentum in the two-point functions $\langle T_{\mu\nu}(x) T_{\rho\sigma}(0)\rangle$ and $\langle J^I_\mu(x) J^M_\nu(0)\rangle$, because two-point functions on vanishing background are related to one-point functions at first order in the background.

If the theory is conformal, the anomaly coefficients $k^{IM}$ and $k$ are related to central terms in the conformal and current algebras (\ie{} in the OPEs) in flat space. When dealing with conformal theories in two dimensions it is convenient to work in Euclidean signature ($x^0 = ix^0_E$) and in radial quantization, using complex coordinates $z,\bar z$. We define
\be
z = x^1 + ix^0_E \;,\qquad \bar z = x^1 - i x^0_E \;,\qquad
\partial_z \equiv \partial = \frac{\partial_1 -i\partial_0^E}2 \;,\qquad \partial_{\bar z} \equiv \bar\partial = \frac{\partial_1 + i \partial_0^E}2 \;.
\ee
Following standard conventions (see \eg{} \cite{DiFrancesco:1997nk}) we define
\be
\label{T(z) def}
T(z) = -2\pi T_{zz}(x) \;, ~~~ \overline T(\bar z) = -2\pi T_{\bar z \bar z}(x) \;,~~~ j^I(z) = -i\pi J^I_z(x) \;,~~~ \bar\jmath^I(\bar z) = -i\pi J^I_{\bar z}(x) \;.
\ee
We will consider CFTs that fall in the following general class: \label{assumptions}
\begin{enumerate}
\item the theory is unitary and the Virasoro generators $L_0, \overline L_0$ are bounded below;
\item the vacuum is normalizable.
\end{enumerate}
Notable exceptions to the second condition are theories with non-compact free bosons that we discuss in Appendix \ref{app: free theories}. These assumptions lead to some standard properties \cite{Polchinski:1998rq} that will be crucial for us. First, in each conformal family there is a primary operator whose conformal weights $(\bar h, h)$ are non-negative.
Second, an operator $\cA$ is holomorphic ($\bar\partial \cA = 0$) if and only if $\bar h = 0$, and it is anti-holomorphic ($\partial\cA = 0$) if and only if $h=0$. The only $(0,0)$ operator is the identity.

In particular, conserved currents have dimension $h+\bar h =1$ and spin $|h - \bar h|=1$, therefore they are either holomorphic (right-moving) or anti-holomorphic (left-moving). Consider the conformal and current algebra OPEs:
\bea
\label{conformal + current OPEs}
T(z) \, T(0) &\sim \frac{c_R}{2z^4} + \frac{2T(0)}{z^2} + \frac{\partial T(0)}{z} \;, \qquad\qquad\qquad &
j^I(z) \, j^J(0) &\sim \frac{k_R^{IJ}}{z^2} \;,
\\
\overline T(\bar z) \, \overline T(0) &\sim \frac{c_L}{2\bar z^4} + \frac{2\overline T(0)}{\bar z^2} + \frac{\bar\partial \overline T(0)}{\bar z} \;, \qquad\qquad &
\bar\jmath^I(\bar z) \, \bar\jmath^J(0) &\sim \frac{k_L^{IJ}}{\bar z^2} \;,
\eea
where $\sim$ is the standard notation for equality up to regular terms. Unitarity constrains $k_R^{IJ}$ and $k_L^{IJ}$ to be semi-positive definite, and to vanish if and only if the global symmetry is trivial. The OPEs between holomorphic and anti-holomorphic fields vanish.
We then have (see Appendix \ref{app: anomalies}):
\be
\label{anomalies CFT}
k^{IJ} = \left\{ \begin{aligned} &k_R^{IJ} &&\text{if $I,J$ are both right-moving} \\ &-k_L^{IJ} \quad &&\text{if $I,J$ are both left-moving} \\ &0 &&\text{otherwise} \end{aligned} \right. \;,\qquad\qquad\qquad k = c_R - c_L \;.
\ee
If the central charges are equal, $c_R = c_L \equiv c$, there is no gravitational anomaly and one can put the theory on a curved background without breaking general covariance. In such renormalization scheme there is a conformal anomaly:
\be
T^\mu_\mu = - \frac c{24\pi} R \;.
\ee

We will be interested in superconformal theories with $\cN=(0,2)$ supersymmetry. Besides the stress tensor $T(z)$, $\overline T(\bar z)$, the superconformal algebra contains two holomorphic \mbox{spin-$\frac32$} operators $T_{F_{1,2}}(z)$ called supercurrents, usually written in terms of the combinations
\be
T_F^\pm(z) = \frac1{\sqrt2} \, \big( T_{F_1}(z) \pm i \, T_{F_2}(z) \big) \;,
\ee
and a holomorphic spin-1 operator $\omega(z)$ called R-symmetry current. Then the right-moving $\cN=2$ superconformal algebra is
\be
\label{N=2 SC algebra}
\begin{aligned}
T(z) \, T(0) &\sim \frac{c_R}{2z^4} + \frac{2T(0)}{z^2} + \frac{\partial T(0)}{z} \\
T(z) \, T_F^\pm(0) &\sim \frac{3T_F^\pm(0)}{2z^2} + \frac{\partial T_F^\pm(0)}{z} \\
T(z) \, \omega(0) &\sim \frac{\omega(0)}{z^2} + \frac{\partial \omega(0)}{z} \\
T_F^+(z) \, T_F^-(0) &\sim \frac{2c_R}{3z^3} + \frac{2\omega(0)}{z^2} + \frac{2T(0)+ \partial\omega(0)}{z}
\end{aligned}\qquad
\begin{aligned}
T_F^+(z) \, T_F^+(0) &\sim T_F^-(z) \, T_F^-(0) \sim 0 \\
\omega(z) \, T_F^\pm(0) &\sim \pm \frac{T_F^\pm(0)}{z} \\
\omega(z) \, \omega(0) &\sim \frac{c_R}{3z^2} \;,
\end{aligned}
\ee
while the left-moving part is as in (\ref{conformal + current OPEs}).
The algebra fixes a relation between the central charge $c_R$ and the R-symmetry anomaly:
\be
\label{central charge anomaly}
c_R = 3k^{RR} \;,
\ee
where $R$ is the value taken by the indices $I,J,\dots$ on the R-symmetry.

\subsection{The exact superconformal R-symmetry}
\label{subsec: exactR}

In a non-conformal $\cN=(0,2)$ supersymmetric theory with a $U(1)_R$ R-symmetry and other Abelian flavor symmetries, \ie{} symmetries under which the supercharges are neutral, the R-current is not uniquely defined. Any linear combination of the Abelian currents under which the supercharges have charge 1 is an equally good R-current. On the contrary in a superconformal theory the R-symmetry is singled out by the $\cN=2$ superconformal algebra.%
\footnote{Under the extra assumption that we adopt a renormalization scheme in which current conservation is not spoiled on a curved background and covariance is not spoiled on a gauge background.}

We would like to characterize the exact superconformal R-symmetry in terms of anomalies,
such that it is invariant under RG flow and independent of a detailed knowledge of the physics at the IR fix point. To this end, we consider a trial R-current $\Omega_\mu^\text{tr}$ constructed by taking a linear combination of the superconformal R-symmetry $\Omega_\mu$ and all Abelian flavor symmetries $J_\mu^I$:
\be\label{Omegatrdef}
\Omega_\mu^\text{tr}(t) = \Omega_\mu + \sum_{I \, (\neq R)} t_I J_\mu^I \;.
\ee
Recall that $\Omega_\mu$ is right-moving, and in line with (\ref{T(z) def}) we have $\omega(z) = -i\pi \Omega_z$.
Then we construct a trial central charge $c_R^\text{tr}(t)$ proportional to the 't~Hooft anomaly of the trial R-symmetry:
\be
c_R^\text{tr}(t) = 3\Big( k^{RR} + 2 \sum_{I\, (\neq R)} t_I k^{RI} + \sum_{I,J\, (\neq R)} t_I t_J k^{IJ} \Big) \;,
\ee
which could be extracted from the two-point function $\langle \Omega_\mu^\text{tr}(x) \, \Omega_\nu^\text{tr}(0)\rangle$, from (\ref{anomaly equations}) or (\ref{anomaly coefficients weak coupling}).

Let us study the constraints imposed on $k^{IJ}$ by superconformal symmetry.
If $J_\mu^I$ is a left-moving flavor current, then $k^{RI} = 0$ because $\Omega_\mu$ is right-moving and (\ref{anomalies CFT}).

If $J_\mu^I$ is a right-moving flavor current, it is part of a supermultiplet. The usual $\cN=(0,2)$ current multiplet $\cJ = (\phi, \psi^\pm, j)$ contains a real scalar, a complex Weyl fermion and a right-moving current.
However in a conformal theory with normalizable vacuum such a multiplet does not exist, because the scalar field should have weight $(0,0)$ and thus be the identity.%
\footnote{On the contrary in a theory with non-normalizable vacuum, a non-holomorphic current can be part of the multiplet $\cJ$. A simple example, discussed in Appendix \ref{app: free theories}, is given by a free non-compact chiral multiplet.}
Indeed the multiplet of $\cN=2$ Kac-Moody currents $\cJ^A_+ = (\psi^A_{1,2}, j^A_{1,2})$ is made of two $\cN=1$ current multiplets $(\psi_a^A, j^A_a)$ with $a=1,2$ \cite{Spindel:1988nh, Spindel:1988sr, Sevrin:1988ps, Kazama:1988qp, Kazama:1988uz}. Here the combined index $(A,a)$ runs over all right-moving flavor currents, covering a subset of the values of the index $I$. In superspace $\cJ^A_+$ is a holomorphic anti-chiral spinor superfield \cite{Hull:1989py}, satisfying $\overline\partial \cJ^A_+ = D_+ \cJ^A_+=0$.
In general there are constraints on the right-moving symmetry algebra (it has to admit a Manin triple \cite{Spindel:1988nh, Kazama:1988uz, Parkhomenko:1992dq, Getzler:1993py}), but in the Abelian case it only has to be even-dimensional. The Abelian current algebra is described by the OPEs
\be
\label{Abelian current algebra}
j_a^A(z) \, j_b^B(0) \sim \delta_{ab} \, \frac{k^{AB}}{z^2} \;,\qquad
j_a^A(z) \, \psi_b^B(0) \sim 0 \;,\qquad
\psi_a^A(z) \, \psi_b^B(0) \sim \delta_{ab} \, \frac{k^{AB}}z \;.
\ee
We have diagonalized the two-point function of the two currents in each $\cN=2$ multiplet (this is convenient, but not necessary), the fermionic two-point function then follows from Jacobi identities involving one supersymmetry generator. Unitarity constrains $k^{AB}$ to be positive definite. The action of supersymmetry is described by the OPEs \cite{Spindel:1988nh}
\bea
T_{F_1}(z) \, \psi^A_a(0) &\sim \frac{iq^A_a}{z^2} + \frac{j^A_a(0)}z \qquad\qquad &
T_{F_2}(z) \, \psi^A_a(0) &\sim - \varepsilon_{ab} \Big( \frac{iq^A_b}{z^2} + \frac{j^A_b(0)}z \Big) \\
T_{F_1}(z) \, j^A_a(0) &\sim \frac{\psi^A_a(0)}{z^2} + \frac{\partial\psi^A_a(0)}z \qquad\qquad &
T_{F_2}(z) \, j^A_a(0) &\sim \varepsilon_{ab} \Big( \frac{\psi^A_b(0)}{z^2} + \frac{\partial\psi^A_b(0)}z \Big)
\eea
where $\varepsilon_{12} = - \varepsilon_{21} = 1$. We have included the central terms $q^A_a$, compatible with Jacobi identities involving two supersymmetry generators. The action of the remaining superconformal generators (including $T_F^\pm$ for completeness) is fixed by Jacobi identities:
\bea
\label{OPEs SCCA}
T(z) \, \psi^A_a(0) &\sim \frac{\psi^A_a(0)}{2z^2} + \frac{\partial \psi^A_a(0)}z \qquad\qquad &
T(z) \, j^A_a(0) &\sim \frac{iq^A_a}{z^3} + \frac{j^A_a(0)}{z^2} + \frac{\partial j^A_a(0)}z \\
T_F^\pm(z) \, \psi^A_a(0) &\sim \frac{\delta_{ab} \mp i \varepsilon_{ab}}{\sqrt2} \, \Big( \frac{iq^A_b}{z^2} + \frac{j^A_b(0)}z \Big) \qquad &
T_F^\pm(z) \, j^A_a(0) &\sim \frac{\delta_{ab} \pm i \varepsilon_{ab}}{\sqrt2} \, \Big( \frac{\psi^A_b(0)}{z^2} + \frac{\partial\psi^A_b(0)}z \Big) \\
\omega(z) \, \psi^A_a(0) &\sim i \varepsilon_{ab} \frac{\psi^A_b(0)}z \qquad\qquad &
\omega(z) \, j^A_a(0) &\sim \varepsilon_{ab} \frac{q^A_b}{z^2} \;.
\eea
The expansion in terms of modes is given in Appendix \ref{app: mode algebra}. The central terms $q^A_a$ are called background charges \cite{Friedan:1985ge} and preserve superconformal symmetry.  Unitarity requires that $q^A_a$ are real.%
\footnote{A well-known application of background charges is to compute correlation functions in Virasoro minimal models from a free real scalar theory \cite{Dotsenko:1984nm}. In that case one turns on an imaginary background charge that explicitly breaks unitarity. Unitarity is then recovered for special values of the background charge such that the central charge coincides with the one of a Virasoro minimal model.}
Note also that the OPE $j^A_a(z) \, T^\pm_F(0)$ does not contain $1/z$ terms, as it should be if $j^A_a$ are flavor currents.

Because of the central terms $q^A_a$ in \eqref{OPEs SCCA}, and in particular in the OPE $T(z)\,  j^A_a(0)$, the currents $j^A_a$ are not primary operators. Taking expectation values, it follows that the time-ordered two-point function is $\langle T(p) \, j^A_a(-p)\rangle_T \sim p_+^2/p_-$. This leads to an anomalous violation of current conservation on a gravitational background and of covariance on a gauge background.
Since there are no mixed gauge-gravitational anomalies in two dimensions \cite{AlvarezGaume:1983ig}, the problem can be cured by the addition of local counterterms to the action while preserving the full superconformal symmetry.
From the point of view of the superconformal algebra, this corresponds to a redefinition of the triplet $(T, T_F^\pm, \omega) \to (T', T_F^{\pm\prime}, \omega')$ that preserves (\ref{N=2 SC algebra}) (up to a shift of the central charge $c_R$) but modifies (\ref{OPEs SCCA}) removing the background charges $q^A_a$. Most importantly for us, supersymmetry relates the various central terms in (\ref{OPEs SCCA}) so that the newly defined R-symmetry $\omega'$ has vanishing mixed gauge anomalies with all right-moving flavor currents $j^A_a$, as follows from the OPE $\omega'(z)\, j^A_a(0)$.

Let us now show how to redefine the stress tensor multiplet. Under the linear shift
\bea
T'(z) &= T(z) + i \alpha^A_a \, \partial j^A_a(z)\;, \\
T_F^{\pm\prime}(z) &= T_F^\pm(z) + \sqrt2\, i \alpha^A_a (\delta_{ab} \pm i \varepsilon_{ab}) \, \partial \psi^A_b(z)\;, \\
\omega'(z) &= \omega(z) + 2 \alpha^A_a \, \varepsilon_{ab} \, j^A_b(z)\;,
\eea
with $\alpha^A_a \in \bR$,
the algebra (\ref{N=2 SC algebra}), (\ref{Abelian current algebra}), (\ref{OPEs SCCA}) is preserved up to the shifts:
\be
q^{\prime A}_a = q^A_a - 2 k^{AB} \alpha^B_a \;,\qquad\qquad c_R' = c_R - 12 \alpha^A_a q^A_a + 12 \alpha^A_a \alpha^B_a k^{AB} \;.
\ee
Since $k^{AB}$ is positive definite, it is always possible to cancel all central terms in (\ref{OPEs SCCA}) by taking $\alpha^A_a = \frac12 (k^{-1})^{AB} q^B_a$. The function $c'_R(\alpha)$ is quadratic and with positive definite second derivative: in fact it is minimized%
\footnote{Since unitarity imposes a lower bound on the central charge and background charges do not break unitarity, the possibility that $c'_R(\alpha)$ is maximized (as opposed to minimized) would be inconsistent.}
precisely at the value of $\alpha^A_a$ for which all right-moving currents are primaries. The central charge at that point, $c_R^* = c_R - 3 q^A_a q^B_a (k^{-1})^{AB}$, is what is usually called the central charge of the theory, and it is constrained by unitarity to be positive. For this choice of $\alpha^A_a$ supersymmetry forbids mixed gauge anomalies between the superconformal R-current and the right-moving flavor currents.

We have proven that at the IR $\cN=(0,2)$ fixed point there are no mixed gauge anomalies between the superconformal R-current and flavor currents:
\be
\label{orthogonality}
k^{RI} = 0\;, \qquad\qquad \forall\, I \neq R \;.
\ee
This statement is true provided we work in a renormalization scheme in which there are no violations of current conservation on a gravitational background and of covariance on a gauge background. At the conformal fixed point this is equivalent to the vanishing of background charges, and to the condition that all flavor currents are primary fields, \ie{} the two-point functions of the stress tensor with flavor currents have no singular terms (apart from possible contact terms). The first condition of current conservation is RG-invariant, because it corresponds to the absence of extra terms in the operator equations (\ref{anomaly equations}), and can be imposed in the UV as well.
We can reformulate (\ref{orthogonality}) as an extremality condition for $c_R^\text{tr}(t)$. The current $\Omega_\mu^\text{tr}(t^*)$ in \eqref{Omegatrdef} is the superconformal R-current for real numbers $t_I^*$ such that
\be
\parfrac{c_R^\text{tr}}{t_I} (t^*) = 0\;, \qquad\qquad \forall\, I \neq R \;.
\ee
Since $c_R^\text{tr}(t)$ is a quadratic function, there is a unique solution $t_I^*$. Because of (\ref{anomalies CFT}) and unitarity, the function is actually maximized along directions $t_I$ that correspond to left-moving currents $J_\mu^I$, and minimized along right-moving ones.
Finally, we note that assumption 2. discussed on page \pageref{assumptions} might be relaxed if we are careful enough not to mix the R-symmetry with non-(anti)holomorphic currents.

\section{Four-dimensional $\cN=4$ SYM on Riemann surfaces}
\label{sec: D3-branes}

Here we study the two-dimensional theories arising at low energy from the (twisted) compactification of four-dimensional $\cN=4$ SYM with gauge group $G$ on a Riemann surface $\Sigma_\fg$ of genus $\fg$.%
\footnote{A description of these theories in the $\cN=(4,4)$ case as non-linear sigma models has been given in \cite{Bershadsky:1995vm}.} We then utilize $c$-extremization to determine the central charges of the IR two-dimensional CFTs.
For gauge group $U(N)$, these theories also arise from $N$ D3-branes wrapped on $\Sigma_\fg$.
To preserve some supersymmetry generically one has to twist the theory, \ie{} to turn on a background gauge field $A_\mu$ coupled to the $SO(6)$ R-symmetry of the four-dimensional theory. The supercharges transform in the representation $\rep{2}\otimes \rep{4}$ of the product of the Lorentz and R-symmetry groups $SO(3,1)\times SO(6)$, therefore one can preserve (at least) 2d $\cN=(0,2)$ supersymmetry by choosing an appropriate Abelian background $A_\mu$.%
\footnote{We will mostly follow the notation and conventions of \cite{Maldacena:2000mw}.}
In terms of the spin connection $\tilde\omega_\mu \equiv \frac12 \omega_\mu^{ab} \varepsilon_{ab}$ on $\Sigma_\fg$, such that $\int \! d\tilde \omega = \frac12 \int \! \sqrt{g} \,R = 4\pi(1-\fg)$, the covariant derivative of a spin-$s$ field is $D_\mu = \partial_\mu + is\tilde\omega_\mu - i A_\mu$.
For $\fg=0$ we choose the 4d R-symmetry background $A_\mu$ such that its field strength is $F = -T \, d\tilde\omega$, for $\fg=1$ we set $F = - T \frac{2\pi}{\vol(\Sigma)}\, \dvol_\Sigma$ (proportional to the volume form), and for $\fg>1$ we set $F = T \, d\tilde\omega$.
The background is taken along the generator
\begin{equation}
T = a_1 T_1 + a_2T_2 + a_3 T_3 \;,
\end{equation}
where $T_{1,2,3}$ are generators of an $SO(2)^3$ embedded block-diagonally into $SO(6)$, and $a_{1,2,3}$ are constants parametrizing the twist (we will find below that $2(\fg-1)a_I \in \bZ$ for $\fg\neq 1$, and $a_I \in \bZ$ for $\fg=1$). To preserve 2d $\cN=(0,2)$ supersymmetry we take
\be
\label{SUSY condition a's}
a_1 + a_2 + a_3 = -\kappa \;,
\ee
where $\kappa=1$ for $\fg=0$, $\kappa=0$ for $\fg=1$ and $\kappa=-1$ for $\fg>1$ (see Appendix \ref{app: SUSY N=4 SYM} for a more detailed discussion).
The meaning of these definitions is that if we choose the metric on $\Sigma_\fg$ to be of constant curvature $R = 2\kappa$
\be
ds^2_\Sigma = e^{2h} (dx^2 + dy^2) \qquad\text{with}\qquad h = \begin{cases} -\log \frac{1 + x^2 + y^2}2 & \text{for } \fg = 0 \\ \frac12 \log 2\pi & \text{for } \fg = 1 \\ - \log y & \text{for } \fg >1\end{cases} \;,
\ee
then the background flux is $F = dA = \sum_{I=1,2,3} F^I T_I$, with $F^I = - a_I \, e^{2h} \, dx \wedge dy$. However we will not commit to a particular metric for now.

For generic $a_I$'s we get $\cN=(0,2)$ supersymmetry, when one of the $a_I$'s is zero we get $\cN=(2,2)$, when two are zero (and $\fg\neq 1$) we get $\cN=(4,4)$ and when all are zero (and $\fg=1$) we get $\cN=(8,8)$ (see Appendix \ref{app: SUSY N=4 SYM}).

The construction has a more geometric character visible in string theory. For gauge group $U(N)$, the twisted theories describe the low-energy dynamics of $N$ D3-branes on a holomorphic 2-cycle  $\Sigma_\fg$ in a local (non-compact) Calabi-Yau fourfold $X$, which is an $\cL_1 \oplus \cL_2 \oplus \cL_3$ line bundle over $\Sigma_\fg$:
$$
\bC^{(1)}\times \bC^{(2)}\times \bC^{(3)} \;\to\; X \;\to\; \Sigma_g \;.
$$
The degree of the line bundle $\cL_I$ is $-2\kappa(\fg-1)a_I$ for $\fg\neq 1$, and $a_I$ for $\fg=1$. The condition that $X$ is Calabi-Yau is \eqref{SUSY condition a's}. When one of the $a_I$'s vanishes then $X = CY_3 \times \bC$ and supersymmetry is enhanced to $\cN=(2,2)$; when two of the $a_I$'s vanish then $X = CY_2 \times \bC^2$ and supersymmetry is enhanced to $\cN=(4,4)$. In the language of wrapped branes the details of the partial topological twist are encoded in the normal geometry to the cycle. The brane picture also facilitates the construction of the supergravity duals discussed below.

The low-energy 2d theory inherits $SO(2)^3$ global symmetry%
\footnote{For special values of $a_I$ the global symmetry is enhanced to a non-Abelian symmetry of the same rank.}
which contains the (yet unknown) 2d superconformal R-symmetry. The trial R-symmetry is a linear combination of the generators of $SO(2)^3$:
\be
\label{TRdef}
T_R = \epsilon_1 T_1 + \epsilon_2 T_2 + (2-\epsilon_1-\epsilon_2) T_3 \;,
\ee
where $\epsilon_{1,2}$ parametrize the mixing, and the R-charge of the 2d supercharges has been fixed to 1. To determine the correct superconformal R-symmetry we can use $c$-extremization, and for that we need the two-dimensional anomalies.

The four-dimensional theory has gaugini in the representation $\rep{2} \otimes \rep{\overline4}$ of $SO(3,1) \times SO(6)$, and the number of 2d massless chiral fermions follows from Kaluza-Klein reduction on $\Sigma_\fg$. The gaugini decompose under $SO(2)^3$ into chiral spinors of charges $A : (-\frac12,-\frac12,-\frac12)$, $B :(\frac12,\frac12,-\frac12)$, $C : (\frac12,-\frac12,\frac12)$, $D:(-\frac12,\frac12,\frac12)$, and by the index theorem the number of right-moving minus left-moving fermions is
\begin{equation}
n_R^{(\sigma)} - n_L^{(\sigma)} = \frac1{2\pi} \int \Tr_\sigma F = - t_\sigma \, \eta_{\Sigma} \;,
\end{equation}
where $\sigma = \{A,B,C,D\}$, $\Tr_\sigma$ is taken in the representation $\sigma$, and $t_\sigma$ are the charges of the fermions in that representation under $A_\mu$, namely $t_A = \frac\kappa2$, $t_B = \frac\kappa2 + a_1 + a_2$, $t_C = -\frac\kappa2 - a_2$, $t_D = - \frac\kappa2 - a_1$. We have also defined
\be
\label{definition eta_Sigma}
\eta_\Sigma = \left\{ \begin{aligned}
&2 |\fg-1| \qquad &&\text{for }\fg \neq 1 \\
&1 &&\text{for } \fg=1 \;,
\end{aligned} \right.
\ee
which equals $\eta_\Sigma = \frac1{2\pi} \int_\Sigma e^{2h} dx\,dy$ in the case of constant curvature metric.
Because of the fermions, for $\fg\neq 1$ well-definiteness of the R-symmetry bundle requires $2(\fg-1)(\pm a_1 \pm a_2 \pm a_3) \in 2\bZ$, which supplemented by the supersymmetry condition (\ref{SUSY condition a's}) is equivalent to $2(\fg-1)a_I \in \bZ$. For $\fg=1$ we simply get $a_I \in \bZ$. We can write the quantization conditions more concisely as:
\be
a_I \, \eta_\Sigma \in \bZ \;.
\ee
Geometrically these conditions correspond to quantization of the fluxes $F^I$, as well as to well-definiteness of the normal bundle to the D3-branes.

The trial right-moving central charge can be computed from (\ref{anomaly coefficients weak coupling}) and (\ref{central charge anomaly}), taking into account that the fermions are in the adjoint representation of the gauge group:
\begin{equation}
c_R^\text{tr}(\epsilon_i) = 3 d_G \sum\nolimits_\sigma \big( n^{(\sigma)}_R - n_{L}^{(\sigma)} \big) \big( q_R^{(\sigma)}(\epsilon_i) \big)^2 \;,
\end{equation}
where $d_G$ is the dimension of the gauge group $G$ and $q_R^{(\sigma)}$ are the charges under $T_R$. We find:
\be
\label{cR trial D3-branes}
c_R^\text{tr} = - 3 \eta_{\Sigma} \, d_G \big( t_A + t_B (\epsilon_1 + \epsilon_2 -1)^2 + t_C (1- \epsilon_2)^2 + t_D (1-\epsilon_1)^2 \big) \;.
\ee
For later convenience we define the following combinations of parameters
\bea
\label{combinations of parameters}
\Theta &= a_1^2 + a_2^2 + a_3^2 - 2(a_1a_2 + a_1a_3 + a_2a_3)\;, \\
\Pi &= (-a_1 + a_2 + a_3)(a_1 - a_2 + a_3)(a_1 + a_2 - a_3) \;,
\eea
which are symmetric under permutations of $a_I$. As a function of $\epsilon_{1,2}$, the trial central charge $c_R^\text{tr}$ is extremized for
\be
\epsilon_i = \frac{2a_i(2a_i+\kappa)}\Theta \qquad\qquad i=1,2 \;,
\ee
as long as $\Theta \neq 0$, and at the critical point it takes the value
\be
\label{cRexact}
c_R = - 12 \eta_{\Sigma} \, d_G \, \frac{a_1a_2a_3}\Theta \;.
\ee
The second derivatives of $c_R^\text{tr}$ are
\be
\label{Hessian}
\partial_{\epsilon_i}^2 c_R^\text{tr} = - 6 \eta_{\Sigma} d_G\, a_1a_2 /a_i \;,\qquad\qquad
\partial_{\epsilon_1}\partial_{\epsilon_2} c_R^\text{tr} = - 3 \eta_{\Sigma} d_G (a_1 + a_2 - a_3 ) \;,
\ee
and the Hessian determinant is: $\det_{ij} \partial_{\epsilon_i} \partial_{\epsilon_j} c_R^\text{tr} = -9 \eta_\Sigma^2 d_G^2 \Theta$. Finally we can compute
\be
c_R - c_L = k = d_G \sum\nolimits_\sigma(n^{(\sigma)}_R - n_L^{(\sigma)}) = 0\;,
\ee
therefore there is no gravitational anomaly. The full matrix of anomalies is
\be
\label{anomaly matrix D3-branes}
k^{IJ} = d_G \sum\nolimits_\sigma \big( n_R^{(\sigma)} - n_L^{(\sigma)} \big) q_I^{(\sigma)} q_J^{(\sigma)} = \frac{\eta_\Sigma\, d_G}2 \mat{ 0 & a_3 & a_2 \\ a_3 & 0 & a_1 \\ a_2 & a_1 & 0}\;, \qquad \text{with $I,J=1,2,3$\;.}
\ee

There are regions in the parameter space of $a_I$'s, with (\ref{SUSY condition a's}) taken into account, where $c_R$ turns out not to be a positive number. This implies that either our assumptions are not valid, \eg{} the ground state is non-normalizable, or that there is no IR fixed point. On the other hand, finding $c_R>0$ does not guarantee the existence of an IR fixed point. With this in mind we postpone the study of the parameter space to Section \ref{sec: sugra solutions} where we shall analyze well-definiteness of the dual supergravity solutions.

The formul\ae{} above for the critical point are not valid when $\fg>1$ and $a_1 = a_2 = \frac12$, $a_3 = 0$ (or permutations thereof) because $\Theta = 0$.%
\footnote{The expressions are not valid on the whole locus $\Theta = 0$. Only the three cases considered in the main text correspond to a good supergravity solution (see Figure \ref{fig: good SUGRA}), therefore we will not study the more general case further.}
In this case, considered in \cite{Maldacena:2000mw}, $c_R^\text{tr}$ is maximized for any $\epsilon_1 + \epsilon_2 = 1$, \ie{} there is a flat direction (for $a_1 = a_3 = \frac12$ it is maximized along $\epsilon_2 = 1$, and for $a_2=a_3 = \frac12$ it is maximized along $\epsilon_1=1$). This is because the symmetry generated by $T_1 - T_2$ (by $T_1$ and by $T_2$ in the other cases) is non-anomalous, and unitarity implies---under our assumptions---that the symmetry is trivial in the IR (no fields are charged under it). Indeed, in this case supersymmetry is enhanced to $\cN=(2,2)$ and the UV global symmetry is enhanced to $SU(2) \times U(1)_L \times U(1)_R$. The whole $SU(2)$, whose Cartan generator is $T_1-T_2$, is trivial in the IR while the two $U(1)$'s are the left and right R-symmetries. As we will review in Section \ref{sec: vectors}, the IR decoupling of the $SU(2)$ symmetry has a nice counterpart in supergravity. From $c$-extremization we obtain that $U(1)_R$ is generated by $T_R = \frac12 (T_1 + T_2) + T_3$ and
\be
\label{cRexact 22}
c_R = c_L = 3 (\fg -1) \, d_G \;.
\ee
It is easy%
\footnote{One imposes $T_L = \delta_1 T_1 + \delta_2 T_2 + (2+\delta_1 + \delta_2) T_3$, then $c_L^\text{tr}(\delta_i)$ is minimized for $\delta_1 + \delta_2 = -1$.}
to repeat $c$-extremization for the left-moving trial central charge $c_L^\text{tr}$, obtaining the same central charges as above and that $U(1)_L$ is generated by $T_L = -\frac12(T_1 + T_2) + T_3$. We can finally compute the matrix $k^{IJ}$ in the basis $(T_R,\, T_L,\, T_1 - T_2)$ finding, as expected, that it is diagonal with a positive, a negative and a vanishing entry. Moreover we notice that the central charges in \eqref{cRexact 22} are integer multiples of 3, and it would be interesting to understand if the SCFTs are non-linear sigma models on Calabi-Yau target spaces of complex dimension $(\fg-1)d_G$.%
\footnote{Note that this $\cN=(2,2)$ SCFT is not the same as the $\cN=(2,2)$ SCFTs obtained in \cite{Kapustin:2006hi} by twisted compactifications of general four-dimensional $\cN=2$ SCFTs on Riemann surfaces.}

\label{(4,4) from D3s}
Another noteworthy case is $\fg>1$, $a_1 = a_2 = 0$ and $a_3=1$ (or permutations thereof), so that the twist is as in \cite{Bershadsky:1995vm, Maldacena:2000mw} and supersymmetry is enhanced to $\cN=(4,4)$. In \cite{Bershadsky:1995vm} the IR 2d field theory was identified as a non-linear sigma model on the Hitchin moduli space on $\Sigma_\fg$ for group $G$, whose complex dimension is $2 d_G (\fg-1)$. The target space is non-compact and the vacuum is non-normalizable (also there is no dual AdS$_3$ supergravity solution), therefore applying $c$-extremization is problematic.
The twist preserves an $SU(2)_L \times SU(2)_R \times U(1)$ global symmetry, where $SU(2)^2$ is part of the superconformal algebra and is Kac-Moody. We can use an $\cN=(0,2)$ subalgebra of the full superconformal algebra, in which the right-moving R-symmetry is generated by the Cartan $T_R = T_1 + T_2$ of $SU(2)_R$, and use (\ref{cR trial D3-branes}) to compute the central charge $c_R = 6 d_G (\fg-1)$. This is  consistent with the dimension of the target \cite{Maldacena:2000mw}.
On the other hand, the $c$-extremization formula (\ref{cRexact}) would give vanishing central charge, which is clearly the wrong result. This must be due to the non-normalizability of the vacuum and the presence of a non-holomorphic current invalidating the procedure. The only candidate is the $U(1)$ symmetry, generated by $T_3$. Indeed from $k^{IJ}$ in (\ref{anomaly matrix D3-branes}) we see that this $U(1)$ is non-anomalous, but cannot be trivial because $c_R^\text{tr}$ varies as the trial R-symmetry is mixed with this $U(1)$. Thus we are lead to the conclusion that the current corresponding to $T_3$ is non-holomorphic. One can check that $c_R^\text{tr}$ is not extremized at $\epsilon_1=\epsilon_2=1$ precisely along the direction of $T_3$. This behavior is very similar to the case of a non-compact free boson, discussed in Appendix \ref{app: free theories}.

For fixed gauge group and $\fg>1$, there is a special twist that minimizes the central charge $c_R$ as a function of the twist parameters $a_I$: it is $a_1=a_2=a_3=\frac13$ with central charge%
\footnote{Notice that these values of $a_I$ are allowed only when $\fg - 1$ is a multiple of $3$ and therefore the theories are well-defined only for such Riemann surfaces. We can relax this condition on $\fg$ if we study a $\mathbb{Z}_3$ orbifold of $\mathcal{N}=4$ SYM.}
\begin{equation}
c_R = \frac83 (\fg-1) d_G \;.
\end{equation}
The twist preserves $SU(3)\times U(1)_R$ global symmetry, and it corresponds to a local $CY_4$ for which the degrees of the three line bundles are equal.

It is worth pointing out that the theories obtained by putting $\mathcal{N}=4$ SYM on $T^2$ and turning on background gauge fields to break supersymmetry to $\cN=(0,2)$ were studied in \cite{Almuhairi:2011ws}. The authors of  \cite{Almuhairi:2011ws} observed a mismatch between a weakly coupled field theory calculation of the central charge and the supergravity result. Computing the central charges using anomalies and $c$-extremization we see that the result in \eqref{cRexact} (with $\kappa=0$) reproduces the supergravity result  for the central charges. We believe that this resolves the puzzle.

We now proceed to construct the AdS$_3$ type IIB supergravity solutions for generic $a_I$, dual to the twisted compactifications of $\cN=4$ $SU(N)$ SYM at large $N$. As we discuss below their central charges agree with (\ref{cRexact}) and (\ref{cRexact 22}) at leading order in $N$.

\subsection{AdS$_3$ vacua from D3-branes}
\label{sec: sugra solutions}

To study holographically the twisted $\cN = 4$ $SU(N)$ SYM on Riemann surfaces at large $N$ we use the approach pioneered in \cite{Maldacena:2000mw}. One can show that to construct the IIB supergravity solutions that describe the backreaction of the wrapped D3-branes one can use the maximal gauged supergravity in five dimensions \cite{Gunaydin:1984qu, Pernici:1985ju, Gunaydin:1985cu}, which is a consistent truncation of IIB supergravity on $S^5$. In fact it is sufficient to work with a simple consistent truncation of the maximal five-dimensional theory which consists of the metric, three Abelian gauge fields $A^{I}_{\mu}$ in the Cartan of $SO(6)$ and two neutral scalars $\phi_1$ and $\phi_2$. This is the same truncation as in \cite{Maldacena:2000mw} and it is sometimes referred to as the STU model. We are interested in solutions of the form
\bea
\label{gravity ansatz}
ds^2_5 &= e^{2f(r)} (-dt^2+dz^2+dr^2)  + e^{2g(r)+2h(x,y)} (dx^2+dy^2) \\
F^I &= - a_I \, e^{2h(x,y)} \, dx \wedge dy \;, \qquad\qquad I=1,2,3\;,
\eea
and with $\phi_{1,2}(r)$ functions only of $r$. We consider closed Riemann surfaces $\Sigma_\fg$ with constant curvature metrics, described by the function $h(x,y)$:
\be
\label{metric function h}
h(x,y) = \begin{cases} - \log \frac{1+x^2+y^2}2 \qquad &\text{for } S^2 \\
\frac12 \log 2\pi &\text{for } T^2 \\      - \log y &\text{for } \mathbb{H}^2 \;,    \end{cases}
\ee
where for $\fg=1$ the range of coordinates is $x,y\in[0,1)$. Notice that $\frac1{2\pi} \int_\Sigma e^{2h} dx\,dy = \eta_\Sigma$, and $F^I$ are identified with the $SO(2)^3$ field strengths in the previous section.
The supersymmetry transformations of fermionic fields in 5d gauged supergravity are given in Appendix \ref{app: 5D sugra}. For the Ansatz of interest, the BPS equations reduce to:
\bea
\label{BPS conditions 5Dsugra}
0 &= g'+ e^f (X^{1}+X^{2}+X^{3})/3 - e^{f-2g}a_IX_{I} \\
0 &= f' + e^f (X^{1}+X^{2}+X^{3})/3 + e^{f-2g}a_I X_I /2 \\
0 &= \phi_1' + \sqrt6 \, e^f (X^{1}+X^{2}-2X^{3})/3 +  \sqrt6\, e^{f-2g} (a_1X_{1}+a_2X_{2}-2a_3X_{3})/2 \\
0 &= \phi_2' + \sqrt2\, e^f (X^{1}-X^{2}) + 3\sqrt{2}\, e^{f-2g}(a_1X_{1}-a_2X_{2})/2 \\
0 &= a_1 + a_2 + a_3 + \kappa
\eea
where $X^I$ and $X_I$ are functions of the two scalars and are defined in (\ref{5d gauged sugra definitions}).
To find AdS$_3$ fixed points we take $f(r) = f_0 - \log r$ and constant $g, \phi_1, \phi_2$.
The solution (see Appendix \ref{app: 5D sugra} for a detailed derivation) is then
\bea
\label{final sugra solution}
e^{6g} &= \frac{a_1^2 \, a_2^2 \, a_3^2}\Pi &
e^{\sqrt{6} \, \phi_1} &= \frac{a_3^2 (a_1 + a_2 - a_3)^2}{a_1 a_2 (-a_1+a_2+a_3)(a_1-a_2+a_3)}  \\
e^{3f_0} &= - \frac{8 \, a_1 a_2 a_3 \Pi}{\Theta^3} \qquad\qquad\qquad &
e^{\sqrt{2}\, \phi_2} &= \frac{a_2\, (a_1 - a_2 + a_3)}{a_1\, (-a_1 + a_2 + a_3)}\;,
\eea
in terms of the combinations in (\ref{combinations of parameters}), and one should remember to impose (\ref{SUSY condition a's}). Some of these solutions with $\kappa\neq 0$ were also obtained in \cite{Cucu:2003bm,Cucu:2003yk}. The solutions with $\kappa=0$ were studied in \cite{Almuhairi:2011ws} (see also \cite{Gauntlett:2007sm}), and the special solution with $a_1=a_2=a_3=1/3$ was discussed in \cite{Naka:2002jz, Gauntlett:2006qw}.

The central charge at leading order in $N$ can be computed using standard holographic techniques \cite{Brown:1986nw, Henningson:1998gx}:
\be
c_R = \frac{3R_{AdS_3}}{2G_{N}^{(3)}}= 6 \eta_\Sigma \, e^{f_0+2g} N^2 = - 12 \eta_{\Sigma} N^2 \, \frac{a_1a_2a_3}\Theta \;.
\ee
This agrees with the field theory calculation in \eqref{cRexact} performed using $c$-extremization.

\begin{figure}[t]
\begin{center}
\includegraphics[height=.30\textwidth]{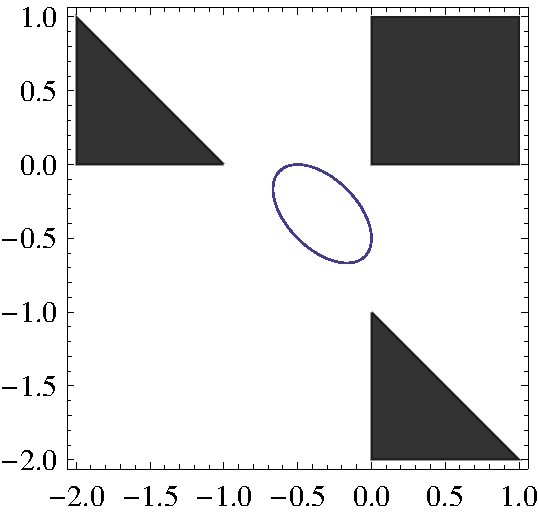}
\hfill
\includegraphics[height=.30\textwidth]{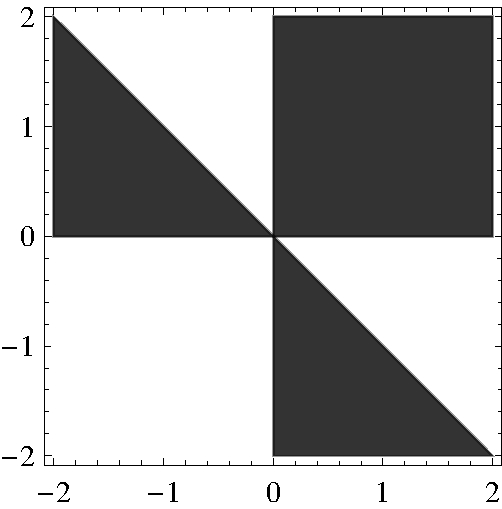}
\hfill
\includegraphics[height=.295\textwidth]{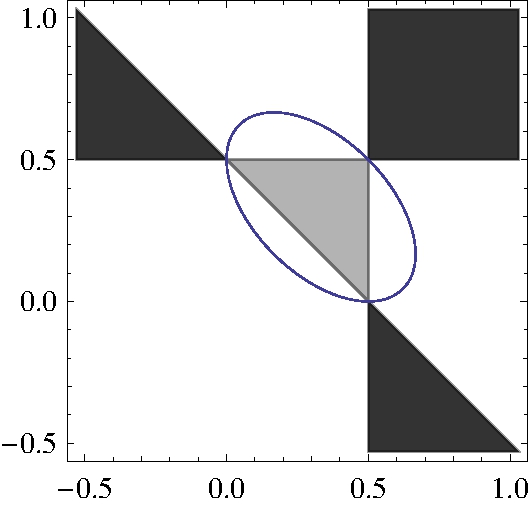}

\caption{Regions of the parameter space $(a_1,a_2)$ (with $a_3 = -a_1-a_2 - \kappa$) where there exist good AdS$_3$ vacua, for genus $\fg=0,1,>1$ respectively. The boundaries are excluded, with the only exception of the three points where the regions touch (together with the circle) in the case $\fg>1$. The blue circle for $\fg=0, >1$ is the $\Theta=0$ locus; for $\fg=1$ it is collapsed at the origin.
\label{fig: good SUGRA}}
\end{center}
\end{figure}

In Appendix \ref{app: 5D sugra} we study the regions in the parameter space of $a_I$'s with (\ref{SUSY condition a's}) where the supergravity solution is well-defined (and in particular the central charge is positive). The result, displayed in Figure \ref{fig: good SUGRA}, is the following: for $\fg=0,1$ there are good AdS$_3$ solutions when two of the parameters obey $a_I > 0$; for $\fg>1$ one needs two of the $a_I > 1/2$, or all three of them to be $a_I < 1/2$. We can thus conclude that for this range of $a_I$ and at large $N$, $\cN=4$ SYM flows to a 2d IR fixed point and the AdS$_3$ supergravity solution is the holographic dual to a normalizable ground state.

In Appendix \ref{app: 5D sugra} we study the signature of the Hessian, which determines the chirality of the currents. In addition to that we study the matrix of Chern-Simons couplings in the effective three-dimensional supergravity on AdS$_3$, and find that it is proportional to the matrix of anomalies $k^{IJ}$ in (\ref{anomaly matrix D3-branes}). As we explain in Section \ref{sec: vectors} this is expected. Moreover for $a_1 = a_2 = \frac12$, $a_3 = 0$ (and permutations) one of the bulk gauge fields does not have a CS term (it does have a Yang-Mills kinetic term). As explained in Section \ref{sec: vectors}, a bulk gauge field with YM kinetic action is not dual to a boundary current, but rather to a path-integrated boundary gauge field (with no kinetic term) that quotients the boundary theory by a vector-like symmetry. This is dual to the fact, discussed above (\ref{cRexact 22}), that in field theory a global symmetry becomes trivial in the IR.

\subsection{Holographic RG flows}
\label{subsec: holoRGD3}

We can study holographic RG flows interpolating between the asymptotically locally
$AdS_5 \times S^5$ solution in the UV and the AdS$_3$ fixed points presented in the previous section in the IR.\footnote{We will assume that the metric on the Riemann surface is the constant curvature one along the whole flow. We expect that when one allows for a more general conformal factor on the Riemann surface the conclusions of \cite{Anderson:2011cz} will still hold.} To construct the supergravity solutions for the RG flows it will be convenient to define a new radial variable
\be
\label{new radial variable}
\rho = f + \frac{1}{2\sqrt{6}} \phi_1+ \frac{1}{2\sqrt{2}} \phi_2 \;,
\ee
such that $\frac{d\rho}{dr} = -e^{f}D$ with
\be
D \equiv X^1 + \frac{3a_1}{2}e^{-2g} X_1 \;.
\ee
One gets $e^{2f}dr^2 = d\rho^2/D^2$ in the metric \eqref{gravity ansatz}.
The system of BPS equations (\ref{BPS conditions 5Dsugra}) then becomes:
\bea
\label{eqn D3 flow}
0 &= \frac{dg}{d\rho} - \frac1D \Big( \frac{X^1 + X^2 + X^3}3 - e^{-2g} a_I X_I \Big) \\
0 &= \frac{d\phi_1}{d\rho} - \frac{\sqrt6}D \Big( \frac{X^1 + X^2 - 2X^3}3 + e^{-2g} \frac{a_1X_1 + a_2X_2 - 2a_3X_3}2 \Big) \\
0 &= \frac{d\phi_2}{d\rho} - \frac{\sqrt2}D \Big( X^1 - X^2 + 3 e^{-2g} \frac{a_1X_1 - a_2X_2}2 \Big) \;,
\eea
and the function $f$ does not appear explicitly anymore. Once (\ref{eqn D3 flow}) has been solved, $f(\rho)$ is determined by (\ref{new radial variable}).

\begin{figure}[t]
\begin{center}
\includegraphics[width=.32\textwidth]{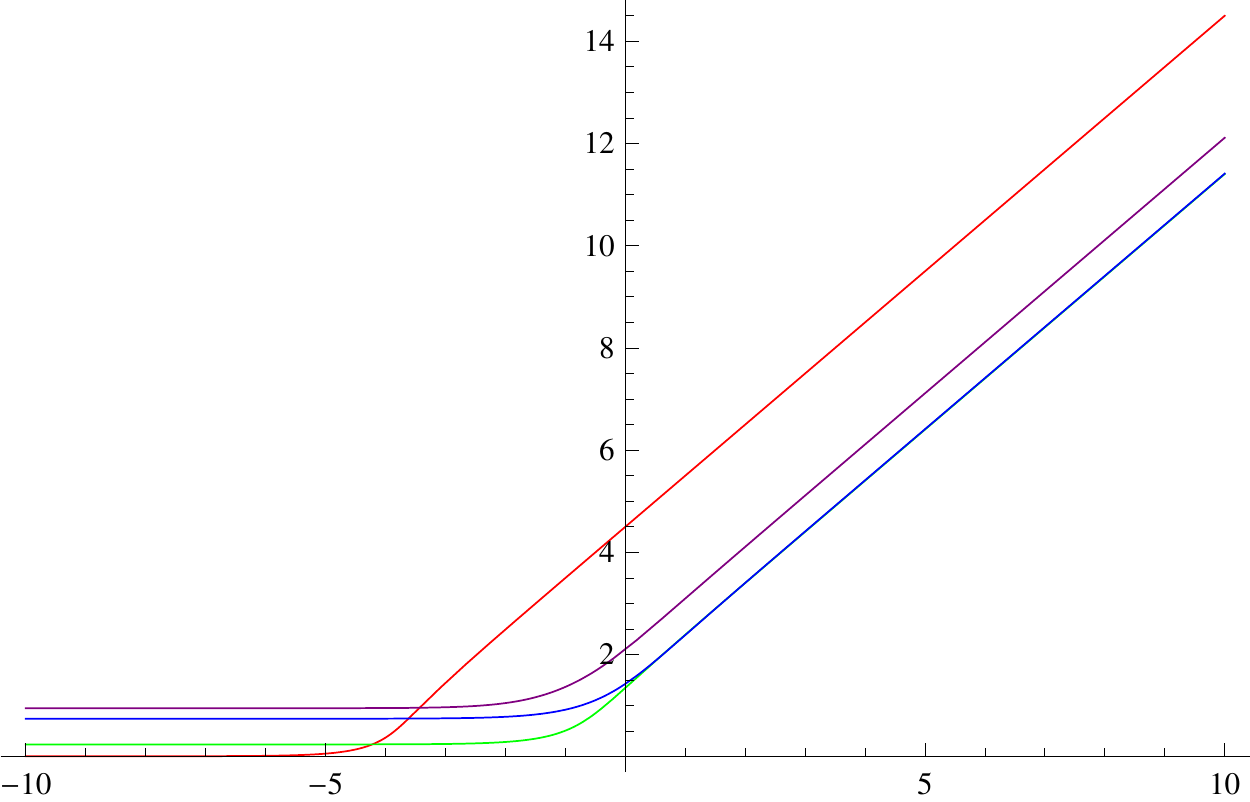}
\hfill
\includegraphics[width=.32\textwidth]{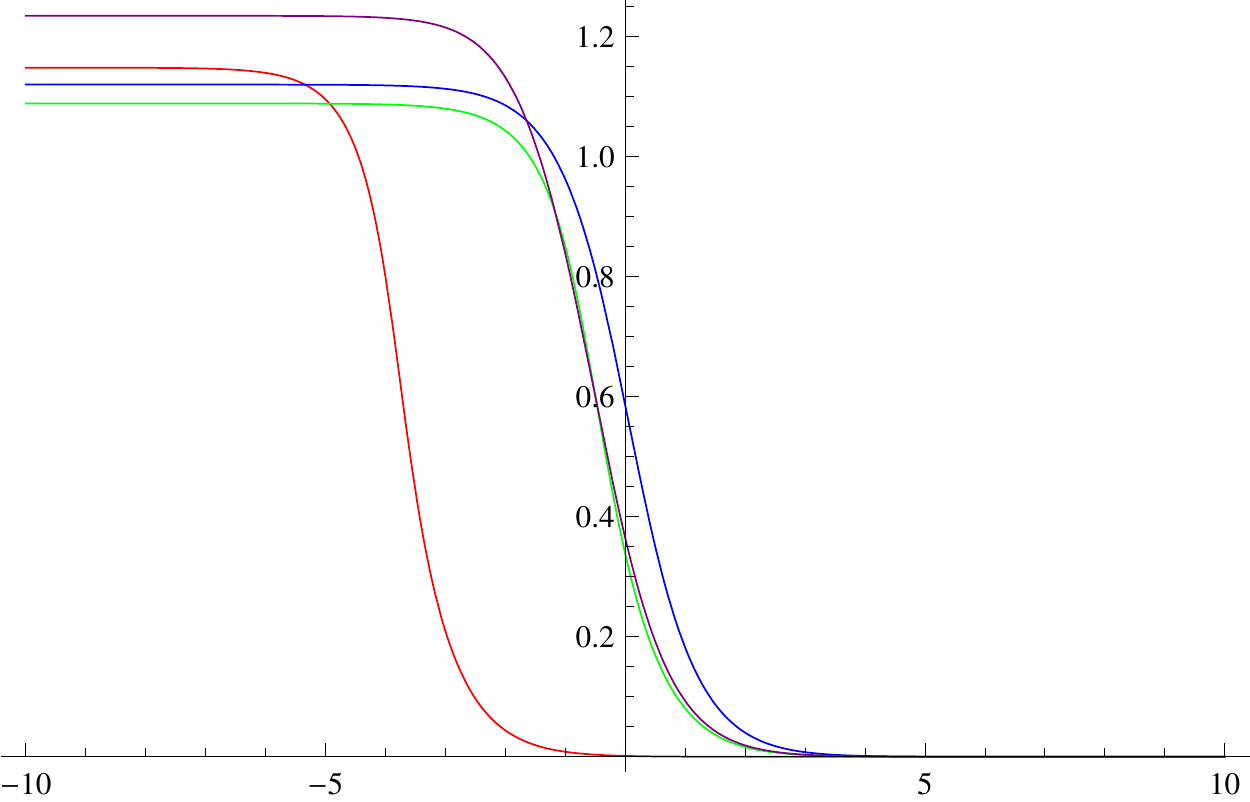}
\hfill
\includegraphics[width=.32\textwidth]{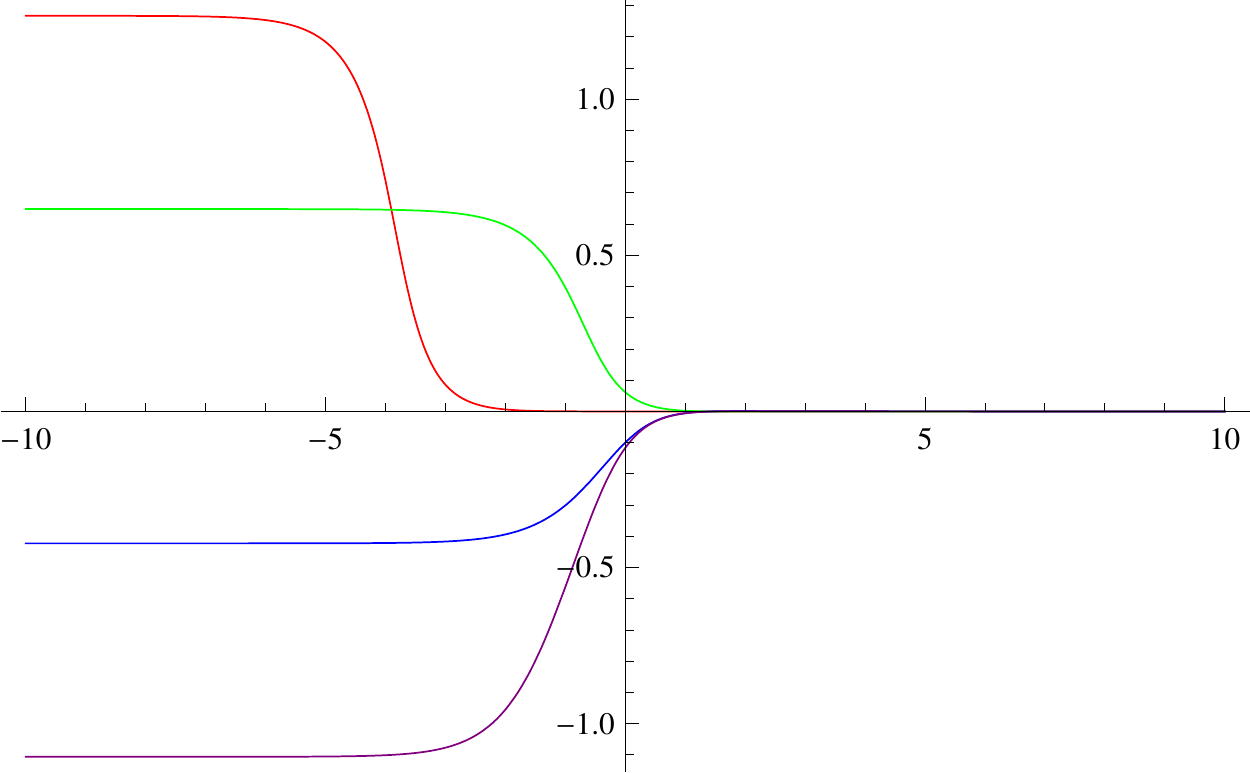}

\caption{Numerical solutions for $g(\rho)$, $\phi_1(\rho)$ and $\phi_2(\rho)$ for some representative values of $a_1$ and $a_2$. The red, green, blue and purple curves refer to $(a_1,a_2) = (1,2), (2,3), (8,6), (15,7)$ respectively.
We have chosen $\kappa=-1$ and $\fg=2$.
\label{flowplotsD3}}
\end{center}
\end{figure}

One can integrate numerically the system (\ref{eqn D3 flow}) and find solutions that interpolate between the AdS$_3$ vacua of the previous section in the IR ($\rho \to - \infty$) and an asymptotically locally AdS$_5$ space in the UV ($\rho \to \infty$). These supergravity domain walls are dual to the RG flows from the twisted $\mathcal{N}=4$ $SU(N)$ SYM at large $N$ in the UV and the two-dimensional $\cN=(0,2)$ SCFTs in the IR.\footnote{Similar solutions in the case of $\kappa=0$ were studied in \cite{Donos:2011pn}.} Some representative solutions are plotted in Figure \ref{flowplotsD3}.

\subsection{Uplift to ten dimensions}
\label{subsec: 10Duplift}

The five-dimensional supergravity solutions constructed above can be uplifted to ten dimensions using the formul\ae{} in \cite{Cvetic:1999xp}, resulting in supersymmetric solutions of type IIB supergravity. The only non-vanishing fields are the ten-dimensional metric and the self-dual five-form flux. The metric is
\be
ds^2_{10} = \Delta^{1/2} \, ds^2_5 + \Delta^{-1/2} \sum_{I=1,2,3} \frac1{X^I} \, \Big( d\mu_I^2 + \mu_I^2 (d\varphi_I + A^I)^2 \Big)\;,
\ee
where $ds^2_5$ is defined in \eqref{gravity ansatz} and $\mu_I$ are constrained as $\sum_{I=1}^3 \mu_I^2 = 1$. A convenient parametrization is
\begin{equation}
\mu_1 = \cos\theta\sin\xi \;, \qquad\qquad \mu_2 = \cos\theta\cos\xi \;, \qquad\qquad \mu_3 = \sin\theta \;,
\end{equation}
with $\theta \in [0,\pi]$, $\xi \in [0,2\pi)$.
 The warp factor $\Delta$ and the one-forms $A^I$ are defined by
\be
\Delta = \sum_{I={1,2,3}} X^I \mu_I^2 \;,\qquad\qquad\qquad
F^I = -a_I \, e^{2h(x,y)}\, dx\wedge dy = dA^I \;,
\ee
as in (\ref{gravity ansatz}). The self-dual five-form flux is $F_{(5)}  = G_{(5)}+ \star_{10}G_{(5)}$ with\footnote{Note that there is a typo in the formula for $G_{(5)}$, eqn. (2.8), in \cite{Cvetic:1999xp}. We are grateful to Chris Pope for useful correspondence on this issue.}
\be
G_{(5)} = \sum_{I=1}^3 \bigg[ 2 X^I \big( X^I \mu_I^2 - \Delta \big) \, \epsilon_{(5)} + \frac1{2(X^I)^2} d(\mu_I^2) \wedge \Big( (d\varphi_I + A^I) \wedge \star_5 F^I + X^{I}\star_5 dX^I \Big) \bigg] \;.
\ee
Here $\epsilon_{(5)}$ is the volume form for $ds^2_5$ and $\star_{5}$ is the Hodge dual with respect to $ds^2_5$.

These ten-dimensional solutions should fit into the general classification of supersymmetric AdS$_3$ solutions of IIB supergravity with five-form flux presented in \cite{Kim:2005ez}. This implies that the seven-dimensional compact manifold, spanned by the coordinates $\{x,y,\theta,\xi,\varphi_I\}$, can be written as a $U(1)$ bundle over a six-dimensional K\"ahler manifold. In particular the solutions with $a_1=a_2=a_3=1/3$ and $a_1=a_2=1/2$, $a_3=0$ were written in this canonical language in \cite{Gauntlett:2006ns} and \cite{Kim:2012ek} respectively. It is worth pointing out that the canonical Killing vector---defined in \cite{Kim:2005ez,Gauntlett:2006ns}---in our solutions is given by
\be
\partial_\psi = \sum_I \frac{X^I}{X^1 + X^2 + X^3} \, \partial_{\varphi_I} \;.
\ee
This matches with the $U(1)$ superconformal R-symmetry computed via $c$-extremization \eqref{TRdef}. More precisely $2\partial_\psi = T_R$. This fact certainly deserves further exploration but we leave this for future work.

\section{Vector fields in AdS$_3$/CFT$_2$}
\label{sec: vectors}

We summarize here some subtle aspects of the AdS$_3$/CFT$_2$ holographic dictionary for bulk vector fields. These subtleties are without a counterpart in higher dimensions since only in three dimensions gauge fields can have both Yang-Mills and/or Chern-Simons kinetic terms. Most of the material presented in the first part is nicely reviewed in \cite{Kraus:2006wn, Jensen:2010em}, while some details were discussed in \cite{Gukov:2004id}.

Consider first a bulk pure CS action, with either $U(1)$ or simple group $G$
\be
S_{CS} = - \frac k{4\pi} \int \Tr \Big( A \wedge dA + \frac23 A \wedge A \wedge A \Big) \;,
\ee
in an asymptotically AdS$_3$ space of radius $\ell$:
\be
ds_3^2 = \frac{r^2}{\ell^2} \big( ds_{\partial M}^2 + \cO(r^{-2}) \big) + \frac{\ell^2}{r^2} dr^2 \;,
\ee
where $ds_{\partial M}^2$ is a representative of a conformal class of boundary metrics. The gauge field is flat and up to gauge transformations it can be taken to have components along the boundary directions; at the boundary it asymptotes to $A^{(0)}$.

A boundary term is needed to get a good variational principle, and to do that we use the complex structure on the boundary. For $k$ positive a good boundary condition (in the sense that the stress tensor is positive and the theory is stable) is to fix $A^{(0)}_{\bar z}$, and for $k$ negative to fix $A^{(0)}_z$. The boundary term is%
\footnote{Recall that $d^2x\, \sqrt{-g^{(0)}} = \frac12 \, d^2z$.}
\be
\label{CS boundary term}
S_\text{bdy} = \frac{|k|}{4\pi} \int_{\partial M} d^2z\, \Tr (A_z A_{\bar z}) = \frac{|k|}{8\pi} \int_{\partial M} d^2x\, \sqrt{-g^{(0)}} \, \Tr A^2 \;.
\ee
While the CS term is topological, the boundary term depends on the complex structure and therefore on the boundary metric. Under a gauge transformation $A \to g^{-1} (d+A)g$, the action transforms by a boundary term \cite{Witten:1988hf, Elitzur:1989nr}, which for $k>0$ is:
\be
\Delta (S_{CS} + S_\text{bdy}) = \frac k{4\pi} \int_{\partial M} d^2z\, \Tr \Big( g^{-1} \partial_z g \, g^{-1} \partial_{\bar z} g + 2 g^{-1} \partial_z g \, A_{\bar z}^{(0)} \Big) + \frac k{12\pi} \int_M \Tr (g^{-1} dg)^3 \;.
\ee
This is the action of the chiral WZW model with a coupling of $j_z = \frac k{2\pi} g^{-1} \partial_z g$ to an external gauge field $A_{\bar z}^{(0)}$. Therefore the path-integral over $A$ includes a chiral WZW model at level $k$, which is the boundary effective theory for a chiral symmetry $G$ \cite{D'Adda:1982es, Witten:1983ar}.

Consider now a small variation of the gauge field $A \to A + \delta A$. The variation of the on-shell action is
\be
\delta(S_{CS} + S_\text{bdy}) = \frac k{2\pi} \int_{\partial M} d^2z\, \Tr \big( A_z^{(0)} \delta A_{\bar z}^{(0)} \big) \;.
\ee
By the standard holographic dictionary, the one-point function of the dual current is
\be
\langle j_z^a \rangle = \frac{g_{z\bar z}^{(0)}}{\sqrt{-g^{(0)}}} \, \frac{\delta S_\text{grav}}{\delta A_{\bar z}^{a\,(0)}} = \frac k{2\pi} \, d^{ab} A_z^{b\, (0)} \;,
\ee
where $d^{ab}$ is the Killing form on the algebra of $G$. We have thus discovered that, for $k>0$, $A_{\bar z}^{(0)}$ is the source and $A_z^{(0)}$ is related to the VEV of a boundary current $j_z$ of conformal weights $(0,1)$.
Taking a derivative $\nabla_{\bar z}$ and using that $A^{(0)}$ is flat, we obtain the 2d chiral anomaly equation. Comparing with (\ref{anomaly equations}) we see that the CS level $k$ equals half of the anomaly coefficient $k^{IM}$.

Since the boundary term depends on the metric, it contributes to the boundary stress tensor. One obtains
\be
T_{ij} = \frac{|k|}{4\pi} \Tr \Big( A_i A_j - \frac12 \, g^{(0)}_{ij} A^2 \Big) \;,
\ee
where $i,j=0,1$ are boundary indices. Changing to holomorphic indices we have for $k>0$:
\be
T_{zz} = \frac\pi k \Tr (j_z j_z) \;,\qquad T_{\bar z\bar z} = \frac k{4\pi} \Tr (A_{\bar z}^{(0)} A_{\bar z}^{(0)} ) \;,\qquad T_{z\bar z} = 0 \;.
\ee

For negative $k$ one finds similar results, but this time $A^{(0)}$ is dual to a left-moving current $j_{\bar z}$ of conformal weights $(1,0)$ with positive definite anomaly proportional to $-k$, and to a right-moving source $A_z^{(0)}$.

So far we only discussed the holographic dictionary for the flat non-propagating mode. In a Yang-Mills-Chern-Simons theory we also have a dynamical propagating mode. In the bulk the vector field is massive, however gauge invariance still reduces the number of degrees of freedom to one. Therefore, as before, out of the holomorphic and anti-holomorphic components one is the source and one is the VEV of a spin-1 operator \cite{Larsen:1998xm}. Consider the YM-CS action with  $k\neq 0$
\be
S = - \frac1{g^2} \int F \wedge \star F - \frac k{4\pi} \int A \wedge F \;+\; S_\text{bdy} \;.
\ee
For simplicity we assume that the gauge group is Abelian. The gauge field has mass $m = |k|g^2/4\pi$.
The asymptotic behavior of the gauge field in the asymptotically AdS$_3$ metric depends on the sign of $k$. For $k>0$ one finds
\be
A_z = A_z^{(0)} + r^{1-\Delta} A_z^{(2)} + \dots \;,\qquad\qquad A_{\bar z} = A_{\bar z}^{(0)} + r^{1+\Delta} A_{\bar z}^{(2)} + \dots
\ee
where $A^{(0)}$ is the non-propagating mode dual to a holomorphic boundary conserved current as discussed before, and $A^{(2)}$ is the propagating mode dual to a dimension $\Delta$ spin-1 boundary primary operator of conformal weights $\big( \frac{\Delta-1}2, \frac{\Delta+1}2 \big)$. The component $A^{(2)}_z$ is dual to the VEV of the operator, and $A^{(2)}_{\bar z}$ to its source. For $k<0$ the role of the $z$ and $\bar z$ components is interchanged, and the dual operator has weights $\big( \frac{\Delta+1}2, \frac{\Delta - 1}2 \big)$. The dimension is
\be
\Delta = 1 + m^2 > 1\;,
\ee
and in no range there is double quantization. For pure CS theory, the gauge field is non-propagating and it is dual to a boundary current only.%
\footnote{Provided $k\neq0$, we can think of adding a YM term $\frac1{g^2} |F|^2$ to the pure CS Lagrangian. We can then eliminate the YM term by sending $g \to \infty$, so that both the mass $m$ of the bulk vector field and the dimension $\Delta$ of the dual operator go to infinity. This decouples the dynamical mode both in the bulk and in the boundary CFT.}

Unitarity constrains the anomaly of the boundary current to be positive definite, and since that is proportional to the bulk Chern-Simons level, the CS term is \emph{required} to describe a symmetry current. In fact the dual to a bulk vector field with only Yang-Mills kinetic action is quite different \cite{Jensen:2010em}. The asymptotic behavior of the gauge field for $k=0$ is
\be
A = \big( \tilde A_i^{(0)} \log r + A_i^{(0)} + \dots \big) \, dx^i\;,
\ee
in a gauge where $A_r = 0$. There is only one possible quantization \cite{Marolf:2006nd} in which $A_i^{(0)}$ is the VEV of a dimension 1 operator and $\tilde A_i^{(0)}$ is its source. Maxwell's equations imply that $\tilde A_i^{(0)}$ is conserved with respect to the boundary metric, $\nabla^i \tilde A_i^{(0)} = 0$. Indeed $A_i^{(0)}$ is dual to a path-integrated gauge field $\cA_i$ in the boundary theory, but without kinetic term (such that it preserves conformal invariance).%
\footnote{Of course we can think of introducing a Yang-Mills kinetic term in the boundary theory, with scale $g_\text{2d}^2$ set by the dimensionful gauge coupling. In the far IR the gauge field can be integrated out, the YM term disappears and the effect is to ``quotient'' the theory by the gauged symmetry.
The example of 2d QED is emblematic: the theory confines, therefore below the scale $g^2_{\text{2d}}$ there are only neutral states. In the far IR there is no current operator due to confinement.}
Such gauging is common in two dimensions (for instance in gauged WZW models \cite{Bardakci:1987ee}) and its effect is to quotient the theory by the $U(1)$ symmetry that is gauged. $\tilde A_i^{(0)}$ is dual to an \emph{external} conserved current $J_i$ coupled to the boundary gauge field. The expectation value of $\cA_i$ is computed by differentiating the holographically renormalized on-shell action (see \cite{Jensen:2010em} for details):
\be
\langle \cA_i \rangle = - \frac1{\sqrt{-g^{(0)}}} \, \delfrac{S_\text{ren}}{\tilde A^{i\,(0)}} = A_i^{(0)} \;.
\ee
Since the derivative is taken with respect to a conserved $\tilde A_i^{(0)}$, the variation is defined only up to a total derivative: $A_i^{(0)} \sim A_i^{(0)} + \partial_i \lambda$. This confirms that $\cA_i$ is a path-integrated boundary gauge field, and only VEVs of gauge invariants (such as its field strength $\cF_{ij}$) are well-defined.

At the quadratic level (or in general when the gauge field is not coupled to matter) there is a simple alternative way to study the system after dualizing the bulk gauge field $A_\mu$ to a massless scalar. By the standard AdS/CFT dictionary $\Delta_\pm = \frac12 \big( d \pm \sqrt{d^2 + 4m^2} \big)$, a massless scalar is dual to a dimension 2 operator and this is the only possible quantization. This operator is precisely the field strength $\cF_{01}$.

If $\cA_i$ is coupled to a conserved boundary current operator $j_i$, the effect of gauging is to remove the current or more precisely to project the Hilbert space onto the subspace of zero or finite charge (depending on the source $J_i$).
Note that if $j_i$ is part of a unitary CFT with normalizable vacuum, then $j_i$ is the vector-like combination of a right-moving and a left-moving current with opposite non-vanishing 't Hooft anomalies. Therefore, the axial current $\star j_i$ is conserved as well. Then the gauging of $j_i$ also lifts the axial current because $\star j_i$ acquires a gauge anomaly.

To summarize, we have shown that the matrix of CS levels for gauge fields in a three-dimensional gravitational theory is proportional to the matrix of 't Hooft anomalies in the dual two-dimensional CFT. This fact is relevant for making the connection between field theory and supergravity discussed in Section \ref{sec: D3-branes}. Furthermore we saw that when there is a bulk gauge field with only a YM kinetic term, in the dual CFT we do not have a corresponding conserved current. This nicely fits with the discussion of the special twist with $a_1=a_2=1/2$ and $a_3=0$ in Section \ref{sec: D3-branes}.

\

The second issue we would like to discuss is supersymmetry. We explained in Section \ref{subsec: exactR} that in a two-dimensional conformal theory with $\cN=(0,2)$ supersymmetry, right-moving Abelian currents are paired into multiplets while left-moving ones remain alone. It is natural to ask how does this happen in the holographic dual.

The 3d $\cN=2$ vector multiplet contains a vector, a Dirac spinor and a real scalar. In the $\cN=2$ pure CS action the spinor and the scalar are auxiliary, so they are not dual by AdS/CFT to boundary operators that can form a multiplet with the boundary current. In the $\cN=2$ YM-CS action the spinor and the scalar are dynamical but massive, and together with the propagating mode of the vector are dual to a boundary multiplet of weights $\big( \frac{\Delta -1}2, \frac{\Delta - 1 + \alpha}2 \big)$ where $\alpha = 0,1,2$ (for $k>0$). Again, we don't find partners for the boundary current.

The partners of the boundary current should somehow come from boundary terms. The interplay between supersymmetry and boundary conditions has been analyzed in great detail in \cite{Belyaev:2008xk} for $\cN=1$ and \cite{Berman:2009kj} for $\cN=2$. The idea is that a boundary orthogonal to the direction $r$ breaks half of the bulk supersymmetry while the other half, $\gamma_r \epsilon_+ = \epsilon_+$, can be realized off-shell by suitable boundary terms without the need of extra boundary conditions. For simplicity and to convey the main idea, we work in three-dimensional  flat space with a boundary, leaving a more detailed analysis in AdS$_3$ for the future.

With 3d $\cN=1$ supersymmetry we have a chiral multiplet $\Phi = a + \bar\theta\psi + \theta^2 f$ (where $\theta$ is the superspace coordinate) with variations
\be
\delta a = \bar\epsilon \psi \;,\qquad \delta\psi = \gamma^\mu \epsilon\, \partial_\mu a + \epsilon \, f \;,\qquad \delta f = \bar\epsilon \gamma^\mu \partial_\mu \psi \;,
\ee
and a spinor multiplet $\Psi = \chi + \theta M + \gamma^\mu \theta\, v_\mu + \theta^2 \big(\lambda - \gamma^\mu \partial_\mu \chi \big)$ with variations
\bea
\delta\chi &= M\epsilon + \gamma^\mu \epsilon \, v_\mu \;,\qquad\qquad &
\delta M &= - \tfrac12 \bar\epsilon \lambda + \bar\epsilon \gamma^\mu \partial_\mu \chi\;, \\
\delta v_\mu &= -\tfrac12 \bar\epsilon \gamma_\mu \lambda + \bar\epsilon \partial_\mu \chi \;,\qquad\qquad &
\delta\lambda &= 2\gamma^{\mu\nu} \epsilon \, \partial_\mu v_\nu \;.
\eea
Gauge transformations are promoted to super-gauge transformations where the local parameter is a chiral superfield $\Phi$; in components:
\be
\delta_g \chi = \psi \;,\qquad \delta_g M = f \;,\qquad \delta_g v_\mu = \partial_\mu a \;,\qquad \delta_g\lambda = 0 \;.
\ee
If this transformation is a symmetry of the action, one can impose Wess-Zumino gauge $\chi = M = 0$.

The 3d $\cN=2$ vector multiplet is formed by one spinor multiplet $\Psi = (\chi, M , v_\mu, \lambda)$ and two chiral multiplets $A = (a, \psi, f)$, $B = (b, \eta, g)$ (see \cite{Berman:2009kj} for details). We can write down a Chern-Simons Lagrangian that preserves 2d $\cN=(0,2)$ supersymmetry along the boundary without boundary conditions:
\be
\cL^{CS}_{(0,2)} = -2 \epsilon^{\mu\nu\rho} v_\mu F_{\nu\rho} + \bar\lambda \lambda + \bar\eta\eta - 2 gb + \partial_r \Big[ -2 \bar\chi_- \lambda_+ - 2 \bar\psi_- \eta_+ - b^2 + 2a \big(g + \partial_r b \big) \Big] \;.
\ee
Notice that not all fields appear, and the only field with a kinetic term is $v_\mu$.
Usually one imposes Wess-Zumino gauge, which in this case corresponds to $A = B = 0$ as well as $\chi = M = 0$. However the CS action is not gauge invariant in the presence of a boundary (a gauge variation gives a boundary term) and likewise it is not invariant under super-gauge transformations, thus Wess-Zumino gauge cannot be imposed. We can further add to $\cL^{CS}_{(0,2)}$ boundary terms, written in terms of boundary superfields, that preserve $\cN=(0,2)$ supersymmetry. In particular we can write
\be
\cL_b = 2\partial_r \Big[ v_j v^j + \bar\chi_- \slashed\partial \chi_- + \bar\psi_- \slashed\partial \psi_- + \partial_j a \, \partial^j a + \bar\chi_- \lambda_+ + \bar\psi_- \eta_+ -ag - a \partial_r b  \Big]\;,
\ee
that contains the boundary term (\ref{CS boundary term}) for the pure CS theory.

In AdS/CFT it is crucial to have a good variational principle and to this end a bulk CS Lagrangian $k \, \cL^{CS}_{(0,2)}$ requires a boundary term $|k| \, \cL_b$ so that we take
\be
\cL = k\, \cL^{CS}_{(0,2)} + |k| \, \cL_b \;.
\ee
For $k > 0$ the total boundary Lagrangian is
\be
\label{boundary L k>0}
\cL_\text{bdy}^{k>0} = 2k \Big[ v_j v^j + \bar\chi_- \slashed\partial \chi_- + \bar\psi_- \slashed\partial \psi_- + \partial_j a \, \partial^j a - \frac12 b^2 \Big] \;.
\ee
In particular the hitherto auxiliary bulk fields $\chi_-$, $\psi_-$, $a$ have become dynamical boundary fields! While the path-integral over $v_\mu$ produces a boundary chiral WZW model describing a right-moving current, we also get a boundary free right-moving Weyl spinor $\chi_- + i \psi_-$ and a free scalar $a$ (as well as the current $\partial a$) to form an $\cN=(0,2)$ current multiplet. Notice that the three extra fields are decoupled. For $k<0$ the total boundary Lagrangian has extra quadratic terms with respect to (\ref{boundary L k>0}), such that the EOMs kill all extra boundary modes besides the current dual to $v_\mu$. Therefore a left-moving current is not accompanied by any other boundary field.

For completeness we would like to point out another possibility, not realized in our supergravity solutions. In \cite{Ivanov:2000tz} a different 3d $\cN=2$ vector multiplet, called ``double-vector multiplet'' was constructed. It contains two gauge fields and two Majorana spinors, and it is made of two  ordinary $\cN=1$ vector multiplets. Indeed, its construction and supersymmetry variations parallel the construction of the $\cN=(0,2)$ current out of  two $\cN=(0,1)$ currents in Section \ref{subsec: exactR}. Although we have not analyzed the details, it is clear that this multiplet may describe a non-trivial pair of boundary currents.

\section{Six-dimensional $\cN=(2,0)$ theory on $\Sigma_1\times \Sigma_2$}
\label{sec: M5-branes}

Let us now move to the study of the two-dimensional theories that arise at low energy from compactifications of the six-dimensional $\cN=(2,0)$ theory on four-manifolds. Here we will focus on four-manifolds which are products of two Riemann surfaces $\Sigma_1 \times \Sigma_2$. To preserve some supersymmetry we twist the six-dimensional theory, in particular we are interested in preserving $\cN=(0,2)$ supersymmetry in two dimensions. We then use the anomaly polynomial of the 6d theory to extract the 't~Hooft and gravitational anomalies of the 2d theories in a way similar to \cite{Benini:2009mz, Alday:2009qq, Bah:2011vv, Bah:2012dg}, and then apply $c$-extremization to obtain the superconformal R-symmetry and the central charges at the IR fixed points, assuming that such fixed points exist.

The 6d theory is really a family of theories classified by the ADE Lie algebras (plus the $\cN=(2,0)$ free tensor multiplet) \cite{Witten:1995zh}. In the $A_{N-1}$ case, we can think of these theories as describing the low-energy dynamics of $N$ M5-branes in M-theory (from a single M5-brane we get the free tensor multiplet). Then the family of $\cN=(0,2)$ partial topological twists we study corresponds to wrapping the M5-branes on $\Sigma_1 \times \Sigma_2$ as a K\"ahler 4-cycle in a Calabi-Yau fourfold. At large $N$ we can use a holographic description of the low-energy theories in eleven-dimensional supergravity. Indeed, we construct an infinite family of novel AdS$_3$ supergravity solutions, preserving 2d $\cN=(0,2)$ supersymmetry, which are gravitational duals to the 2d SCFTs. The central charges can be computed holographically, and again they nicely agree with $c$-extremization.

\subsection{Field theory}
\label{sec: M5 field theory}

The R-symmetry of the six-dimensional $\cN=(2,0)$ theory is $Sp(2) \cong SO(5)$. On a four-manifold $M_4 = \Sigma_1 \times \Sigma_2$ (with genera $\fg_1$ and $\fg_2$) the holonomy group is generically $SO(2)_1 \times SO(2)_2$, and we can preserve 2d $\cN=(0,2)$ supersymmetry by turning on an Abelian background $A_\mu$ coupled to an $SO(2)^2$ subgroup of $SO(5)$, embedded block-diagonally. The construction is very similar to the one in Section \ref{sec: D3-branes}. The supercharges transform in the representation $\rep{4} \otimes \rep{4}$ (with symplectic Majorana condition) of the product of the Lorentz and R-symmetry groups $SO(5,1) \times SO(5)$. We will turn on a background gauge field $A$ with field strength $F = dA = \sum_{\sigma=1,2} F_\sigma$ which is the sum of two components $F_\sigma$, living on $\Sigma_\sigma$ ($\sigma=1,2$).%
\footnote{This is not the most general background, since $\dim H^2(\Sigma_1 \times \Sigma_2) = 2+4\fg_1\fg_2$. We will not analyze the most general situation here.}
In terms of the two spin connections $\tilde \omega_\mu^{(\sigma)} \equiv \frac12 \omega_\mu^{(\sigma)\,ab} \varepsilon_{ab}$ on $\Sigma_\sigma$, such that $\int d\tilde\omega^{(\sigma)} = \frac12 \int \sqrt{g_\sigma}\, R_\sigma = 4\pi(1-\fg_\sigma)$, the field strength components are $F_\sigma = - T_\sigma d\tilde\omega^{(\sigma)}$ for $\fg_\sigma = 0$, $F_\sigma = - T_\sigma \frac{2\pi}{\vol(\Sigma_\sigma)} \dvol_{\Sigma_\sigma}$ for $\fg_\sigma=1$, and $F_\sigma = T_\sigma d\tilde\omega^{(\sigma)}$ for $\fg_\sigma > 1$. The two generators $T_\sigma$ are taken as
\be
\label{M5 generators}
T_\sigma = a_\sigma T_A + b_\sigma T_B \;,\qquad\qquad \sigma = 1,2
\;,
\ee
where $T_{A,B}$ are the generators of $SO(2)^2 \subset SO(5)$, and $a_\sigma, b_\sigma$ are real constants parametrizing the twist. To preserve 2d $\cN=(0,2)$ supersymmetry we take
\be
\label{M5 SUSY condition}
a_\sigma + b_\sigma = - \kappa_\sigma \;,
\ee
where $\kappa_\sigma = 1$ for $\fg_\sigma = 0$, $\kappa_\sigma = 0$ for $\fg_\sigma = 1$, and $\kappa_\sigma = -1$ for $\fg_\sigma >1$. We can parametrize the solutions to these constraints in terms of two real numbers $z_\sigma$:
\be
\label{parametrization z}
a_\sigma = \frac{-\kappa_\sigma + z_\sigma}2 \;,\qquad\qquad b_\sigma = \frac{-\kappa_\sigma - z_\sigma}2 \;.
\ee
The fermions of the 6d theory transform in the representation $\overline{\rep{4}} \otimes \rep{4}$ of $SO(5,1) \times SO(5)$, and under $SO(2)_A \times SO(2)_B \subset SO(5)$ they have charges $\big( \pm \frac12, \pm\frac12 \big)$. For $\fg_\sigma \neq 1$, well-definiteness of the bundles over $\Sigma_\sigma$ requires $2(\fg_\sigma-1)(\pm a_\sigma \pm b_\sigma) \in 2\bZ$, which supplemented by the supersymmetry condition \eqref{M5 SUSY condition} is equivalent to $a_\sigma, b_\sigma \in \bZ/2(\fg_\sigma-1)$; for $\fg_\sigma=1$ we simply get $a_\sigma, b_\sigma \in \bZ$. More succinctly we can write:
\be
\eta_\sigma\, a_\sigma \,,\, \eta_\sigma\, b_\sigma \in \bZ \qquad\qquad \sigma = 1,2 \;,
\ee
where $\eta_\sigma$ is defined in (\ref{definition eta sigma}).
The quantization of the alternative parameters is $z_\sigma \in \bZ/(\fg_\sigma-1)$ for $\fg_\sigma\neq1$, and $z_\sigma \in 2\bZ$ for $\fg_\sigma=1$.

As explained in detail in Appendix \ref{app: SUSY (2,0)}, for generic non-zero values of the parameters $a_\sigma, b_\sigma$ we preserve 2d $\cN=(0,2)$ supersymmetry. When $a_\sigma = 0$ or $b_\sigma = 0$ (and $\fg_{1,2} \neq 1$) we get $\cN=(0,4)$, while when one $a_\sigma$ and one $b_\sigma$ vanish we get $\cN=(2,2)$. When three of the parameters vanish we have $\cN=(4,4)$. Finally, when $a_\sigma = b_\sigma = 0$ and $\fg_{1,2} = 1$ we have maximal supersymmetry $\cN=(8,8)$.

The low-energy 2d theory inherits $SO(2)^2$ global symmetry (possibly enhanced to a non-Abelian symmetry of the same rank) which contains the 2d superconformal R-symmetry. The trial R-symmetry is a linear combination of the generators of $SO(2)^2$:
\be
T_R = (1+\epsilon) T_A + (1-\epsilon) T_B \;,
\ee
where the real number $\epsilon$ parametrizes the mixing and we have fixed the R-charge of the complex supercharge to 1.

The exact superconformal R-symmetry is found through $c$-extremization, and for that we need the two-dimensional anomalies. In the spirit of \cite{Benini:2009mz, Alday:2009qq, Bah:2011vv, Bah:2012dg}, we compute them by integrating the anomaly polynomial of the 6d theory on $\Sigma_1 \times \Sigma_2$. Let us denote by $t_\sigma$ the Chern roots of the tangent bundles on $\Sigma_\sigma$, and by $n_{A,B}$ the Chern roots of the R-symmetry bundle. The condition for supersymmetry is
\be
t_1 + t_2 + n_A + n_B = 0 \;.
\ee
Now we add to the background for the normal bundle a flux $F_R$ that couples to the 2d trial R-symmetry, in order to detect anomalies. We take:
\be
n_A = a_1 f_1 + a_2 f_2 + (1+\epsilon)\, c_1(F_R) \;,\qquad\qquad n_B = b_1 f_1 + b_2 f_2 + (1-\epsilon)\, c_1(F_R) \;,
\ee
where $c_1$ is the first Chern class.
We have introduced the two classes $f_\sigma$ on the surfaces $\Sigma_\sigma$, along which the field strength components $F_\sigma$, defined above (\ref{M5 generators}), are taken. In particular
\be
\label{definition eta sigma}
\int_{\Sigma_\sigma} f_\sigma \equiv \eta_\sigma = \begin{cases} 2|\fg_\sigma-1| &\text{for } \fg_\sigma \neq 1 \\ 1 &\text{for } \fg_\sigma = 1 \end{cases}
\ee
as in (\ref{definition eta_Sigma}), and $\int t_\sigma = 2(1-\fg_\sigma)$. The anomaly eight-form of the Abelian 6d $\cN=(2,0)$ theory, \ie~a free tensor multiplet or a free M5-brane, is \cite{Witten:1996hc}
\be
\label{1 M5 anomaly}
I_8 [1] = \frac1{48} \Big[ p_2(NW) - p_2(TW) + \frac14 \big( p_1(TW) -p_1(NW) \big)^2 \Big] \;,
\end{equation}
where $p_k$ is the $k$-th Pontryagin class, $TW$ denotes the tangent bundle to the 6d worldvolume and $NW$ denotes the R-symmetry bundle (which is the same as the normal bundle to the M5-branes). For a generic $\cN=(2,0)$ theory of type $G$ the anomaly polynomial is \cite{Harvey:1998bx, Intriligator:2000eq, Yi:2001bz}
\be
\label{G M5 anomaly}
I_8[G] = r_G I_8[1] + \frac{d_G h_G}{24}\, p_2(NW) \;,
\ee
where $r_G$, $d_G$ and $h_G$ are the rank, dimension and Coxeter number of $G$ (see Table \ref{tab: groups}). It will be useful to remember that $d_G h_G$ is always a multiple of 6.
\begin{table}[t]
$$
\begin{array}{|c|c|c|c|}
\hline
G & r_G & d_G & h_G \\[0.2ex]
\hline
\tabs A_{N-1} & N-1 & N^2-1 & N \\
D_N & N & N(2N-1) & 2N-2 \\
E_6 & 6 & 78 & 12 \\
E_7 & 7 & 133 & 18 \\
E_8 & 8 & 248 & 30 \\ [0.2ex]
\hline
\end{array}
$$
\caption{Rank, dimension and Coxeter number of simply-laced Lie algebras. Notice that  $d_G = r_G (h_G+1)$ and $d_G h_G$ is a multiple of 6.
\label{tab: groups}}
\end{table}
In terms of the Chern roots $n_i$, the Pontryagin classes are
\be
p_1 = \sum\nolimits_i n_i^2 \;,\qquad\qquad p_2 = \sum\nolimits_{i<j} n_i^2 n_j^2 \;.
\ee
With this at hand we can expand out (\ref{G M5 anomaly}) and integrate over $\Sigma_1 \times \Sigma_2$. In the integration only terms linear in $t_1$ or $f_1$, and linear in $t_2$ or $f_2$, survive. We obtain a 4-form:
\begin{multline}
\int_{\Sigma_1 \times \Sigma_2} I_8[G] = \frac{\eta_1\eta_2}{24}\bigg[  \bigg( d_G h_G \big( \kappa_1\kappa_2(3-\epsilon^2) + 2(\kappa_1z_2 + \kappa_2z_1)\epsilon - z_1z_2(1-3\epsilon^2)  \big) + \phantom{\,} \\
+ 3r_G (\kappa_1\kappa_2 + z_1z_2\epsilon^2) \bigg)  c_1(F_R)^2 - \frac{r_G}4 (\kappa_1\kappa_2 + z_1z_2)\, p_1(TW_2) \bigg] \;,
\end{multline}
where $TW_2$ is the tangent bundle to the worldvolume of the two-dimensional theory, and we have used the parametrization of the twist in terms of $z_\sigma$ (\ref{parametrization z}). This result has to be compared with the anomaly polynomial of a two-dimensional theory:
\be
\label{anomaly polynomial 2d}
I_4 = \frac{c_R}6 \, c_1(F_R)^2 - \frac{c_R - c_L}{24}\, p_1(TW_2) \;.
\ee
In this way we extract the trial right-moving central charge $c_R^\text{tr}$ and the gravitational anomaly:
\bea
\label{cR trial and grav anomaly}
c_R^\text{tr}(\epsilon) &= \frac{\eta_1\eta_2}4 \bigg[ d_G h_G \Big( \kappa_1\kappa_2(3-\epsilon^2) + 2(\kappa_1z_2 + \kappa_2z_1)\epsilon - z_1z_2(1-3\epsilon^2)  \Big) + 3r_G (\kappa_1\kappa_2 + z_1z_2\epsilon^2) \bigg] \\
k &= c_R - c_L=\frac{\eta_1\eta_2}4 r_G (\kappa_1 \kappa_2 + z_1z_2) \;.
\eea
The function $c_R^\text{tr}(\epsilon)$ is extremized at
\be
\epsilon = \frac{d_G h_G (\kappa_1z_2 + \kappa_2 z_1)}{d_G h_G(\kappa_1 \kappa_2 - 3z_1 z_2) - 3r_G z_1 z_2} \;,
\ee
when the denominator does not vanish. Plugging this back into $c_R^\text{tr}(\epsilon)$, we find the left and right central charges to be:
\bea
\label{ccexactM5}
c_R &= \frac{\eta_1\eta_2}4\; \frac{d_G^2 h_G^2 \cP + 3 d_G h_G r_G \big( z_1^2 z_2^2 - 6\kappa_1\kappa_2 z_1z_2 + \kappa_1^2 \kappa_2^2 \big) - 9 r_G^2 \kappa_1 \kappa_2 z_1 z_2}{d_G h_G(\kappa_1 \kappa_2 - 3z_1 z_2) - 3r_G z_1 z_2}\;, \\
c_L &=  \frac{\eta_1\eta_2}4\; \frac{d_G^2 h_G^2 \cP + 2 d_G h_G r_G \big( 3 z_1^2 z_2^2 - 8\kappa_1\kappa_2 z_1z_2 + \kappa_1^2 \kappa_2^2 \big) + 3 r_G^2 z_1z_2( z_1z_2 - 2 \kappa_1 \kappa_2)}{d_G h_G(\kappa_1 \kappa_2 - 3z_1 z_2) - 3r_G z_1 z_2},
\eea
where we have defined
\be
\cP \equiv 3z_1^2z_2^2 + \kappa_1^2 z_2^2 + \kappa_2^2 z_1^2 - 8\kappa_1\kappa_2z_1z_2 + 3\kappa_1^2 \kappa_2^2 \;.
\ee
The second derivative of the trial central charge is
\be
\label{d2 epsilon c}
\parfrac{^2 c_R^\text{tr}(\epsilon)}{\epsilon^2} = \frac{\eta_1\eta_2}2 \, r_G \Big( h_G (h_G+1) (3z_1z_2 - \kappa_1\kappa_2) + 3z_1z_2 \Big) \;.
\ee
Its sign determine the chirality of the extra flavor current. There are regions in parameter space where the central charges are not both positive. As discussed in Section \ref{sec: D3-branes}, this means that one of our assumptions is not met or there is no IR fixed point at all. In Section \ref{sec: M5 sugra} we use holography to determine the range of parameters where, at least at large $N$, there are good IR CFTs with normalizable vacuum.

For the $A_N$ theory, the large $N$ limit of the central charges is
\be
\label{cR large N M5}
c_R^{A_N} \,\simeq\, c_L^{A_N} \,\simeq\, \frac{\eta_1 \eta_2 N^3}4 \; \frac{3z_1^2z_2^2 + \kappa_1^2 z_2^2 + \kappa_2^2 z_1^2 - 8\kappa_1\kappa_2z_1z_2 + 3\kappa_1^2 \kappa_2^2}{\kappa_1 \kappa_2 - 3z_1 z_2}
\ee
where the numerator is $\cP$,
and for the $D_N$ theory the result is $c_R^{D_N} \simeq c_L^{D_N} \simeq 4c_R^{A_N}$. Both large $N$ results nicely match with the supergravity computations in Section \ref{sec: M5 sugra}.

We emphasize that the gravitational anomaly in (\ref{cR trial and grav anomaly}) for this class of two-dimensional SCFTs does  not vanish. In fact, the 6d $\cN=(2,0)$ theory itself has a gravitational anomaly, which is canceled by anomaly inflow when we place M5-branes in M-theory. Moreover a CFT with modular invariant partition function has $c_R - c_L = 0~\text{mod}~24$. This is not generically the case for the CFTs in question.

The anomaly calculation is also valid for the Abelian $\cN=(2,0)$ theory after setting $r_G=1$ and $h_G =0$ as follows from (\ref{G M5 anomaly}). We find:
\be
c_R = \frac34 \, \eta_1 \eta_2 \, \kappa_1 \kappa_2 \;,\qquad\qquad c_L = \frac14 \, \eta_1\eta_2 (2\kappa_1\kappa_2 - z_1z_2) \;.
\ee
Since in a supersymmetric theory $c_R>0$, there can be a fixed point with normalizable vacuum only for $\fg_{1,2} > 1$ or $\fg_{1,2}=0$. Note that for this class of theories $\epsilon=0$ and $c_R$ is independent of the twist parameters $z_{\sigma}$.

\subsubsection{Geometric engineering and special twists}
\label{sec: M5-brane construction}

For the $A_{N-1}$ theories we can give a geometric interpretation of the construction.%
\footnote{The geometric interpretation is also valid for the $D_N$ series after performing a $\bZ_2$ orbifold $\vec x \to - \vec x$ of the $\bR^5$ normal bundle.}
The twisted 6d $\cN=(2,0)$ theory describes the low-energy dynamics of $N$ M5-branes%
\footnote{More precisely, the $A_{N-1}$ theory describes the dynamics once the decoupled center of mass is removed.}
on a manifold $M_4 = \Sigma_1 \times \Sigma_2$, and the family of twists corresponds to embedding $M_4$ into a Calabi-Yau fourfold as a K\"ahler 4-cycle. More precisely, the geometry is a local model of two line bundles $\cL_A \oplus \cL_B$ over $M_4$. The line bundle $\cL_A$ has degrees $p_\sigma$ on $\Sigma_\sigma$ ($\sigma = 1,2$), and $\cL_B$ has degrees $q_\sigma$. The total space is a local $CY_4$ for
\be
p_\sigma + q_\sigma = 2\fg_\sigma - 2 \;,\qquad\qquad \sigma=1,2 \;.
\ee
The degrees $p_\sigma, q_\sigma$ of the line bundles are associated to the parameters $a_\sigma,b_\sigma$ as:
\be
p_\sigma = a_\sigma\, \eta_\sigma \;,\qquad\qquad\qquad q_\sigma = b_\sigma \, \eta_\sigma \;.
\ee

There are three special twists, corresponding to special $CY_4$ manifolds, which we now discuss in more detail. We will assume $\fg_{1,2} > 1$ for simplicity.
\begin{itemize}

\item $z_1=1$ and $z_2=-1$ (or viceversa). Supersymmetry is enhanced to $\cN=(2,2)$ and the $CY_4$ simplifies to $HK_1 \times HK_1$, where each hyper-K\"ahler factor is the canonical bundle over one of the Riemann surfaces. This is the twist studied in Section 3 of \cite{Gauntlett:2001jj}. The central charges are
\be
c_R = c_L = (\fg_1-1)(\fg_2-1) (4d_Gh_G + 3r_G) \;.
\end{equation}
They are an integer multiple of 3 for any simply-laced group $G$, and it would be interesting to understand if the SCFT is a non-linear sigma model on a Calabi-Yau target.
Moreover $c_R^\text{tr}(\epsilon)$ is maximized, see (\ref{d2 epsilon c}), as it should because the extra current is the left-moving R-symmetry.

\item \label{(0,4) from M5s}
$z_1 = z_2 = 1$ (or $-1$). Supersymmetry is enhanced to $\cN=(0,4)$, the global symmetry is enhanced to $SO(2)_A \times SO(3)_B$ and the $CY_4$ simplifies to $CY_3 \times \bC$, where $CY_3$ is the canonical bundle over $M_4$. The central charge $c_R$ can be computed from an $\cN=(0,2)$ subalgebra, with R-symmetry the Cartan of $SO(3)_B$. This corresponds to $\epsilon = -1$ for $z_1=z_2=1$ (and $\epsilon=1$ for $z_1=z_2=-1$) and gives
\be
c_R = 2(\fg_1-1)(\fg_2-1)(4d_Gh_G+3r_G) \;,\qquad c_L = 4(\fg_1-1)(\fg_2-1)(2d_Gh_G + r_G) \;.
\ee
As it should, $c_R$ is an integer multiple of 6 because it is proportional to the level of the $SU(2)_B$ current algebra.

We notice that $c$-extremization would give different, and incorrect, values for $\epsilon$ and the central charges. This is likely due to the fact that the vacuum is non-normalizable and the extra flavor current is non-holomorphic. Some evidence for this is provided by the absence of an AdS$_3$ vacuum at large $N$, as we find in Section \ref{sec: M5 sugra} and was emphasized in Section 4.1 of \cite{Gauntlett:2000ng}. We observe an analogous behavior for a free non-compact boson in Appendix \ref{app: free theories}. Note that this 2d SCFT is not the same as the MSW $\cN=(0,4)$ SCFT  \cite{Maldacena:1997de, Minasian:1999qn} that one obtains by wrapping M5-branes on a very ample divisor in a \emph{compact} $CY_3$.

\item $z_1=z_2=0$. Supersymmetry is still $\cN=(0,2)$ but the global symmetry is enhanced to $SU(2) \times U(1)_R$. The R-symmetry, diagonally embedded in $SO(2)_A \times SO(2)_B$, cannot mix with $SU(2)$, and in fact $c$-extremization gives $\epsilon = 0$. This twist is studied in Section 4.2 of \cite{Gauntlett:2000ng}, and can be performed for any K\"ahler 4-cycle in a $CY_4$. The central charges are
\be
c_R = 3(\fg_1-1)(\fg_2-1) (d_G h_G+r_G) \;,\qquad c_L = (\fg_1-1)(\fg_2-1) (3d_G h_G+2r_G) \;.
\ee
Moreover $c_R^\text{tr}(\epsilon)$ is maximized, therefore the extra current is left-moving.

Let us point out that for fixed $G$ and $\fg_{1,2}$, this twist leads to the minimal value of $c_R$ as a function of $z_{1,2}$.

\end{itemize}

\subsection{Supergravity}
\label{sec: M5 sugra}

Let us now construct the supergravity solutions that describe the backreaction of $N$ M5-branes on $\Sigma_1 \times \Sigma_2$ at large $N$. Such solutions, when they exist, contain a warped AdS$_3$ factor and are dual to the 2d IR fixed points. To construct them we again follow the approach of \cite{Maldacena:2000mw} and use the fields of maximal seven-dimensional gauged supergravity. In fact we will only need a consistent truncation of the full seven-dimensional theory described in \cite{Liu:1999ai}. The action and supersymmetry variations were derived in \cite{Pernici:1984xx}, but here we use the conventions of \cite{Gauntlett:2000ng}. Further details can be found in Appendix \ref{app: 7D sugra}.
The non-zero components of the $SO(5)$ gauge field are in the Cartan of $SO(5)$ and their Abelian field strengths will be denoted $F^A$ and $F^B$. The Ansatz is
\bea\label{7DsugraAnsatz}
ds_7^2 &= e^{2f(r)}(-dt^2 + dz^2 + dr^2) + \sum_{\sigma=1,2} e^{2g_\sigma(r) + 2h_\sigma(x_\sigma,y_\sigma)}(dx_\sigma^2 + dy_\sigma^2)\;, \\
F^A &= - \sum_{\sigma=1,2} \frac{a_\sigma}4 e^{2h_\sigma(x_\sigma, y_\sigma)} dx_\sigma \wedge dy_\sigma \;,\qquad
F^B = - \sum_{\sigma=1,2} \frac{b_\sigma}4 e^{2h_\sigma(x_\sigma, y_\sigma)} dx_\sigma \wedge dy_\sigma\;, \\
\lambda_1 &= \lambda_1(r) \;,\qquad\qquad \lambda_2 = \lambda_2(r) \;,
\eea
where the metric functions $h_{1,2}$ are defined as in (\ref{metric function h}), depending on the genera $\fg_{1,2}$ of the Riemann surfaces. For $\fg_\sigma = 0$ the coordinates $x_\sigma, y_\sigma$ span $\bR^2$; for $\fg_\sigma = 1$ they span $[0,1)^2$; for $\fg_\sigma > 1$ the range $x_\sigma \in \bR$, $y_\sigma > 0$ describes the hyperbolic plane $\bH^2$ which we then quotient by a discrete subgroup of $PSL(2,\bR)$ to obtain a compact Riemann surface. Recall that $\frac1{2\pi} \int_\Sigma e^{2h} dx\,dy = \eta_\sigma$, as in (\ref{definition eta_Sigma}) and (\ref{definition eta sigma}).
The constants $a_{1,2}, b_{1,2}$ are identified with those in (\ref{M5 generators}) and are quantized in the same way. Notice that in 7d supergravity we have fixed the gauge coupling $g=4$ (see Appendix \ref{app: 7D sugra}), therefore the fluxes are normalized differently than in Section \ref{sec: M5 field theory} and the quantization condition reads here: $\frac1{2\pi} \int F^{A,B} \in \bZ/4$. There is also a three-form gauge potential in the seven-dimensional theory. For our Ansatz it takes the form:
\be
S_5= - \frac1{32\sqrt3} \, (a_1b_2 + a_2b_1) \, e^{-4(\lambda_1+\lambda_2)-2(g_1+g_2)} \, dt\wedge dz\wedge dr \;.
\ee
The supersymmetry transformations of fermionic fields in 7d gauged supergravity are presented in Appendix \ref{app: 7D sugra}. We impose the following projectors on the supergravity spinor $\epsilon$:
\be
\label{7D projectors}
\gamma_3 \epsilon = \epsilon \;,\qquad\qquad \gamma_{45}\epsilon = \gamma_{67}\epsilon = \Gamma^{12}\epsilon = \Gamma^{34}\epsilon = \pm \epsilon \;,
\ee
for a choice of sign in the last equation,
where $\gamma_i$ are tangent space and $\Gamma^i$ are $SO(5)$ gamma matrices. These projectors are motivated by the brane construction of Section \ref{sec: M5-brane construction}. The BPS equations impose the same differential constraints on the metric functions for both signs in \eqref{7D projectors}, therefore the background generically preserves $1/8$ of the supersymmetry, \ie{} one complex supercharge.
In addition to that the BPS equations set
$$
a_\sigma + b_\sigma = - \kappa_\sigma\;,
$$
as in (\ref{M5 SUSY condition}), so that $a_\sigma,b_\sigma$ can be parametrized in terms of $z_\sigma$ as in (\ref{parametrization z}) and are quantized as explained there.
The rest of the BPS equations yield the following unwieldy differential system for the metric functions and the scalars:
\bea
\label{7D BPS eqns}
e^{-f}f' &=  -\frac15 \big( 2e^{-2\lambda_1}+2e^{-2\lambda_2}+e^{4\lambda_1 + 4\lambda_2} \big) - \frac3{80}(a_1b_2+a_2b_1) \, e^{-2\lambda_1-2\lambda_2-2g_1-2g_2} \\
&\qquad\qquad\qquad\qquad - \frac1{20} \big( a_1 e^{2\lambda_1-2g_1} + a_2 e^{2\lambda_1-2g_2} + b_1 e^{2\lambda_2-2g_1} + b_2 e^{2\lambda_2-2g_2}) \;, \\
e^{-f}g_1' &=  -\frac15 \big( 2e^{-2\lambda_1}+2e^{-2\lambda_2}+e^{4\lambda_1+4\lambda_2} \big) + \frac1{40} (a_1b_2+a_2b_1) \, e^{-2\lambda_1-2\lambda_2-2g_1-2g_2} \\
&\qquad\qquad\qquad\qquad + \frac1{20} \big( 4a_1 e^{2\lambda_1-2g_1}+4b_1 e^{2\lambda_2-2g_1}-a_2 e^{2\lambda_1-2g_2}-b_2 e^{2\lambda_2-2g_2} \big) \;, \\
e^{-f}g_2' &=  -\frac15 \big( 2e^{-2\lambda_1}+2e^{-2\lambda_2}+e^{4\lambda_1+4\lambda_2} \big) + \frac1{40} (a_1b_2+a_2b_1) \, e^{-2\lambda_1-2\lambda_2-2g_1-2g_2} \\
&\qquad\qquad\qquad\qquad + \frac1{20} \big( 4a_2 e^{2\lambda_1-2g_2}+4b_2 e^{2\lambda_2-2g_2}-a_1 e^{2\lambda_1-2g_1}-b_1 e^{2\lambda_2-2g_1} \big) \;, \\
e^{-f}\lambda_1' &= - \frac25 \big( 3e^{-2\lambda_1}-2e^{-2\lambda_2}-e^{4\lambda_1+4\lambda_2} \big) + \frac1{80} (a_1b_2+a_2b_1) \, e^{-2\lambda_1 - 2\lambda_2 - 2g_1 - 2g_2} \\
&\qquad\qquad\qquad\qquad - \frac1{20} \big( 3a_1 e^{2\lambda_1-2g_1} + 3a_2 e^{2\lambda_1-2g_2} - 2b_1 e^{2\lambda_2-2g_1} - 2b_2 e^{2\lambda_2-2g_2} \big) \;, \\
e^{-f}\lambda_2' &=  -\frac25 \big( 3e^{-2\lambda_2} - 2e^{-2\lambda_1} - e^{4\lambda_1+4\lambda_2} \big) + \frac1{80} (a_1b_2+a_2b_1) \, e^{-2\lambda_1-2\lambda_2-2g_1-2g_2} \\
&\qquad\qquad\qquad\qquad - \frac1{20} \big( 3b_1 e^{2\lambda_2-2g_1} + 3b_2 e^{2\lambda_2-2g_2} - 2a_1 e^{2\lambda_1-2g_1} - 2a_2 e^{2\lambda_1-2g_2} \big) \;.
\eea
To get an AdS$_3$ vacuum we further constrain the various radial functions:
\be
e^{f(r)} = e^{f_0}/r \;,\qquad g_1 , g_2 = \text{const} \;,\qquad X_1 \equiv e^{2\lambda_1} = \text{const} \;,\qquad X_2 \equiv e^{2\lambda_2} = \text{const} \;.
\ee
The differential system reduces to an algebraic system for $(f_0,g_1,g_2,X_1,X_2)$. The solution is worked out in detail in Appendix \ref{app: 7D sugra}. Here we present the end result. The scalars are given by
\bea
X_1^5=e^{10\lambda_1} &=& \dfrac{(a_1b_2^2+a_2^2 b_1)(a_1^2b_2+a_2b_1^2)(a_1b_2+a_2b_1-b_1b_2)^2}{(a_1^2b_2^2+a_2^2b_1^2+a_1a_2b_1b_2)(a_1b_2+a_2b_1-a_1a_2)^3} \;, \\
X_2^5=e^{10\lambda_2} &=& \dfrac{(a_1b_2^2+a_2^2 b_1)(a_1^2b_2+a_2b_1^2)(a_1b_2+a_2b_1-a_1a_2)^2}{(a_1^2b_2^2+a_2^2b_1^2+a_1a_2b_1b_2)(a_1b_2+a_2b_1-b_1b_2)^3} \;.
\eea
The warp factors of the Riemann surfaces are
\be
e^{2g_1} = \frac{a_1 X_1 + b_1 X_2}{4X_1^2 X_2^2} \;,\qquad\qquad e^{2g_2} = \frac{a_2 X_1 + b_2 X_2}{4X_1^2X_2^2} \;,
\ee
and the AdS$_3$ warp factor is
\be
e^{f_0} = \frac{b_1b_2 - a_1b_2 - a_2b_1}{a_1a_2 + b_1b_2 - 2a_1b_2 - 2a_2b_1} \, X_2 \;.
\ee

Given such AdS$_3$ supergravity solutions, the central charges of the dual field theories can be extracted at leading order in $N$ in a standard way (we use the conventions of \cite{Maldacena:2000mw}):
\be
\label{7dsugracc}
c_R \,\simeq\, c_L \,\simeq\, \frac{8N^3}{\pi^2} \, e^{f_0+2g_1+2g_2} \, \vol(\Sigma_1\times \Sigma_2)=2 \eta_1\eta_2 N^3  \frac{a_1^2b_2^2 + a_2^2b_1^2 + a_1a_2b_1b_2}{2a_1b_2 + 2a_2b_1 - a_1a_2 - b_1b_2}\,.
\ee
After expressing the twist parameters $a_\sigma, b_\sigma$ in terms of $\kappa_\sigma, z_\sigma$, this expression exactly agrees with the field theory result (\ref{cR large N M5}).

In Appendix \ref{app: 7D sugra} we study in what range of parameters $(a_\sigma, b_\sigma)$ there are good AdS$_3$ solutions. The result is plotted there in Figure \ref{fig: good SUGRA M5}. We find regular supergravity solutions only when at least one of the Riemann surfaces is hyperbolic. For $\fg_{1,2}>1$ there is an infinite connected region in the $(z_1,z_2)$ parameter space, while for $\fg_1>1$ and $\fg_2=0,1$ (or viceversa) there are two disconnected infinite regions. In the appendix we also compute the matrix $k^{IJ}$ of 't~Hooft anomalies in field theory, and check that at large $N$ it matches with the Chern-Simons matrix of the effective three-dimensional supergravity on AdS$_3$.

\subsection{Holographic RG flows}
\label{subsec: 7Dto3Dflows}

\begin{figure}[t]
\begin{center}
\hfill
\includegraphics[width=.39\textwidth]{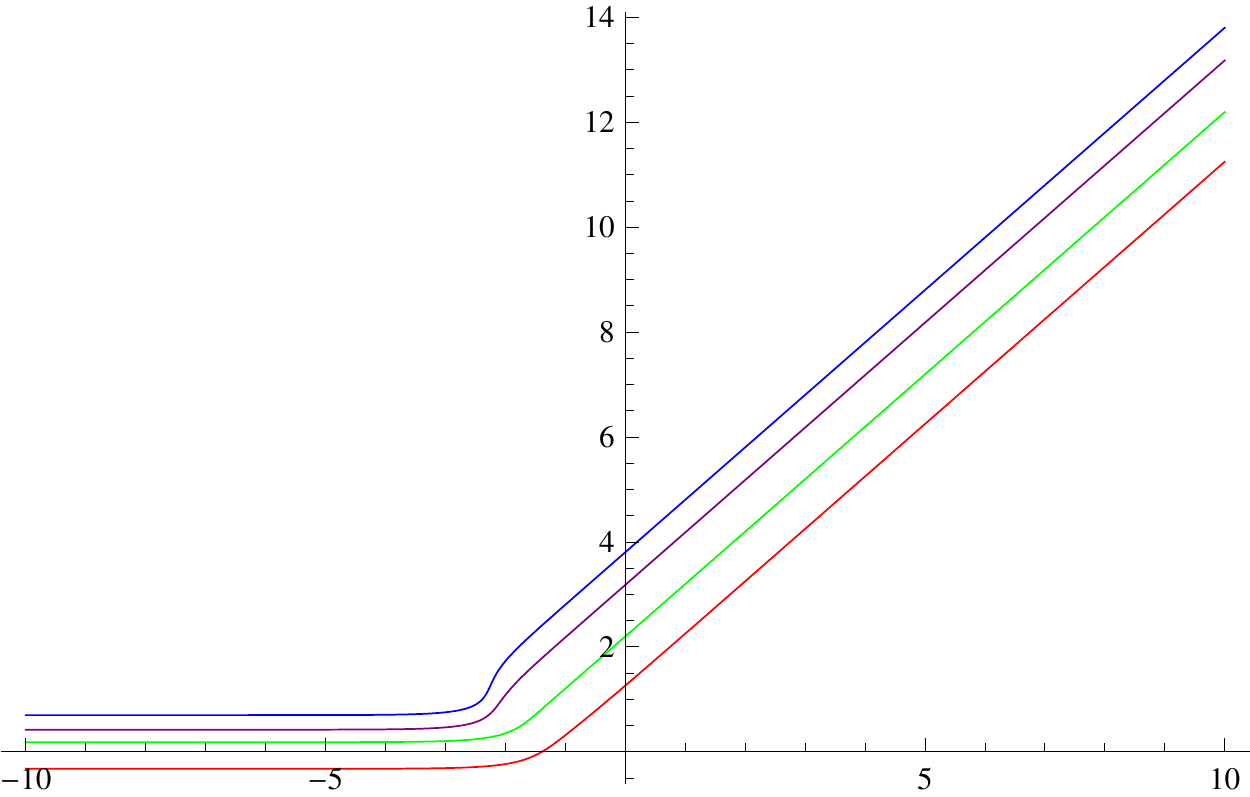}
\hfill
\includegraphics[width=.39\textwidth]{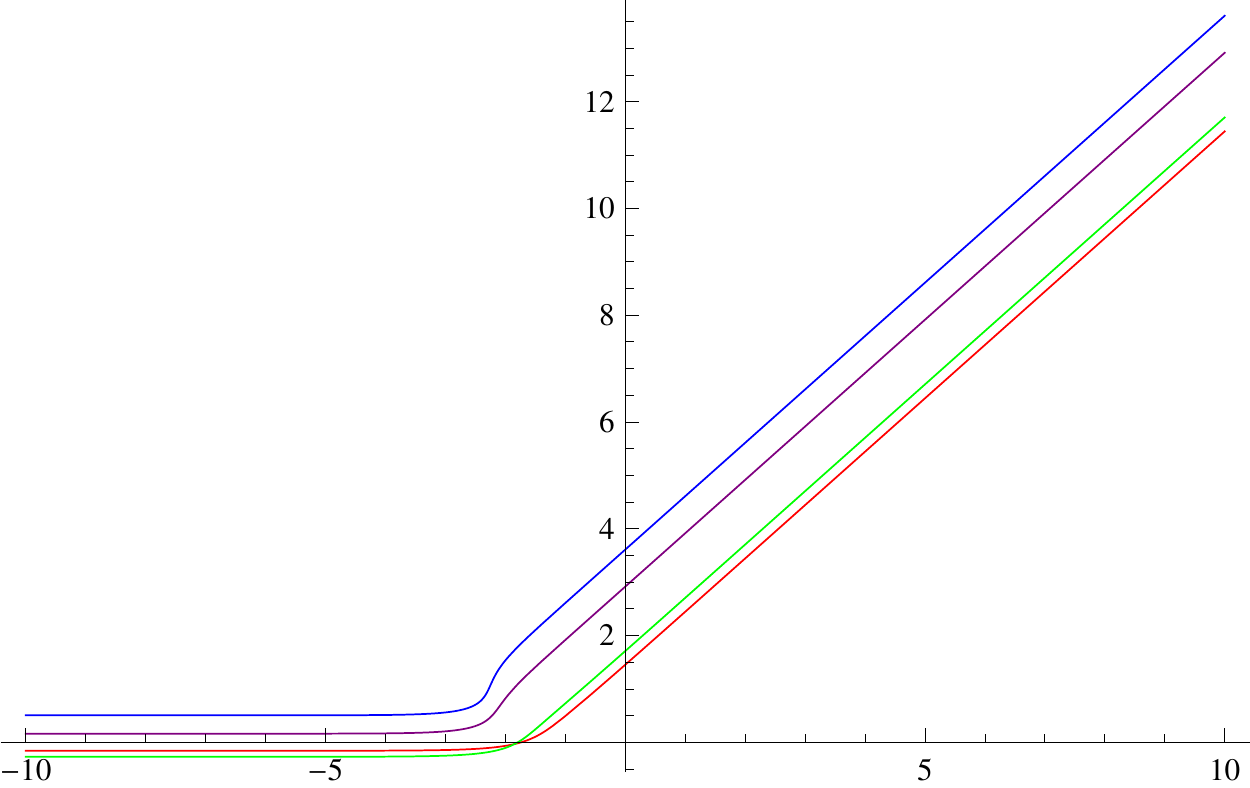}
\hfill

\vspace{.5cm}
\hfill
\includegraphics[width=.39\textwidth]{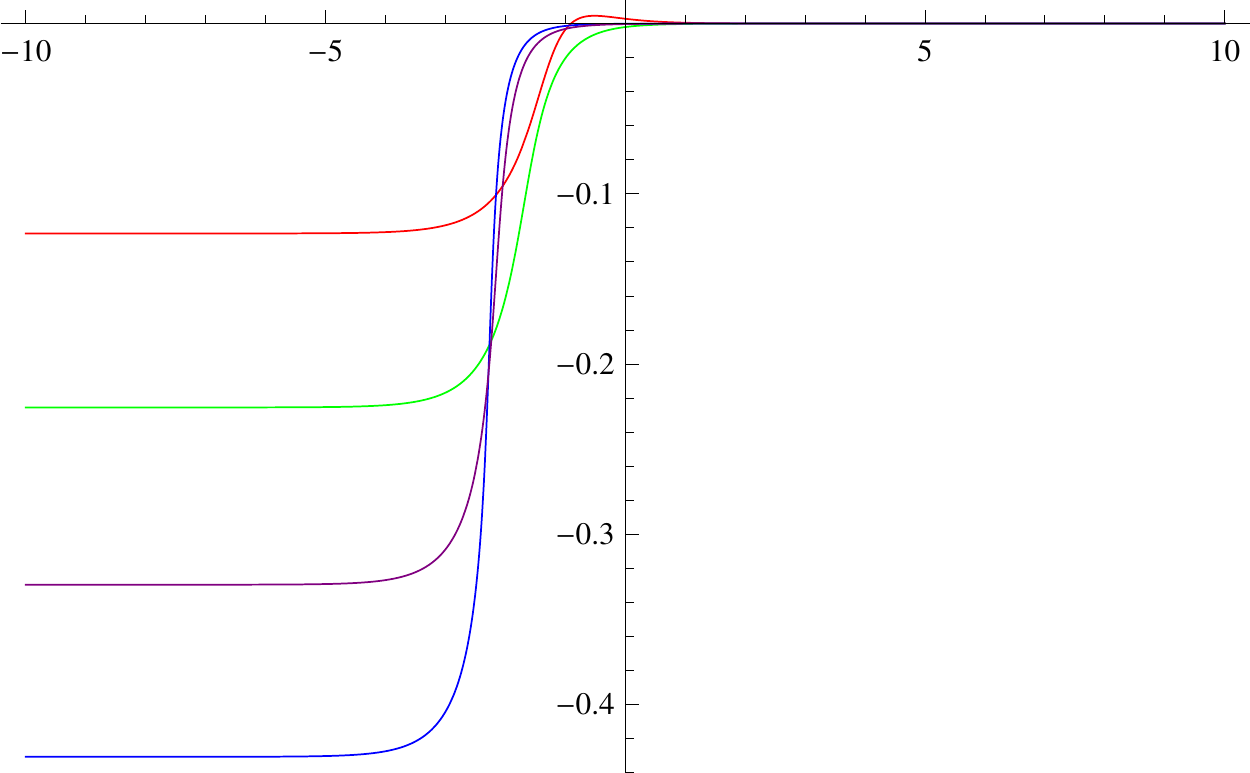}
\hfill
\includegraphics[width=.39\textwidth]{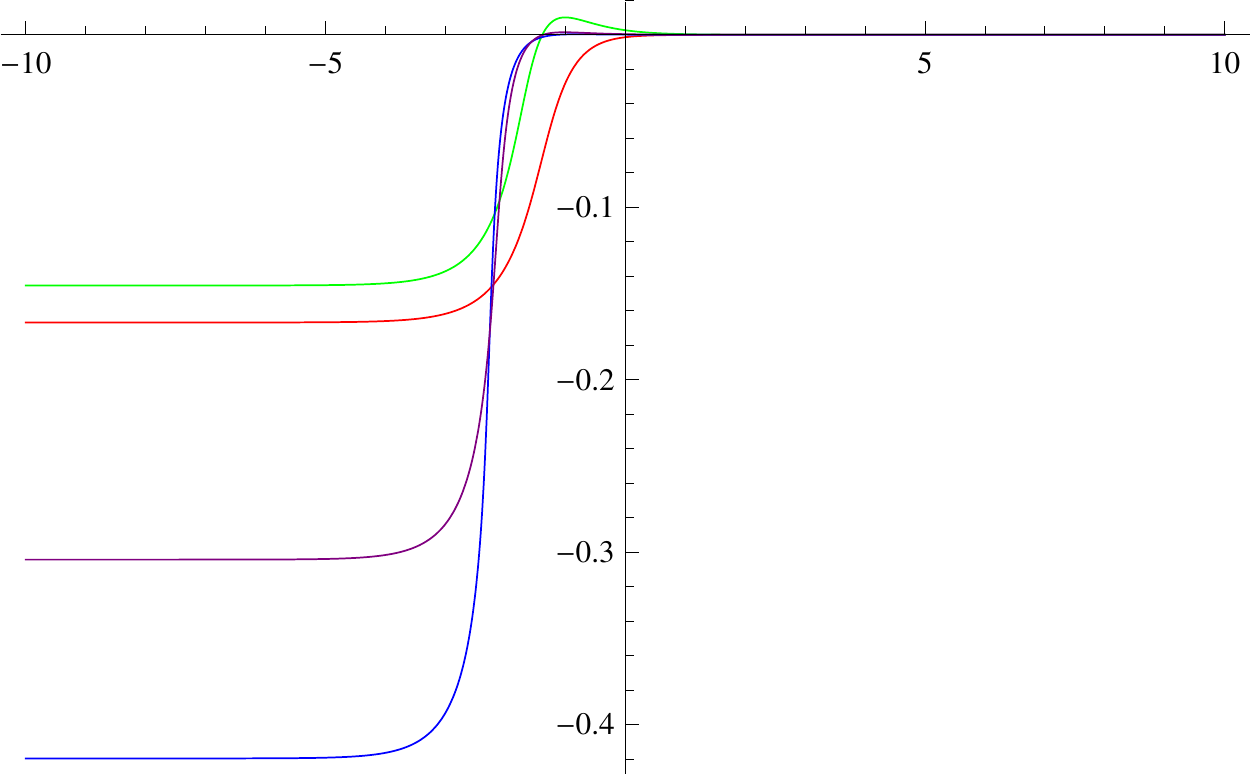}
\hfill

\caption{Numerical solutions for $g_1(\rho)$, $g_2(\rho)$, $\lambda_1(\rho)$ and $\lambda_2(\rho)$ (clockwise from upper-left) for some representative values of $a_{\sigma}$ and $b_{\sigma}$. The red, green, purple and blue curves refer to $(z_1,z_2)=(3,-5), (11,-3), (15,-7), (23,-13)$, respectively.
We have chosen $\kappa_1=\kappa_2=-1$ and $\fg_1 = \fg_2 =2$.
\label{flowplotsM5}}
\end{center}
\end{figure}

To study holographic RG flows between an asymptotically locally AdS$_7$ and the AdS$_3$ vacua in the IR one has to integrate the equations (\ref{7D BPS eqns}). It is convenient to define a new radial variable%
\footnote{Again we assume that the metric on the four-manifold is invariant under the flow. The generalization along the lines of \cite{Anderson:2011cz} would be very interesting.}
\be
\rho = f(r) -\lambda_1(r)-\lambda_2(r) \;,
\ee
such that $\frac{d\rho}{dr} = -e^f D$ with
\be
D \equiv e^{4(\lambda_1 + \lambda_2)} + \dfrac{1}{16}(a_1b_2+a_2b_1)e^{-2(\lambda_1+\lambda_2+g_1+g_2)} \;.
\ee
Then the BPS equations read:
\bea
Dg_1'(\rho) &= \frac15 \big( 2e^{-2\lambda_1}+2e^{-2\lambda_2}+e^{4\lambda_1+4\lambda_2} \big) - \frac1{40} (a_1b_2+a_2b_1) \, e^{-2\lambda_1-2\lambda_2-2g_1-2g_2} \\
&\qquad\qquad\qquad\qquad - \frac1{20} \big( 4a_1 e^{2\lambda_1-2g_1}+4b_1 e^{2\lambda_2-2g_1}-a_2 e^{2\lambda_1-2g_2}-b_2 e^{2\lambda_2-2g_2} \big) \;, \\
Dg_2'(\rho)&= \frac15 \big( 2e^{-2\lambda_1}+2e^{-2\lambda_2}+e^{4\lambda_1+4\lambda_2} \big) - \frac1{40} (a_1b_2+a_2b_1) \, e^{-2\lambda_1-2\lambda_2-2g_1-2g_2} \\
&\qquad\qquad\qquad\qquad - \frac1{20} \big( 4a_2 e^{2\lambda_1-2g_2}+4b_2 e^{2\lambda_2-2g_2}-a_1 e^{2\lambda_1-2g_1}-b_1 e^{2\lambda_2-2g_1} \big) \;, \\
D\lambda_1'(\rho) &=  \frac25 \big( 3e^{-2\lambda_1}-2e^{-2\lambda_2}-e^{4\lambda_1+4\lambda_2} \big) - \frac1{80} (a_1b_2+a_2b_1) \, e^{-2\lambda_1 - 2\lambda_2 - 2g_1 - 2g_2} \\
&\qquad\qquad\qquad\qquad + \frac1{20} \big( 3a_1 e^{2\lambda_1-2g_1} + 3a_2 e^{2\lambda_1-2g_2} - 2b_1 e^{2\lambda_2-2g_1} - 2b_2 e^{2\lambda_2-2g_2} \big) \;, \\
D\lambda_2'(\rho) &=  \frac25 \big( 3e^{-2\lambda_2} - 2e^{-2\lambda_1} - e^{4\lambda_1+4\lambda_2} \big) - \frac1{80} (a_1b_2+a_2b_1) \, e^{-2\lambda_1-2\lambda_2-2g_1-2g_2} \\
&\qquad\qquad\qquad\qquad + \frac1{20} \big( 3b_1 e^{2\lambda_2-2g_1} + 3b_2 e^{2\lambda_2-2g_2} - 2a_1 e^{2\lambda_1-2g_1} - 2a_2 e^{2\lambda_1-2g_2} \big) \;.
\eea
Note that $e^{2f}dr^2 = d\rho^2/D^2$. One can integrate this system of non-linear ODEs numerically and some representative flow solutions are presented in Figure \ref{flowplotsM5}. With our choice of conventions the UV asymptotic region is at $\rho \to \infty$ and the IR AdS$_3$ region is at $\rho \to -\infty$.

\subsection{The uplifted solutions in eleven dimensions}
\label{subsec: 7Dto11Duplift}

Since the seven-dimensional supergravity we used is a consistent truncation of eleven-dimensional supergravity, all solutions discussed above can be uplifted. For this purpose, we exploit the uplifting formul\ae{} in \cite{Cvetic:1999xp} (see also \cite{Nastase:1999cb,Nastase:1999kf}). The eleven-dimensional solution is presented most succinctly in terms of the functions $X_1 \equiv e^{2\lambda_1}$, $X_2 \equiv e^{2\lambda_2}$, $X_0 \equiv \left(X_1X_2\right)^{-2}$. The metric takes the form
\be
\label{11dmetric}
ds^2_{11}= \Delta^{1/3} ds_7^2 + \frac14 \Delta^{-2/3} \bigg[ X_0^{-1} d\mu_0^2 + \sum\nolimits_{i=1,2} X_{i}^{-1} \big( d\mu_i^2 + \mu_i^2 (d\phi_i + 4A^{(i)})^2 \big) \bigg] \;,
\ee
where $ds^2_7$ is defined in \eqref{7DsugraAnsatz}, $A^{(i)}$ are the two seven-dimensional gauge fields that we called $A^A, A^B$ in (\ref{7DsugraAnsatz}), and \be
\Delta = \sum\nolimits_{\alpha=0,1,2} X_\alpha \mu_\alpha^2 \;,\qquad\qquad\qquad \sum\nolimits_{\alpha=0,1,2} \mu_\alpha^2 = 1 \;.
\ee
The periods of the angular coordinates $\phi_i$ are $2\pi$, and for $\mu_\alpha$ we can use the parametrization
\be
\mu_0 =\cos\alpha \;,\qquad\qquad \mu_1 = \sin\alpha\, \cos\beta \;,\qquad\qquad \mu_2 = \sin\alpha\, \sin\beta \;.
\ee
The four-form flux of the eleven-dimensional solution is
\begin{multline}
\star_{11} F_{(4)} = 4 \sum_{\alpha=0,1,2} \big( X_\alpha^2 \mu_\alpha^2 -\Delta \, X_\alpha \big) \, \epsilon_{(7)} + 2 \Delta \, X_0 \, \epsilon_{(7)} + \frac14 \sum_{\alpha=0,1,2} X_\alpha^{-1}(\star_7 dX_\alpha) \wedge d(\mu_\alpha^2) \\
+ \frac14 \sum_{i=1,2} X_i^{-2} d(\mu_i^2) \wedge (d\phi_i+4A^{(i)}) \wedge \star_7 F^{(i)} \;,
\end{multline}
where $F^{(i)}=dA^{(i)}$, $\epsilon_{(7)}$ is the volume form of $ds^2_7$, while $\star_7$ and $\star_{11}$ are the Hodge operators for $ds^2_7$ and $ds^2_{11}$ in \eqref{11dmetric} respectively. It is worth pointing out that this solution captures the full holographic RG flows from AdS$_7$ to AdS$_3$ and not only the IR AdS$_3$ vacua. At the IR fixed points the eleven-dimensional metric is a warped product of AdS$_3$ with an eight-dimensional compact manifold which is a squashed $S^4$ fibration over $\Sigma_1 \times \Sigma_2$. The isometry of the internal manifold for generic values of $\lambda_i$ is $U(1)_A \times U(1)_B$. It would be interesting to understand in some detail the geometry of this class of compact eight-dimensional manifolds and how they fit in the more general class of manifolds with flux which yield supersymmetric warped AdS$_3$ compactifications of eleven-dimensional supergravity \cite{Gauntlett:2006ux,Figueras:2007cn}.

There are some interesting orbifolds of the eleven-dimensional solutions above that we would like to comment on. Modding out by
\be
\phi_1 \,\to\, -\phi_1 \;,\qquad\qquad \phi_2 \,\to\, -\phi_2 \;,\qquad\qquad \mu_0 \,\to\, -\mu_0 \;,
\ee
the topological $S^4$ is replaced by $\mathbb{RP}^4$ which again is fibered over the two Riemann surfaces. These smooth solutions preserve the same amount of supersymmetry and provide the holographic dual to twisted compactifications of the 6d $\cN=(2,0)$ theory of type $D_N$ on $\Sigma_1\times \Sigma_2$. The central charges of the solutions (at leading order in $N$) can be easily computed: the volume of $\bR\bP^4$ is half of that of $S^4$, and the number of units of $F_{(4)}$ flux on it is half as much as on the covering space. As a result, the central charge is four times larger than in \eqref{7dsugracc}, in perfect agreement with the field theory result mentioned after \eqref{cR large N M5}.

Another interesting orbifold is given by
\be
\phi_1 \,\to\, e^{\frac{2\pi i}{k}} \phi_1 \;,\qquad\qquad \phi_2 \,\to\, e^{-\frac{2\pi i}{k}} \phi_2\;, \qquad\qquad k\in \mathbb{Z} \;.
\ee
Modding out by such $\mathbb{Z}_k$ one obtains a solution with $A_{k-1}$ type singularities at the north and south poles of the topological $S^4$. For all values of $k$, $\fg_i$ and $z_i$ the solutions preserve $\cN=(0,2)$ supersymmetry. These mildly singular supergravity solutions are dual to twisted compactifications of the 6d $\cN=(1,0)$ theory of type $(A_{N-1}, A_{k-1})$ on $\Sigma_1 \times \Sigma_2$.

\section{Six-dimensional $\cN=(2,0)$ theory on other four-manifolds}
\label{sec: other 4manifolds}

Here we consider compactifications of the 6d $\cN=(2,0)$ theory on more general four-manifolds than in Section \ref{sec: M5-branes}. First we compute the central charges of the IR 2d SCFTs, exploiting $c$-extremization when needed. Second we compare the results for the $A_N$ and $D_N$ series at large $N$ with the dual supergravity solutions. Some of these solutions have been constructed in \cite{Gauntlett:2000ng, Gauntlett:2001jj} and some are new. As in Section \ref{sec: M5-branes}, for the $A_{N-1}$ case we can adopt a geometric point of view and think of the construction as arising from $N$ M5-branes wrapping 4-cycles in higher-dimensional manifolds.

\subsection{K\"ahler 4-cycle in CY$_4$}
\label{subsec: K4inCY4}

Consider a K\"ahler four-manifold $M_4$, whose holonomy group is (contained into)
$U(2)_s = SU(2)_s \times U(1)_s$. To preserve 2d $\cN=(0,2)$ supersymmetry, we turn on a background Abelian gauge field along an $SO(2)_A \times SO(2)_B$ embedded block-diagonally into the R-symmetry group $SO(5)$,
and proportional to the $U(1)_s$ spin connection.%
\footnote{When $b^{(1,1)}>1$, a more general background might be considered.}
In particular the background is taken along the generator
\be
\label{KE4 twist generator}
T = \frac{1+z}2 T_A + \frac{1-z}2 T_B\;,
\ee
where $z$ parametrizes the twist.
From the geometric point of view, in order to preserve supersymmetry the total space must be a local $CY_4$ (times a spectator $\bR$) and $M_4$ is a holomorphic 4-cycle in it. Equivalently, the normal bundle is taken as a rank-two line bundle $\cL_A \oplus \cL_B$ whose total degree equals the one of the canonical bundle of $M_4$. Denoting by $t_{1,2}$ the Chern roots of the tangent bundle to $M_4$ and by $n_{A,B}$ those of the normal bundle, the $CY_4$ condition is
\be
t_1 + t_2 + n_A + n_B = 0 \;.
\ee
We get a one-parameter family of twists. The background breaks the global symmetry from $SO(5)$ to $SO(2)_A \times SO(2)_B$, and generically the IR R-symmetry has to be determined with $c$-extremization.

To determine the anomalies of the 2d theory, we also turn on a background for the 2d trial R-symmetry along the generator
\be
 T_R = (1+\epsilon) T_A + (1-\epsilon)T_B\;.
\ee
This amounts to the following shifts of the Chern roots of the normal bundle:
\be
n_A = - \frac{1+z}2 (t_1 + t_2) + (1+\epsilon) c_1(F_R) \;,\qquad n_B = - \frac{1-z}2 (t_1 + t_2) + (1-\epsilon)c_1(F_R) \;.
\ee
Here $\epsilon$ parametrizes the mixing of the R-symmetry with the flavor symmetry.
The R-charge of the supercharges has been fixed to 1. For $z=0$ the twist preserves an enhanced $SU(2)_I \times U(1)_R$ global symmetry: the R-symmetry must be the Abelian factor represented by $n_A + n_B$, \ie{} it must be $\epsilon = 0$.

We integrate the anomaly polynomial $I_8[G]$ (\ref{G M5 anomaly}) of the 6d $\cN=(2,0)$ theory of type $G$ on $M_4$, as we did in Section \ref{sec: M5 field theory}. Only the classes that contain a 4-form on the tangent bundle of $M_4$ contribute. The first Pontryagin number $P_1(M_4)$ and the Euler number $\chi(M_4)$ are
\be
P_1(M_4) = \int_{M_4} (t_1^2 + t_2^2) \;,\qquad\qquad \chi(M_4) = \int_{M_4} t_1t_2 \;.
\ee
Comparing with the anomaly polynomial of a 2d theory (\ref{anomaly polynomial 2d}) we extract the trial central charge $c_R^\text{tr}(\epsilon)$ and the gravitational anomaly $k=c_R-c_L$:
\bea
c_R^\text{tr}(\epsilon) &= \frac{P_1 + 2\chi}8 \Big[ (3-z^2) d_G h_G + 2r_G -4d_G h_G z \epsilon + (d_G h_G + r_G)(3z^2-1)\epsilon^2 \Big] + \frac{r_G}4 \chi(1+\epsilon^2) \\
k &= \frac{r_G}8 \Big( P_1(3+z^2) + 2\chi(1+z^2) \Big) \;.
\eea
The trial central charge is extremized for
\be\label{epsK4CY4}
\epsilon = \frac{2z (P_1+2\chi) d_G h_G}{(3z^2-1)(P_1+2\chi)(d_G h_G+r_G) +2\chi r_G} \;,
\ee
when the denominator does not vanish. This result correctly reproduces the expectation we mentioned before that $\epsilon=0$ for $z=0$.
Plugging the value of $\epsilon$ \eqref{epsK4CY4} in the expression for $c_R^\text{tr}(\epsilon)$ we find the exact right-moving central charge
\begin{multline}
c_R = \bigg[ 3d_G^2 h_G^2 (z^2-1)^2 (P_1 + 2\chi)^2 + d_G h_G r_G (P_1 + 2\chi) \Big( P_1 (5-16z^2+3z^4) + 6\chi(1-6z^2+z^4) \Big) \\
- 2r_G^2(P_1 + 3\chi) \big( P_1(3z^2-1) + 6z^2\chi \big) \bigg] / \Big[ 8(1-3z^2)(P_1+2\chi)(d_Gh_G+r_G) - 16 \chi r_G \Big] \;.
\end{multline}
The left-moving central charge is $c_L = c_R-k$. The second derivative of $c_R^\text{tr}(\epsilon)$ is
\be
\parfrac{^2 c_R^\text{tr}(\epsilon)}{\epsilon^2} = \frac{P_1 + 2\chi}4 (d_G h_G + r_G)(3z^2-1) + \frac{r_G}2 \chi \;,
\ee
which is positive when the extra flavor symmetry is right-moving, and negative when left-moving.

\label{(0,4) from M5s Kahler}
For the special twist $z=1$ (or $z=-1$), the total space is a K\"ahler 4-cycle in a $CY_3$ (times a spectator $\bR^3$), supersymmetry is enhanced to $\cN=(0,4)$ and the global symmetry is enhanced to $SO(2)_A \times SO(3)_B$. The central charges can be found by exploiting an $\cN=(0,2)$ subalgebra, whose R-symmetry is the Cartan of $SO(3)_B$. This corresponds to $\epsilon=-1$ and yields
\be
c_R = \big( P_1 + 2\chi \big) d_G h_G + \frac{P_1 + 3\chi}2\, r_G \;,\qquad\qquad k = c_R - c_L = \frac{r_G}2\, (P_1 + \chi) \;.
\ee
On a K\"ahler manifold the Betti numbers are $b_2^+ = 2b^{(2,0)} + 1$ and $b_2^- = b^{(1,1)} - 1$, therefore using $P_1 = 3 (b_2^+ - b_2^-)$ and $\chi = b_2^+ + b_2^- - 4 b^{(1,0)} + 2$ we get $(P_1 + 3\chi)/2 = 6 \big( b^{(2,0)} - b^{(1,0)} + 1)$.%
\footnote{One also obtains
$$
c_R - c_L = r_G \big( 4 b^{(2,0)} - b^{(1,1)} - 2b^{(1,0)} + 4 \big) \;.
$$
This matches with what found in \cite{Ganor:1996xg} for the $A_1$ $\cN=(2,0)$ theory on K\"ahler four-manifolds with $b^{(1,0)} = 0$.
}
This implies that the central charge $c_R$ is a multiple of 6, as it should be because of $\cN=4$ superconformal symmetry. On the other hand $c$-extremization would give a different (and wrong) result: this suggests that the vacuum is non-normalizable, and indeed we show below that at large $N$ there are no AdS$_3$ solutions. This discussion parallels the one on page \pageref{(0,4) from M5s} and in Appendix \ref{app: free theories}.

As we have emphasized throughout the paper, finding positive central charges does not guarantee the existence of an IR fixed point with normalizable vacuum. Indeed supergravity (analyzed for the $z=0$ twist in \cite{Gauntlett:2000ng}) suggests that a normalizable IR fixed point exists only for K\"ahler 4-cycles that admit a negative-curvature Einstein metric.

Let us now consider the large $N$ limit. For $G = A_{N-1}$ we obtain
\be
\label{KE4 FT central charge large N}
c_R \;\simeq\; c_L \;\simeq\; \frac{3(1-z^2)^2}{8(1-3z^2)} (P_1 + 2\chi) \, N^3 \;.
\ee
As we will see below, this nicely matches with the supergravity result.
Although positivity of the central charge requires $z^2<\frac13$, we find from supergravity that actual IR fixed points exist only for $|z|<\frac13$. Whether examples besides $z=0$ exist depends on the quantization of $z$ which in turn depends on the choice of four-manifold.

The central charges of the Abelian M5-brane theory are obtained by setting $r_G=1$, $h_G=0$:
\be
c_R = \frac14 (P_1 + 3\chi) \;,\qquad c_L = \frac18 \Big( 2\chi(2-z^2) - P_1(1+z^2) \Big)\;,
\ee
and the IR R-symmetry is determined by $\epsilon = 0$ (whenever $2\chi + (3z^2-1)(P_1+2\chi)\neq0$).

\paragraph{The supergravity dual.} Here we construct the supergravity solutions for the backreacted configuration of $N$ M5-branes wrapped on $M_4$, dual to the twisted compactified 6d $\cN=(2,0)$ theory of type $A_{N-1}$. We will specialize to the case of 4-cycles $M_4$ admitting an Einstein metric, leaving the study of the most general case to the future. For the special twist $z=0$, the solution has been constructed in \cite{Gauntlett:2000ng}. We work with the same set of fields of maximal 7d gauged supergravity as in Section \ref{sec: M5 sugra}. We consider the metric Ansatz
\be
ds^2 = e^{2f(r)} \big( -dt^2+dz^2+dr^2 \big) + e^{2g(r)} ds^2_{M_4}
\ee
where $ds^2_{M_4}$ is a constant-curvature (\ie{} Einstein) metric on $M_4$, normalized as
\be
R^{(4)}_{\mu\nu} = \kappa \, g^{(4)}_{\mu\nu}
\ee
in terms of its Ricci tensor $R^{(4)}_{\mu\nu}$ and $\kappa \in \{ 1,0,-1\}$. The two supergravity scalars $\lambda_{1,2}$ are functions of $r$ only, and the two Abelian gauge fields $A^{A,B}$ have field strengths proportional to the K\"ahler form $\omega$ on $M_4$ with coefficients $a$ and $b$ respectively:
\be
F^A = - \frac a4 \, \omega \;,\qquad\qquad F^B = - \frac b4 \, \omega \;.
\ee
The Einstein condition is equivalently $\cR = \kappa \, \omega$ in terms of the Ricci form $\cR$.
From this Ansatz one derives the following supergravity BPS equations:
\bea
\label{BPSeqnK4CY4}
e^{-f}f' &= -\frac15 \big( 2e^{-2\lambda_1} + 2e^{-2\lambda_2} + e^{4\lambda_1+4\lambda_2} \big) - \frac3{40} ab \, e^{-2\lambda_1-2\lambda_2-4g} - \frac1{10} e^{-2g} \big( a \, e^{2\lambda_1} + b \, e^{2\lambda_2} \big) \\
e^{-f}g' &= -\frac15 \big( 2e^{-2\lambda_1} + 2e^{-2\lambda_2} + e^{4\lambda_1+4\lambda_2} \big) + \frac1{20} ab \, e^{-2\lambda_1-2\lambda_2-4g} + \frac3{20} e^{-2g} \big( a \, e^{2\lambda_1} + b \, e^{2\lambda_2} \big) \\
e^{-f}\lambda_1' &= -\frac25 \big( 3e^{-2\lambda_1} - 2e^{-2\lambda_2} - e^{4\lambda_1+4\lambda_2} \big) + \frac1{40} ab \, e^{-2\lambda_1 - 2\lambda_2 - 4g} - \frac1{10} e^{-2g} \big( 3a \, e^{2\lambda_1} - 2b \, e^{2\lambda_2} \big) \\
e^{-f}\lambda_2' &= -\frac25 \big( 3e^{-2\lambda_2} - 2e^{-2\lambda_1} - e^{4\lambda_1+4\lambda_2} \big) + \frac1{40} ab \, e^{-2\lambda_1 - 2\lambda_2 - 4g} - \frac1{10} e^{-2g} \big( 3b \, e^{2\lambda_2} - 2a \, e^{2\lambda_1} \big) \;.
\eea
In addition to these differential equations one also has to impose
\begin{equation}
a+b = -\kappa \;.
\end{equation}
To obtain AdS$_3$ solutions we further fix $e^{f(r)} = e^{f_0}/r$ and take $f_0, g, \lambda_{1,2}$ to be constant. Proceeding in a way parallel to Appendix \ref{app: 7D sugra}, we find the solution:
\bea\label{K4CY4sol}
e^{5f_0} &= \frac{a^2b^2(a+b)^2(2a-b)^2(2b-a)^2}{3(4ab - a^2 - b^2)^5}\;, \qquad\qquad &
e^{10\lambda_1} &= \frac{b^2(a+b)^2(2a-b)^2}{3a^3(2b-a)^3}\;, \\
e^{10g} &= \frac{27a^4b^4}{1024 (a+b)(2a-b)(2b-a)}\;, &
e^{10\lambda_2} &= \frac{a^2(a+b)^2(2b-a)^2}{3b^3(2a-b)^3} \;.
\eea
Taking into account the constraint $a+b=-\kappa$, we find good supergravity solutions (where all right-hand-sides in \eqref{K4CY4sol} above are positive) only for $\kappa = -1$. We can adopt the parametrization
\be
a = \frac{1+z}2 \;,\qquad\qquad b = \frac{1-z}2\;,
\ee
where the parameter $z$ has to be identified with that in (\ref{KE4 twist generator}). Good supergravity solutions only for the limited range
\be
\frac13 < a < \frac23 \qquad\text{that is}\qquad |z|<\frac13 \;,\qquad\text{and}\qquad \kappa = -1 \;.
\ee
The central charge for these AdS$_3$ vacua is computed in the standard way:
\be
c = \frac{8N^3}{\pi^2} \, e^{f_0+4g} \vol(M_4) = \frac{a^2 b^2}{4ab-a^2-b^2} \, \frac{3\vol(M_4) N^3}{2\pi^2} = \frac{(1-z^2)^2}{1-3z^2} \, \frac{3\vol(M_4)N^3}{16\pi^2} \;.
\ee
As we review in Appendix \ref{app: some geometry}, for a negatively-curved K\"ahler-Einstein four-manifold the volume is related to the first Pontryagin number $P_1(M_4)$ and Euler number $\chi(M_4)$ by the formula (\ref{volume KE4}):
\be
P_1 + 2\chi = \frac1{2\pi^2} \, \vol(M_4) \;.
\ee
With this, the holographic central charge above perfectly agrees with the field theory result at large $N$ (\ref{KE4 FT central charge large N}).

\subsection{K\"ahler 4-cycle in HK$_2$}
\label{subsec: K4inHK2}

Let us consider again the K\"ahler four-manifold $M_4$, with holonomy $U(2)_s \cong SU(2)_s \times U(1)_s$, but this time with a different twist. We switch on a non-Abelian R-symmetry background equal to the whole $U(2)_s$ part of the spin connection; this background is embedded in the $SO(5)_R$ R-symmetry in the natural way, identifying $SO(4)_s \supset U(2)_s$ with $SO(4) \subset SO(5)_R$. By decomposing the supercharges $Q$ in the representation $\rep{4} \otimes \rep{4}$ of $SO(5,1) \times SO(5)_R$ under $SO(1,1) \times SU(2)_s \times U(1)_s \times SU(2)_I \times U(1)_I$:
\be
\rep{4} \otimes \rep{4} \;\to\; \big[ \tfrac i2 \otimes \rep{1}_\frac12 + \tfrac i2 \otimes \rep{1}_{-\frac12} + \big( - \tfrac i2 \big) \otimes \rep{2}_0 \big] \otimes \big[ \rep{1}_\frac12 + \rep{1}_{-\frac12} + \rep{2}_0 \big]\;,
\ee
it is easy to see that the twist preserves $\cN=(1,2)$ supersymmetry. It also leaves an unbroken $U(1)_I$ symmetry, which has to be identified with the 2d IR R-symmetry.
By looking at the transformation properties of the scalars $\Delta \in \rep{1} \otimes \rep{5}$ after the twist, one finds that the total space is $T^*M_4$, the cotangent bundle to $M_4$.
Therefore the total geometry is a local hyper-K\"ahler manifold of quaternionic dimension 2, and $M_4$ is a complex special Lagrangian 4-cycle (complex with respect to one complex structure, Lagrangian with respect to the other two).

We notice that one could have considered a twist by an $SU(2)_I$ background only, equal to the $SU(2)_s$ part of the spin connection. Or one could have switched on the $U(1)_I$ part as well, proportional to the $U(1)_s$ spin connection but by a constant different than $\pm 1$. These twists lead to $\cN=(1,0)$ supersymmetry, and will not be considered further.

The R-symmetry $U(1)_I$ is singled out without the need to apply $c$-extremization. To compute its 't~Hooft anomaly and the central charges, we resort again to integration of the anomaly polynomial \cite{Benini:2009mz, Alday:2009qq, Bah:2011vv, Bah:2012dg}. Denoting by $t_{1,2}$ the Chern roots of the tangent bundle to $M_4$, and by $n_{A,B}$ those of the normal bundle, the twist corresponds to
\be
n_A = -t_1 + c_1(F_R) \;,\qquad\qquad n_B = -t_2 + c_1(F_R)\;,
\ee
and the condition for supersymmetry is
\be
t_1 + t_2 + n_A + n_B = 0 \;.
\ee
The R-symmetry generator is $T_R = T_A + T_B$. By integrating $I_8[G]$ on $M_4$ and comparing with $I_4$ we extract
\be
c_R = \frac14 \Big( d_G h_G (P_1 + 4\chi) + r_G (P_1 + 3\chi) \Big) \;,\qquad\qquad k = c_R - c_L = \frac{r_G P_1}2 \;.
\ee

For $G = A_{N-1}$ and at large $N$, we obtain the leading behavior
\be
\label{HK2 central charge sugra}
c_R \;\simeq\; c_L \;\simeq\; \frac14 (P_1 + 4\chi) N^3 \;.
\ee
Let us compare this formula with supergravity. The solutions for the near-horizon geometry of $N$ M5-branes on $M_4$ have been constructed in \cite{Gauntlett:2001jj}. A necessary condition for supersymmetry is that $M_4$ admits a K\"ahler-Einstein metric of constant sectional curvature, implying that $M_4$ is $\bC\bP^2$ for $\kappa=1$, flat space for $\kappa=0$, the Bergman metric on the open unit ball $D^2$ in $\bC^2$ (also called the complex hyperbolic plane $\bC\bH^2$) for $\kappa = -1$, or a quotient of these spaces by a discrete group of isometries. The AdS$_3$ supergravity solutions then exist only for $\kappa = -1$.
The central charge, computed holographically, is
\be
c = \frac{5N^3}{24\pi^2} \, \vol(M_4) \;.
\ee
Now we can use the volume formula (\ref{volume KE4}) for negatively-curved K\"ahler-Einstein four-manifolds, supplemented by the extra constraint $P_1 = \chi$ (\ref{volume extra condition CH2}) for complex hyperbolic space-forms $\bC\bH^2/\Gamma$, to get:
\be
P_1 + 4\chi = 5 \chi = \frac{5}{6\pi^2} \vol(M_4) \;.
\ee
With this, the holographic central charge exactly matches with (\ref{HK2 central charge sugra}).

For the Abelian M5-brane theory ($r_G=1$, $h_G=0$) we get
\begin{equation}
c_R = \frac14 \, (3\chi + P_1) \;,\qquad\qquad  c_L = \frac14 \, (3\chi- P_1) \;.
\end{equation}

\subsection{Co-associative 4-cycle in $G_2$-holonomy manifolds}
\label{subsec: M4inG2}

Finally, let us consider a generic compact four-manifold $M_4$ with holonomy $SO(4)_s \simeq SU(2)_{s\ell} \times SU(2)_{sr}$. To preserve $\cN=(0,2)$ supersymmetry we take the following twist: decompose the R-symmetry $SO(5)_R \to SO(3)_I \times SO(2)_I$, then turn on an R-symmetry background equal to the $SU(2)_{s\ell}$ part of the spin connection and identified with $SO(3)_I$. This leaves an unbroken $U(1)_I$ global symmetry, which is identified with the 2d IR R-symmetry. By looking at the transformation properties of the scalars $\Delta \in \rep{1} \otimes \rep{5}$ after the twist, one realizes that the total space is $\Lambda_- M_4$, the bundle of anti-self-dual 2-forms on $M_4$. Therefore the total geometry is a local $G_2$-holonomy manifold (of real dimension 7), and $M_4$ is a co-associative 4-cycle.

The R-symmetry is $U(1)_I$ and to compute anomalies and central charges we integrate the anomaly polynomial $I_8[G]$. The twist corresponds to
\be
n_A = -t_1 - t_2 \;,\qquad\qquad n_B = 2c_1(F_R)\;,
\ee
and the condition for supersymmetry is $t_1 + t_2 + n_A + n_B = 0$. The properly normalized R-symmetry generator is $T_R = 2T_B$. We extract
\be
c_R = d_H h_G (P_1 + 2\chi) + \frac{r_G}2 (P_1 + 3\chi) \;,\qquad\qquad k=c_R - c_L = \frac{r_G}2 (P_1 + \chi) \;.
\ee

For $G = A_{N-1}$ and at large $N$, we obtain the leading behavior
\be
\label{G2 central charges sugra}
c_R \;\simeq\; c_L \;\simeq\; (P_1 + 2\chi) \, N^3 \;.
\ee
Let us compare this formula with supergravity. The solutions for the near-horizon geometry of $N$ M5-branes on $M_4$ have been constructed in \cite{Gauntlett:2000ng}.  A necessary condition for supersymmetry is that $M_4$ admits an Einstein metric with purely anti-self-dual Weyl tensor.%
\footnote{In 2 and 3 dimensions an Einstein metric is locally completely fixed, leading for negative curvature to the hyperbolic planes $\bH^2$ and $\bH^3$ respectively. In 4 dimensions the Einstein condition leaves the Weyl tensor free. Of course $\bH^4$ is one such space.}
Then solutions exist only for negative curvature $\kappa = -1$. The central charge, computed holographically, is
\be
c = \frac{N^3}{6\pi^2} \, \vol(M_4) \;.
\ee
Exploiting the volume formula (\ref{volume HFE4}) for $W_+=0$, $\vol(M_4) = 6\pi^2(P_1 + 2\chi)$, this exactly matches with (\ref{G2 central charges sugra}).

For the Abelian M5-brane theory (set $r_G=1$, $h_G = 0$) we get
\be
c_R = \frac12 (P_1 + 3\chi) \;,\qquad\qquad c_L = \chi \;.
\ee

\section{Conclusions}
\label{sec: conclusions}

In this paper we provided a detailed proof of the $c$-extremization principle for 2d $\cN=(0,2)$ SCFTs uncovered in \cite{Benini:2012cz}. We also studied a plethora of examples where $c$-extremization is essential to calculate the correct R-symmetry and central charges of interacting $\cN=(0,2)$ SCFTs. All our examples arise at the end of RG flows from higher dimensional theories on compact manifolds and one may wonder whether other applications can be found. Natural places to look at are non-linear sigma models and gauged linear sigma models. In particular, the $S^2$ partition function of $\cN=(2,2)$ GLSMs, as function of the R-symmetry, has been computed in \cite{Benini:2012ui, Doroud:2012xw} and it would be interesting to understand if it is related to $c$-extremization.

It would also be interesting to explore if $c$-extremization can teach us anything about the $c$-theorem for $\cN=(0,2)$ theories. As emphasized in \cite{Freedman:2005wx}, it has not been proven that the RG flow in two dimensions is a gradient flow. It might be possible to use ideas similar to the one in \cite{Kutasov:2003ux} to address this point.

An interesting question is to understand what is the gravity dual of $c$-extremization. It should be possible to answer this question in three-dimensional gauged supergravity, along the lines of \cite{Tachikawa:2005tq, Szepietowski:2012tb}. There might also exist a corresponding geometric statement for the compact seven or eight-dimensional manifolds on which we compactify string or M-theory to AdS$_3$, similarly to \cite{Martelli:2005tp, Martelli:2006yb}.

After having established $a$-maximization in four dimensions \cite{Intriligator:2003jj}, $F$-maximization in three dimensions \cite{Jafferis:2010un, Closset:2012vg} and $c$-extremization in two dimensions \cite{Benini:2012cz}, it is natural to wonder whether a similar principle exists in other dimensions. In five and six, the minimal R-symmetry group of a superconformal algebra is non-Abelian and therefore there cannot be mixing.\footnote{There are no superconformal theories in more than six dimensions.} The only unexplored case seems to be superconformal quantum mechanics. It is conceivable that an extremization principle exists for models with Abelian R-symmetry, and a natural guess for the quantity to extremize is the Euclidean path-integral on a circle, which is a sort of Witten index. A concrete class of models where extremization might be needed is twisted compactifications of M2-branes on Riemann surfaces. Such constructions should have holographic duals similar to the ones studied in this paper (see for example \cite{Gauntlett:2001qs}) and it would be interesting to study them.

The 2d $\cN=(0,2)$ theories discussed in this work are interesting in their own right, and it would be desirable to have a two-dimensional description of them. This should be feasible for 4d $\cN=4$ SYM on (punctured) Riemann surfaces, because we have a Lagrangian description to begin with. A possible approach is to proceed along the lines of \cite{Benini:2010uu} and study 3d $\cN=8$ SYM on a graph (whose ``fattening'' is the punctured Riemann surface), likely to produce ``star-shaped'' quivers.
On the other hand the construction of 2d SCFTs from $\cN=4$ SYM on Riemann surface might lead to a web of dualities corresponding to different pants decompositions of the Riemann surface as in \cite{Gaiotto:2009we}.
The theories from M5-branes are more mysterious, but at least the case of the 6d $A_1$ $\cN=(2,0)$ theory on (punctured) $\Sigma_1 \times \Sigma_2$ should be accessible using the four-dimensional Lagrangian description of Gaiotto's theories \cite{Gaiotto:2009we, Benini:2009gi, Gaiotto:2009hg}. Another possible approach, in the case of K\"ahler 4-cycles in toric $CY_4$'s, could be to reduce M-theory to type IIA along a toric isometry as in \cite{Aganagic:2009zk, Benini:2009qs, Benini:2011cma} to obtain D4-branes on a toric $CY_3$ fibered over $\bR$.

The twisted compactifications of $\cN=4$ SYM suggest a vast generalization where 2d $\cN=(0,2)$ SCFTs are obtained by placing generic 4d $\cN=1$ SCFTs (possibly with Abelian flavor symmetries) on Riemann surfaces with various partial topological twists, and flowing to the IR. A natural starting place is to consider four-dimensional SCFTs arising from D3-branes placed on toric $CY_3$ singularities. These theories should also admit a holographic description. Work along these lines is currently in progress.

\section*{Acknowledgments}

We are grateful to Mike Anderson, Chris Beem, Marcos Crichigno, Jacques Distler, Henriette Elvang, Jaume Gomis, Chris Herzog, Kristan Jensen, Ilarion Melnikov, Tim Olson, Chris Pope, Leonardo Rastelli,  Martin Ro\v cek, Sav Sethi, Yuji Tachikawa, Balt van Rees, and Brian Wecht for sharing their insights with us during the course of this work. FB would like to thank the Newton Institute (Cambridge), KIAS (Seoul), IPMU (Tokyo), and the University of Kyoto; NB would like to thank KITP, USC, Caltech, the University of Michigan and Texas A\&M for warm hospitality while parts of this work were completed. This work is supported in part by DOE grant DE-FG02-92ER-40697.

\appendix

\section{The anomaly polynomial}
\label{app: anomaly polynomial}

The anomaly polynomial of a $2n$-dimensional theory is a formal $(2n+2)$-form characteristic class constructed out of a fiber bundle (in our case corresponding to the global symmetry group) and the tangent bundle, which encodes the (continuous) anomalies of the theory. In two dimensions and with $U(1)^N$ fiber bundle, we can write the characteristic class
\be
\label{anomaly poly 2d}
I_4 = \frac12 \sum_{I,M} k^{IM} \, c_1(F^I) \wedge c_1(F^M) - \frac k{24} \, p_1(R) = - \sum_{I,M} \frac{k^{IM}}{8\pi^2} F^I \wedge F^M + \frac k{192\pi^2} \Tr R^2 \;,
\ee
where $F^I = dA^I$ are the two-form field strengths of $U(1)^N$, $R = d\Gamma + \Gamma\wedge \Gamma$ is the matrix-valued curvature two-form constructed out of the connection one-form $\Gamma\ud{\mu}{\nu} \equiv \Gamma^\mu_{\rho\nu} dx^\rho$, $c_1$ are first Chern classes and $p_1$ is the first Pontryagin class.
The anomalies are extracted with the descent formalism \cite{Wess:1971yu, AlvarezGaume:1983cs, AlvarezGaume:1984dr} (see also \cite{Ginsparg:1985qn}). First we write
\be
I_4 = dI_3 \;,\qquad\text{with}\qquad I_3 = - \sum_{I,M} \frac{k^{IM}}{8\pi^2} \, A^I \wedge F^M + \frac k{192\pi^2} \Tr \Gamma \wedge R \;.
\ee
Consider gauge variations $\delta_\lambda A^I = d\lambda^I$ and coordinate transformations $x^\mu \to x^\mu - \xi^\mu(x)$. The connection one-form transforms as $\delta_\xi\Gamma = \nabla v$ with $v\ud{\alpha}{\beta} = \partial_\beta \xi^\alpha$. The variation of the Chern-Simons form $I_3$ is locally exact:
\be
\label{I3I2}
\delta I_3 = dI_2^{(1)} \;,\qquad\text{with}\qquad I_2^{(1)} = - \sum_{I,M} \frac{k^{IM}}{8\pi^2}\, \lambda^I F^M + \frac k{192\pi^2} \Tr (v\, d\Gamma) \;.
\ee
By the anomaly inflow argument \cite{Callan:1984sa} the variation of the quantum action $S$ is equal to $\delta_\lambda S = 2\pi \int d^2x\, I_2^{(1)}$. Therefore
\begin{multline}
\label{delS}
\delta S = \int d^2x\, \Big( \partial_\mu\lambda^I\, \delfrac{S}{A_\mu^I} + 2 \nabla_{(\mu} \xi_{\nu)} \, \delfrac{S}{g_{\mu\nu}} \Big) = - \int d^2x\, \sqrt{-g} \Big( \lambda^I \nabla_\mu j^{I\mu} + \xi_\nu \nabla_\mu T^{\mu\nu} \Big) \\
= \int d^2x\, \sqrt{-g} \Big( - \frac{k^{IM}}{8\pi} \, \lambda^I F^M_{\mu\nu} \varepsilon^{\mu\nu} - \frac k{96\pi} \, \xi^\alpha \varepsilon^{\mu\rho} \partial_\mu \partial_\beta \Gamma^\beta_{\alpha\rho} \Big) \;,
\end{multline}
where repeated indices are summed over. Comparison of the two lines leads to (\ref{anomaly equations}). Notice that we have written the gravitational part of the Chern-Simons form solely in terms of the Levi-Civita connection $\Gamma$, without the appearance of the spin connection $\omega$. This corresponds to a choice of scheme in which local Lorentz rotations are non-anomalous and the stress tensor $T^{\mu\nu}$ is symmetric, while diffeomorphisms are anomalous and the stress tensor is not conserved. We could have chosen (by expressing $I_3$ in terms of $\omega$) a different scheme in which local Lorentz rotations are anomalous (and the stress tensor is not symmetric) but diffeomorphisms are not (and $T^{\mu\nu}$ is conserved).

The anomaly polynomial of a $2n$-dimensional right-moving (complex) Weyl fermion in representation \rep{r} of a gauge group $G$ takes an elegant form:
\be
\label{anomaly poly Weyl generic}
I_{2n+2} = \text{ch}_\rep{r}(F) \, \hat A(R) \big|_{2n+2} \;,
\ee
which is the same as the Dirac index density in $2n+2$ dimensions. Here $\text{ch}_\rep{r}(F)$ is the Chern character and $\hat A$ is the Dirac genus:
\be
\text{ch}_\rep{r}(F) = \Tr_\rep{r} e^{iF/2\pi} = \dim\rep{r} + c_1 + \frac{c_1^2 - 2c_2}2 + \dots \;,\qquad \hat A(R) = 1 - \frac{p_1}{24} + \dots \;.
\ee
For a two-dimensional Weyl fermion of charge $q$ under a $U(1)$ symmetry, the polynomial (\ref{anomaly poly Weyl generic}) takes the form (\ref{anomaly poly 2d}) with $k^{II} = q^2$, $k=1$.

\section{Gauge, gravitational and conformal anomalies in two dimensions}
\label{app: anomalies}

In two dimensions anomalies are computed exactly by one-loop diagrams with two current insertions. To fix our notation, let us review the computation following \cite{AlvarezGaume:1983ig}. We consider Lorentz signature $(-,+)$ and take gamma matrices $\{\gamma^a,\gamma^b\} = 2\eta^{ab}$:
\be
\gamma^{\hat0} = - \gamma_{\hat0} = \mat{0&1\\-1&0} \;,\qquad \gamma^{\hat1} = \gamma_{\hat1} = \mat{0&1\\1&0} \;,\qquad \gamma^{\hat0} \gamma^{\hat1} =\gamma^3 = \mat{1&0\\0&-1}\;,
\ee
where hatted indices are flat. The chirality matrix is $\gamma^3$, satisfying $\gamma_a \gamma^3 = - \varepsilon_{ab} \gamma^b$ where we take the covariant antisymmetric tensor $\varepsilon^{\hat0\hat1} = - \varepsilon^{\hat1\hat0} = - \varepsilon_{\hat0\hat1} = \varepsilon_{\hat1\hat0} = 1$ in terms of flat indices. We take vielbein $e^a_\mu$ such that $g_{\mu\nu} = e^a_\mu e^b_\nu \eta^{ab}$, so that $\det e^a_\mu \equiv e = \sqrt{-g}$. On flat space (or in the linearized approximation) it will be convenient to introduce light-cone coordinates
\be
x^\pm = \frac{x^1 \pm x^0}{\sqrt2} = x_{\mp} = \frac{x_1 \mp x_0}{\sqrt2} \;,\qquad x^0 = \frac{x^+ - x^-}{\sqrt2} \;,\qquad  x^1 = \frac{x^+ + x^-}{\sqrt2}\;,
\ee
so that $dx^0 \wedge dx^1 = dx^+ \wedge dx^-$. The metric and the antisymmetric tensor in the coordinates $\{+,-\}$ are:
\be
g_{\mu\nu} = g^{\mu\nu} = \mat{0&1\\1&0} \;,\qquad\qquad \varepsilon^{\mu\nu} = - \varepsilon_{\mu\nu} = \mat{0 & 1 \\ -1 & 0} \;.
\ee
Notice in particular that $(\gamma^+)^2 = (\gamma^-)^2 = 0$.

\begin{figure}[t]
\begin{center}
\includegraphics[width=.6\textwidth]{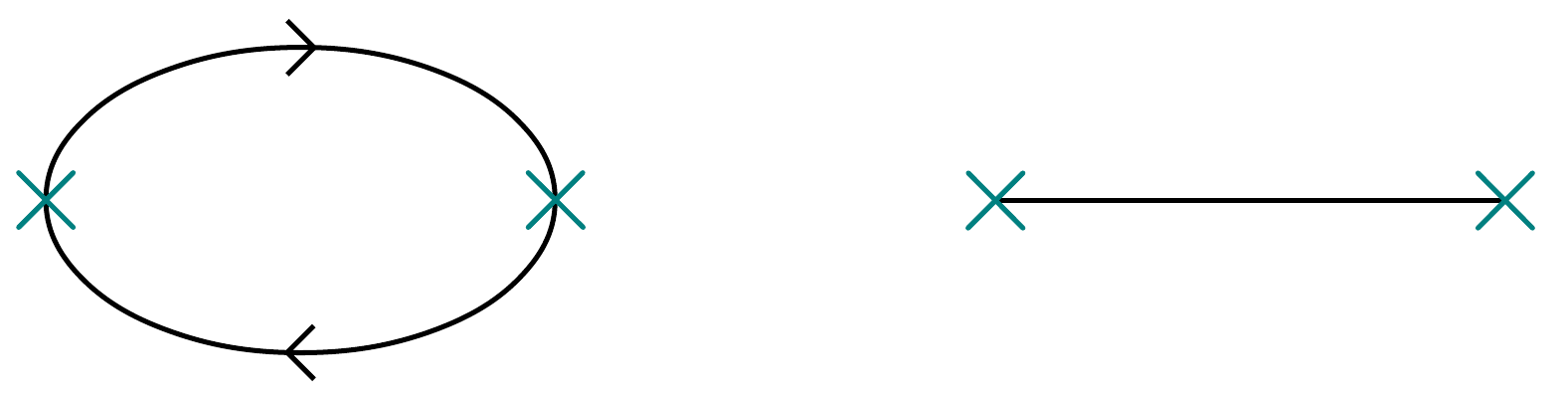}
\caption{Left: one-loop diagram for two-point functions, with a Weyl fermion running in the loop. When a scalar runs in the loop, there are no arrows and therefore there are two possible ways of contracting the operators. Right: tree level diagram for the two-point function.
\label{fig: one-loop}}
\end{center}
\end{figure}

\paragraph{Gauge anomaly.} Consider a spin-$\frac12$ (complex) Weyl fermion of positive chirality on flat space which is coupled to the external gauge field $A_\mu$:
\be
S = \int d^2x\, i \bar\psi \gamma^\mu (\partial_\mu - iA_\mu) \psi \;,
\ee
where $\psi$ is subject to the constraint $\gamma^3\psi = \psi$. This implies $0 = \gamma^+\psi = \gamma_-\psi$, and by the Dirac equation $0 = \partial_-\psi = \frac1{\sqrt2}(\partial_1 - \partial_0)\psi$, \ie{} the fermion is right-moving. The current operator is $J^\mu(x) = \delfrac{S}{A_\mu(x)} = \bar\psi \gamma^\mu \psi$, and the only non-vanishing component is $J_+$. We compute the two-point function
\be
U_{++}(p) = \int d^2x\, e^{-ipx} \, \langle J_+(x) \, J_+(0) \rangle_T \;,
\ee
where $\langle \; \rangle_T$ is time-ordered, at one-loop as in Figure \ref{fig: one-loop}. Using that the fermion propagator is $\frac{1+\gamma^3}2 \frac{i \pslash}{p^2 + i\epsilon}$ we find:
\be
U_{++}(p) = \int \frac{dk_+ dk_-}{(2\pi)^2} \Tr \big[ \bar\psi \gamma_+ \psi \bar\psi \gamma_+ \psi \big] = \int \frac{dk_+ dk_-}{4\pi^2} \, \frac1{\big[ k_- + p_- + \frac{i\epsilon}{p_+ + k_+} \big]\big[ k_- + \frac{i\epsilon}{k_+} \big]} \;.
\ee
One can perform the contour integral in $k_-$, noticing that there is non-vanishing contribution only when the poles at $k_- = - \frac{i\epsilon}{k_+}$ and $k_- = - p_- - \frac{i\epsilon}{p_+ + k_+}$ are on opposite sides of the real axis. If, for instance, $p_+>0$ one has $-p_+ < k_+ < 0$, and therefore
\be
U_{++}(p) = \frac{2\pi i}{4\pi^2} \int_{-p^+}^0 \frac{dk_+}{p_-} = \frac i{2\pi} \, \frac{p_+}{p_-} \;.
\ee
Coupling each vertex to $A_-$ (and taking into account a factor $\frac12$ from Bose symmetry of the vertices), $U_{++}(p)$ gives the effective action $iS_\text{eff}(A_\mu)$, where
\be
S_\text{eff}(A_\mu) = \frac1{4\pi} \int d^2p\, \frac{p_+}{p_-} \, A_-(p) A_-(-p) \;.
\ee
A similar computation for a left-moving Weyl fermion gives $U_{--}(p) = \frac{i}{2\pi} p_-/p_+$.

We can repeat the same analysis for a real chiral boson linearly coupled to the external gauge field. Consider a real scalar, with action and current
\be
S = \int d^2x\, \Big( \frac12 \partial_\mu\phi\, \partial^\mu \phi + \frac1{\sqrt\pi} A_\mu \partial^\mu \phi \Big) \;,\qquad\qquad J_\mu = \frac1{\sqrt\pi} \, \partial_\mu \phi \;,
\ee
where the normalization is chosen for later convenience. Since only $\partial_\mu \phi$ appears in the current, we can impose the right-moving constraint $\partial_-\phi = 0$ so that the only non-vanishing component is $J_+$. This time the two-point function $U_{++}(p)$ is computed at tree level as in Figure \ref{fig: one-loop} (with the propagator $\frac{i}{p^2 + i\epsilon}$):
\be
U_{++}(p) = \frac1\pi \, p_+ \frac{i}{p^2 + i\epsilon} p_+ = \frac i{2\pi} \, \frac{p_+}{p_-} \;.
\ee
The result is the same as for a Weyl fermion. This could have been anticipated since the two are equivalent via bosonization.

The quadratic effective action $S_\text{eff}(A_\mu)$ encodes important information about anomalies. After differentiating once one obtains the one-point function of the current $J_\mu$ on a background, at first order in $A_\mu$. This gives the gauge anomaly. Taking second derivatives yields the two-point function of the current on a vanishing background. In a conformal theory this is related to the Schwinger term in the current algebra.

We now consider a more general theory with right-moving spinors $\psi_{Ri}$ and left-moving spinors $\psi_{La}$, labeled by indices $i,a$. We assume also that in the theory there are Abelian currents $J_\mu^I$, with right and left charges $Q_i^I$ and $Q_a^I$, coupled to external vector fields $A_\mu^I$. The total Lagrangian is
\be
\cL = \sum_{j,I} i \bar\psi_{Rj} \gamma^\mu (\partial_\mu - i Q_j^I A_\mu^I) \psi_{Rj} + \sum_{a,I} i \bar\psi_{La} \gamma^\mu (\partial_\mu - i Q^I_a A_\mu^I) \psi_{La} \;.
\ee
Let us define the symmetric positive-definite matrices
\be
k_R^{IJ} = \sum\nolimits_i Q_i^I Q_i^J \;,\qquad\qquad k_L^{IJ} = \sum\nolimits_a Q_a^I Q_a^J \;.
\ee
Following the same steps as above one finds the effective action $S_\text{eff}(A_\mu^I)$. In the process of renormalization we are free to add to the effective action local counterterms which are polynomial functions of momenta. The most general renormalized effective action with the correct dimension and compatible with Lorentz symmetry is
\be
S_\text{eff}(A_\mu^I) =  \int \frac{d^2p}{4\pi} \, \Big[ k_R^{IJ} \frac{p_+}{p_-} A_-^I(p) A_-^J(-p) + k_L^{IJ} \frac{p_-}{p_+} A_+^I(p) A_+^J(-p) + B^{IJ} A_-^I(p) A_+^J(-p) \Big] \;,
\ee
where repeated indices are summed and the real matrix $B^{IJ}$ does not have to be symmetric. We can compute the one-point functions $\langle J^{I\mu}(p) \rangle_A = \delfrac{S}{A_\mu^I(-p)}$, where $\langle \; \rangle_A$ is linearized at first order in the background, and the divergences
\be\label{anomcounter}
\langle \partial_\mu J^{I\mu} \rangle_A = \frac i{4\pi} \Big[ (2k_R^{IJ} + B^{JI}) p_+ A_-^J + (2k_L^{IJ} + B^{IJ}) p_- A_+^J \Big] \;,
\ee
using $\partial_\mu = i p_\mu$. Unless $k^{IJ}_R = k^{IJ}_L$, there is no choice of $B^{IJ}$ that sets the right hand side of \eqref{anomcounter} to zero, and this is the manifestation of an anomaly. To fix a renormalization scheme we require that the anomalous divergences take a covariant form, imposing
\be
B^{(IJ)} = -k_R^{IJ} - k_L^{IJ} \;.
\ee
This leaves the antisymmetric part of $B^{IJ}$ free. We also require that the anomalies are \emph{symmetric}, \ie{} $B^{[IJ]} = 0$. In this renormalization scheme
\be
\partial_\mu J^{I\mu} = \sum_M \frac{k^{IM}}{8\pi} \, F_{\mu\nu}^M \epsilon^{\mu\nu} \;,
\ee
where the symmetric matrix $k^{IM}$ is defined as
\be
k^{IM} = k_R^{IM} - k_L^{IM} = \Tr_\text{Weyl fermions} \gamma^3 Q^I Q^M \;.
\ee

Notice that our renormalization scheme is not the one usually adopted in the computation of the chiral anomaly in a parity invariant theory. To illustrate this we consider the example of a Dirac fermion: the vector current $J_\mu^V = \bar\psi \gamma_\mu \psi$ and the axial current $J_\mu^A = \bar\psi \gamma_\mu \gamma^3 \psi$ are classically related by $J^A_\mu = - \varepsilon_{\mu\nu} J^{V\mu}$. In the basis $\{V,A\}$ we have $k^{IJ}_R = \smat{ 1 & 1 \\ 1 & 1}$, $k^{IJ}_L = \smat{ 1 & -1 \\ -1 & 1}$. One can insist on preserving the relation between the currents at the quantum level by choosing an asymmetric $B^{IJ} = \smat{-2 & 2 \\ -2 & -2}$ (as can be checked comparing the one-point functions) leading to
\be
\partial_\mu J^{V\mu} = 0 \;,\qquad\qquad \partial_\mu J^{A\mu} = \frac1{2\pi} \, F^V_{\mu\nu} \varepsilon^{\mu\nu} \;,
\ee
however one has to renounce a symmetric anomaly matrix.

\paragraph{Gravitational and conformal anomaly.} Consider now a Weyl fermion of positive chirality coupled to a gravitational background:
\be
S = \int d^2x\, \sqrt{-g}\, i \bar\psi \, e^{\mu a} \gamma_a \Big( \partial_\mu + \frac14 \omega_\mu^{bc} \gamma_{bc} \Big) \psi = \frac i2 \int d^2x\, (\det e)\, e^{\mu a} \bar\psi \gamma_a \overset{\leftrightarrow}\partial_\mu \psi \;.
\ee
Here we used the two-dimensional identity: $ \frac i4 (\det e) \, e^{\mu a} \gamma_a \omega_\mu^{bc} \gamma_{bc} = \frac i2 (\det e) \, \varepsilon^{\rho\sigma} \partial_\rho e^a_\sigma \, \varepsilon_{ab} \gamma^b = \frac i2 \, \partial_\mu \big( e^{\mu a} \det e) \, \gamma_a $. The stress tensor is $T^{\mu\nu}(x) = - \frac{2}{\sqrt{-g}} \, \delfrac{S}{g_{\mu\nu}(x)}$. In a weak gravitational field, $g_{\mu\nu} = \eta_{\mu\nu} + h_{\mu\nu}$, the linearized coupling to gravity is $\cL^{(1)} = - \frac12 h^{\mu\nu} T_{\mu\nu}$ where (using the equations of motion)
\be
T_{\mu\nu} = \frac i4 \bar\psi (\gamma_\mu \overset{\leftrightarrow}\partial_\nu + \gamma_\nu \overset{\leftrightarrow}\partial_\mu) \psi \;,
\ee
and indices are raised with $\eta^{\mu\nu}$. Imposing the chirality constraint $\gamma^3\psi = \psi$, the only non-vanishing component is $T_{++}$. We compute the two-point function
\be
U_{++++}(p) = \int d^2x\, e^{-ipx} \, \langle T_{++}(x) \, T_{++}(0) \rangle_T \;,
\ee
at one-loop as in Figure \ref{fig: one-loop}. Proceeding as before and performing the contour integral in $k_-$, we find:
\begin{multline}
U_{++++}(p) = \frac14 \int \frac{dk_+dk_-}{(2\pi)^2} \, (p_+ + 2k_+)^2 \Tr \big[ \bar\psi \gamma_+ \psi \bar\psi \gamma_+ \psi \big] = \, \\
= \int \frac{dk_+ dk_-}{4\pi^2} \, \frac{(p_+ + 2 k_+)^2}{\big[ k_- + p_- + \frac{i\epsilon}{p_+ + k_+} \big]\big[ k_- + \frac{i\epsilon}{k_+} \big]} = \frac{i}{24\pi} \, \frac{p_+^3}{p_-} \;.
\end{multline}
A left-moving Weyl fermion gives $U_{----}(p) = \frac{i}{24\pi} p_-^3/p_+$.

Let us again repeat the computation for a real chiral boson. Consider a real scalar, with action and stress tensor
\be
S = \frac12 \int d^2x\, \sqrt{-g} \, g^{\mu\nu} \partial_\mu \phi \, \partial_\nu \phi \;,\qquad\qquad T_{\mu\nu} = \partial_\mu \phi \, \partial_\nu \phi - \frac12 g_{\mu\nu} \partial_\rho \phi \, \partial^\rho \phi \;.
\ee
Since only $\partial_\mu\phi$ appears in the stress tensor, we can impose the right-moving constraint $\partial_-\phi = 0$ so that the only non-vanishing component is $T_{++} = \partial_+\phi\, \partial_+\phi$. The one-loop computation of the two-point function $U_{++++}(p)$ as in Figure \ref{fig: one-loop} (including a factor of 2 from the two possible Wick contractions) gives:
\be
U_{++++}(p) =  - 2 \int \frac{dk_+ dk_-}{(2\pi)^2} \, \frac{k_+^2 (p_+ + k_+)^2}{\big[ (p+k)^2 + i\epsilon \big]\big[ k^2 + i\epsilon \big]} = \frac i{24\pi} \, \frac{p_+^3}{p_-} \;.
\ee
Again the result is identical to the one for a Weyl fermion.

Coupling each vertex of $U_{\mu\nu\rho\sigma}$ to $-\frac12 h^{\alpha\beta}$ (and including a factor $\frac12$ from Bose symmetry) one finds the quadratic effective action $S_\text{eff}(h_{\mu\nu})$. To be general, consider a theory with $c_R$ right-moving Weyl fermions (or chiral bosons) and $c_L$ left-movers. After adding local counterterms, the most general renormalized effective action is
\begin{multline}
S_\text{eff} = \frac1{192\pi} \int d^2p\, \Big[ c_R \frac{p_+^3}{p_-} h_{--}(p) h_{--}(-p) + c_L \frac{p_-^3}{p_+} h_{++}(p) h_{++}(-p) + \phantom{\,} \\
+ A \, p_+^2 h_{--}(p) h_{+-}(-p) + B \, p_+ p_- h_{+-}(p) h_{+-}(-p) + \phantom{\,} \\
+ C \, p_+ p_- h_{++}(p) h_{--}(-p) + D \, p_-^2 h_{++}(p) h_{+-}(-p) \Big] \;.
\end{multline}
We compute the one-point functions $\langle T_{++}(p) \rangle_h = -2 \delfrac{S}{h_{--}(-p)}$, $\langle T_{--}(p) \rangle_h = -2 \delfrac{S}{h_{++}(-p)}$ and $\langle T_{+-}(p) \rangle_h = - \delfrac{S}{h_{+-}(-p)}$, where $\langle \; \rangle_h$ is at first order in the background, and the divergences:
\bea
\langle \partial_\mu T\ud{\mu}{+}\rangle_h &= - \frac{ip_+}{192\pi} \Big[ (4c_R + A) p_+^2 h_{--} + 2(A+B) p_+ p_- h_{+-} + (2C+D) p_-^2 h_{++} \Big] \;, \\
\langle \partial_\mu T\ud{\mu}{-} \rangle_h &= - \frac{ip_-}{192\pi} \Big[ (2C+A) p_+^2 h_{--} + 2(B+D) p_+ p_- h_{+-} + (4c_L + D) p_-^2 h_{++} \Big] \;.
\eea
If $c_R = c_L \equiv c$, one can choose a renormalization scheme ($A = -B = -2C = D = -4c$) in which the stress tensor is conserved: $\nabla_\mu T^{\mu\nu}=0$. In this scheme one finds a trace anomaly:%
\footnote{We have taken into account that in momentum space the two-dimensional linearized scalar curvature is $R = -p_+^2 h_{--} - p_-^2 h_{++} + 2p_+p_- h_{+-}$.}
\be
T^\mu_\mu = 2T_{+-} = - \frac{c_R + c_L}{48\pi} R = - \frac c{24\pi} R \;.
\ee
If $c_R \neq c_L$, conservation of the stress tensor cannot be achieved, thus there is an anomaly. We can use a renormalization scheme ($A=-3c_R - c_L$, $B = 2C = 2(c_R + c_L)$, $D = -c_R - 3 c_L$) in which local Lorentz rotations are non-anomalous and the stress tensor is symmetric. At linearized order the non-conservation is
\be
\partial_\mu T^{\mu\nu} \simeq - \frac{c_R - c_L}{192\pi} \varepsilon^{\nu\rho} \partial_\rho R \;.
\ee
The full consistent anomaly, derived in Appendix \ref{app: anomaly polynomial}, is
\be
\nabla_\mu T^{\mu\nu} = \frac{c_R - c_L}{96\pi} \, g^{\nu\alpha} \varepsilon^{\mu\rho} \partial_\mu \partial_\beta \Gamma^\beta_{\alpha\rho} \;.
\ee

\paragraph{Anomalies and central charges.} Let us now focus on conformal theories in flat space, and compare the renormalized one-loop two-point functions $U_{\mu\nu\rho\sigma}(p)$ and $U_{\mu\nu}(p)$ (from which anomalies can be extracted) with the stress tensor and current OPEs. The latter are usually expressed in Euclidean signature ($x^0 = ix^0_E$) and radial quantization, using complex coordinates:
\bea
z &= x^1 + ix^0_E = \sqrt2\, x^+ \;,\qquad\qquad &
\partial_z &= \partial = \tfrac12 (\partial_1 -i\partial_0^E) = \tfrac1{\sqrt2} \partial_+ \;,\\
\bar z &= x^1 - i x^0_E = \sqrt2\, x^- \;,\qquad\qquad &
\partial_{\bar z} &= \bar\partial = \tfrac12 (\partial_1 + i \partial_0^E) = \tfrac1{\sqrt2} \partial_- \;.
\eea
Making use of the Gauss theorem one can show that $\delta^2(x^0,x^1) = \frac1{2\pi i} \partial_- \frac1{x^+} = \frac1{2\pi i} \partial_+ \frac1{x^-}$.
Let us focus on the right-moving sector. Taking the expectation value of the stress tensor and current OPEs in (\ref{conformal + current OPEs}), using (\ref{T(z) def}) and remembering that on a vanishing background one-point functions vanish, one finds:
\be
\label{TT+jj+}
\langle T_{++}(x) \, T_{++}(0) \rangle_T = \frac{c_R}{8\pi^2 (x^+)^4} \;,\qquad\qquad \langle j_+^I(x) \, j_+^J(0) \rangle_T = - \frac{k_R^{IJ}}{4\pi^2(x^+)^2} \;,
\ee
where $\langle \; \rangle_T$ is now in radial quantization.
Using the identity $\partial_- (x^+)^{-1-n} = \frac1{n!} (-\partial_+)^n \partial_- (x^+)^{-1} = \frac{2\pi i}{n!} (-\partial_+)^n \delta^2(x)$, one can compute the Fourier transform of \eqref{TT+jj+} to find
\be
U_{++++}(p) = \frac{ic_R}{24\pi} \, \frac{p_+^3}{p_-} \;,\qquad\qquad U_{++}^{IJ}(p) = \frac{ik_R^{IJ}}{2\pi} \, \frac{p_+}{p_-} \;.
\ee
$U_{----}(p)$ and $U^{IJ}_{--}(p)$ are computed in a similar way. This explicitly shows how the two-point functions of the stress tensor and conserved currents on a vanishing background determine the anomaly coefficients $c_{R,L}$ and $k_{R,L}^{IJ}$.

\section{Free theories}
\label{app: free theories}

In this section we analyze the simple example of a free chiral multiplet, to understand to what extent normalizability of the vacuum is necessary for $c$-extremization to work.

The chiral multiplet $\Phi$ contains a complex scalar $\phi$ and a right-moving Weyl fermion $\psi$, see for example \cite{Witten:1993yc}. The expansion in $\cN=(0,2)$ superspace is:
\be
\Phi = \phi + \sqrt2\, \theta_+ \psi - 2i \theta_+ \bar\theta_+ \partial\phi \;.
\ee
The multiplet $\Phi$ has zero R-charge (therefore $R[\phi]=0$ and $R[\psi]=-1$) because the scalar has vanishing dimension. Indeed this reproduces the correct central charge $c_R = 3k^{RR} = 3$.

If the scalar is compact (target space $T^2$) the vacuum is normalizable and all currents are Kac-Moody. For instance, there are right-moving flavor currents%
\footnote{Recall that for a non-compact complex scalar $\phi$ and a Weyl fermion $\psi$ we have
\be
\phi(z) \, \bar\phi(w) \,\sim\, - \log |z-w|^2 \;,\qquad\qquad \psi(z) \, \bar\psi(w) \,\sim\, \frac1{z-w} \;.
\ee
}
\be\label{flavorcurrents}
J^1 = i  \, \frac{\partial\phi + \partial\bar\phi}{\sqrt2} \;,\qquad\qquad\qquad J^2 =  \, \frac{\partial\phi - \partial\bar\phi}{\sqrt2}\;,
\ee
which act by constant shifts of the scalar. Since the mixed anomalies are $k^{R1} = k^{R2} = 0$ and $k^{IJ} = \smat{ 1 & 0 \\ 0 & 1}$ with $I,J=1,2$, $c$-extremization confirms that these flavor currents do not mix with $U(1)_R$. For special values of the radii the flavor currents in \eqref{flavorcurrents} become the Cartan generators of larger $SU(2)$ current algebras, which indeed cannot mix with $U(1)_R$.

If the scalar is non-compact (target $\mathbb{R}^2$) we have an additional flavor symmetry $J^\ell_\mu$ under which both $\phi$ and $\psi$ have charge 1:
\be
J^\ell = - \phi \, \partial \bar\phi + \partial\phi \, \bar\phi - \bar\psi \psi \;,\qquad\qquad \bar J^\ell = - \phi \, \bar\partial \bar\phi + \bar\partial\phi\, \bar\phi \;.
\ee
The matrix of anomalies between $R$ and $\ell$ is (the contributions of bosons cancel out):
\be
k^{IJ} = \mat{1 & -1 \\ -1 & 1} \qquad\qquad\text{with $I,J = R, \ell$} \;.
\ee
The trial central charge (neglecting the currents $J^{1,2}$) is $c_R^\text{tr}(t) = 3(1-t)^2$ which is minimized at $t^*=1$ with $c_R^\text{tr}(t^*) = 0$. This is clearly the wrong result. The reason is that when the boson is non-compact the vacuum is non-normalizable and there can be non-holomorphic currents. Indeed the current $J^\ell$ is non-normalizable and non-holomorphic (it of course obeys the conservation equation $\bar\partial J^\ell + \partial \bar J^\ell = 0$). Non-holomorphic currents can have non-vanishing two-point function with the R-current, therefore $c_R^\text{tr}(t)$ is not extremized along these directions.

This simple example serves as an illustration to the limitations of the $c$-extremization procedure for SCFTs without a normalizable vacuum. Indeed we found very similar behavior on page \pageref{(4,4) from D3s} for $\cN=(4,4)$ theories from wrapped D3-branes, and on pages \pageref{(0,4) from M5s} and \pageref{(0,4) from M5s Kahler} for $\cN=(0,4)$ theories from wrapped M5-branes.

\section{$\cN=2$ superconformal current algebra}
\label{app: mode algebra}

For completeness, let us write down the superconformal and Abelian current algebras in terms of modes. This is useful to check Jacobi identities. A holomorphic operator $\cO(z)$ of conformal weight $(0,h)$ is decomposed in modes $\cO_m$ according to:
\be
\cO(z) = \sum_{m \, \in \,\bZ + \alpha} \frac{1}{z^{m+h}} \, \cO_m \;,\qquad\qquad \cO_m = \frac1{2\pi i} \oint dz\, z^{m+h-1} \, \cO(z) \;,
\ee
where $\alpha \in [0,1)$ depends on the boundary conditions in radial quantization. In particular the supercharges are the modes $G\makebox[0pt][l]{$^\pm$}_{-\frac12}$ of the supercurrents $T^\pm_F(z)$, and $\omega_0$ is the R-charge.

The $\cN=2$ superconformal algebra (\ref{N=2 SC algebra}) reads:
\be
\begin{aligned}
\, [L_m, L_n ] &= \frac{c_R}{12} \, (m^3 -m) \, \delta_{m+n,0} + (m-n) \, L_{m+n} \\
[L_m, G_r^\pm] &= \Big( \frac m2 - r \Big) G^\pm_{m+r} \\
[L_m, \omega_n] &= - n \, \omega_{m+n} \\
\{ G_r^+, G_s^- \} &= \frac{c_R}3 \Big( r^2 - \frac14 \Big) \delta_{r+s,0} + 2 L_{r+s} + (r-s) \omega_{r+s}
\end{aligned}\qquad\begin{aligned}
\{ G_r^+, G_s^+ \} &= \{ G_r^-, G_s^- \} = 0 \\
[\omega_n, G_r^\pm ] &= \pm G_{r+n}^\pm \\
[\omega_m, \omega_n] &= \frac{c_R}3 \, m \, \delta_{m+n,0} \;.
\end{aligned}
\ee
The Abelian current algebra (\ref{Abelian current algebra}) is
\be
[j^A_{a,m}, j^B_{b,n}] = \delta_{ab} k^{AB} m\, \delta_{m+n,0} \;,\qquad [j^A_{a,m}, \psi^B_{b,r}] = 0 \;,\qquad \{\psi^A_{a,r}, \psi^B_{b,s}\} = \delta_{ab} k^{AB} \delta_{r+s,0} \;.
\ee
Finally, the action of the superconformal algebra on the current multiplet (\ref{OPEs SCCA}) is:
\bea
\,[L_m, \psi^A_{a,r}] &= - \Big( \frac m2 + r \Big)\, \psi^A_{a,m+r} &
[\omega_m, \psi^A_{a,r}] &= i \varepsilon_{ab} \, \psi^A_{b,m+r} \\
[L_m, j^A_{a,n}] &= \frac i2 \, q^A_a (m^2+m) \, \delta_{m+n,0} - n \, j^A_{a,m+n} \qquad\qquad&
[\omega_m, j^A_{a,n}] &= \varepsilon_{ab} \, q^A_b \, m \, \delta_{m+n,0} \\
\{ G_r^\pm, \psi^A_{a,s}\} &= \frac{\delta_{ab} \mp i \varepsilon_{ab}}{\sqrt2} \Big[ i q^A_b \Big( r + \frac12 \Big) \delta_{r+s,0} + j^A_{b,r+s} \Big] \qquad &
[ G_r^\pm, j^A_{a,n}] &= - \frac{\delta_{ab} \pm i \varepsilon_{ab}}{\sqrt2} \, n \, \psi^A_{b,r+n} \;.
\eea

\section{Supersymmetry of the twisted theories}
\label{app: SUSY}

In this appendix we study the amount of supersymmetry preserved by the compactified and twisted theories discussed in the text, namely 4d $\cN=4$ SYM and the 6d $\cN=(2,0)$ theory.

\subsection{$\cN=4$ SYM}
\label{app: SUSY N=4 SYM}

Let us analyze the amount of 2d supersymmetry preserved by 4d $\cN=4$ SYM compactified on a Riemann surface $\Sigma_\fg$ of genus $\fg$. The supercharges transform in the representation $\rep{2} \otimes \rep{4}$ under $SO(3,1) \times SO(6)$, and decompose into
\be
Q \,\to\, \Big[ \big( \tfrac i2 , \tfrac12\big) \oplus \big( -\tfrac i2 , -\tfrac12 \big) \Big] \otimes \Big[ \big(\tfrac12, \tfrac12, \tfrac12 \big) \oplus \big(\tfrac12, -\tfrac12, -\tfrac12 \big) \oplus \big(-\tfrac12, \tfrac12, -\tfrac12 \big)  \oplus \big(-\tfrac12, -\tfrac12, \tfrac12 \big)  \Big]\;,
\ee
under $SO(1,1) \times SO(2)_{\Sigma} \times SO(2)_1 \times SO(2)_2 \times SO(2)_3$.

Let us discuss the non-flat case $\fg\neq 1$ first. For generic $a_I$'s (satisfying $a_1 + a_2 + a_3 = -\kappa$) only the complex supercharge $\big(\tfrac i2, \tfrac12 \big) \otimes \big( \frac12, \frac12, \frac12 \big)$ is covariantly constant and is a candidate preserved supercharge. Since this supercharge is charged under the flux $F$, we can actually decompose the global symmetry $U(1)^3$ in such a way that the R-symmetry couples exactly to $F$. Therefore turning on $F$ corresponds to giving VEV to an external off-shell gravity multiplet, and the only thing we have to check is that the supercharge is covariantly constant. In the generic case we preserve 2d $\cN=(0,2)$ supersymmetry and $U(1)^3$ global symmetry. If in addition two of the non-vanishing $a_I$'s are equal the global symmetry is enhanced to $SU(2) \times U(1)^2$, and if all of them are equal it is enhanced to $SU(3)\times U(1)$. It is easy to check that if one of the $a_I$'s is zero there is another covariantly constant complex supercharge of opposite 2d chirality and we have $\cN=(2,2)$. If two $a_I$'s are zero (as in \cite{Bershadsky:1995vm}) then the supersymmetry is $\cN=(4,4)$ and the global symmetry is $SU(2)^2 \times U(1)$.

The flat case $\fg=1$ requires a different discussion because the flux $F$ does not give charge to the covariantly constant supercharges (as $a_1 + a_2 + a_3 = 0$) and it is thus coupled to a flavor symmetry. The symmetry gives charges $a_I$ to the three complex chiral multiplets $\Phi_I$ of $\cN=4$ SYM in $\cN=1$ notation, and indeed it leaves the superpotential $\Tr \big( \Phi_1[\Phi_2,\Phi_3]\big)$ invariant.
A flavor current couples to an external vector multiplet and we should then check the vanishing of the external gaugino variation. For generic $a_I$'s there are two covariantly constant supercharges: $\big( \pm \tfrac i2, \pm\tfrac12 \big) \otimes \big( \frac12, \frac12, \frac12 \big)$. Indeed the gauging preserves 4d $\cN=1$ supersymmetry, and the BPS equation from the vanishing of the (off-shell) gaugino variation in the $\cN=1$ vector multiplet reads
\be
\label{SUSY 1st eqn}
0 = \Big( \frac12 \gamma^{\mu\nu} F_{\mu\nu} + iD \Big) \xi \;,
\ee
where $D$ is the auxiliary scalar and $\xi$ is the (4-component) Weyl spinor parameter in the \rep{2} of $SO(3,1)$. Choosing $D = \pm |F_{23}|$ preserves 2d $\cN=(0,2)$ supersymmetry (or $\cN=(2,0)$ for the opposite sign). In particular turning on $D$ corresponds to turning on a coupling $D\sum_I a_I|\phi_I|^2$ in the SYM Lagrangian, besides the magnetic flux $F$. In the curved case $\fg\neq1$, this coupling is automatically turned on by the conformal coupling of scalars to the scalar curvature. See \cite{Almuhairi:2011ws} for a similar discussion.

If $\mathfrak{g}=1$ and one of the $a_I$'s is zero, say $a_3 = 0$, then $a_1 = -a_2$ and there are four covariantly constant supercharges. Indeed in this case $F$ couples to the matter fields of $\mathcal{N}=4$ SYM as to a hypermultiplet (and the global symmetry is enhanced to $SU(2)_R \times U(1)^2$), therefore one can make $F_{\mu\nu}$ part of an $\cN=2$ vector multiplet and preserve 4d $\cN=2$ supersymmetry in the gauging. The BPS equations from the vanishing of the two (off-shell) gaugini variations are (see \eg{} \cite{Sohnius:1985qm}):
\be
0 = \frac12 \sigma^{[\mu} \bar\sigma^{\nu]} F_{\mu\nu} \xi_i + i \sigma^\mu D_\mu X \varepsilon_{ij} \xi^{jc} + i \vec D \cdot \vec\tau\du{i}{j} \xi_j \;,
\ee
where $\xi_{1,2}$ are an $SU(2)_R$-doublet of (2-component) Weyl spinor parameters, $X$ is a complex scalar, $\vec D$ are a triplet of auxiliary fields, the 4-component gamma matrices are $\gamma^\mu = \smat{ 0 & \sigma^\mu \\ \bar\sigma^\mu & 0}$, $\xi^c$ are the charge conjugates of $\xi$, and $\vec\tau$ are the Pauli matrices. To preserve supersymmetry we set $X=0$ and turn on $\vec D$, without loss of generality along the third component $D^{(3)}$. Notice that this breaks $SU(2)_R$ to a $U(1)$ subgroup. Switching back to 4-component Weyl spinor parameters $\xi_i$, the BPS equations reduce to
\be
\label{SUSY 2nd eqn}
0 = \frac12 \gamma^{\mu\nu} F_{\mu\nu} \xi_i + i D^{(3)} (\tau_3)\du{i}{j} \xi_j \;.
\ee
Choosing $D^{(3)} = \pm |F_{23}|$ preserves 2d $\cN=(2,2)$ supersymmetry (for both signs) because the two equations from (\ref{SUSY 2nd eqn}) are equal to (\ref{SUSY 1st eqn}) but with opposite signs in front of $D^{(3)}$.

Finally, if all $a_I$'s are zero the 2d theory has $\cN=(8,8)$ supersymmetry and is the same one as the gauge theory on the worldvolume of D1-branes.

\subsection{$\cN=(2,0)$ theory}
\label{app: SUSY (2,0)}

Let us now consider the 6d $\cN=(2,0)$ theory on a four-manifold $M_4 = \Sigma_1 \times \Sigma_2$, where the two Riemann surfaces $\Sigma_\sigma$, $\sigma=1,2$, have genera $\fg_\sigma$. The supercharges transform in the representation $\rep{4} \otimes \rep{4}$, with symplectic Majorana condition, under $SO(5,1) \times SO(5)$, and decompose into
\begin{multline}
Q \,\to\, \Big[ \big( \tfrac i2 , \tfrac12, \tfrac12 \big) \oplus \big( \tfrac i2 , -\tfrac12, -\tfrac12 \big) \oplus \big( -\tfrac i2 , \tfrac12, -\tfrac12 \big) \oplus \big( -\tfrac i2 , -\tfrac12, \tfrac12 \big) \Big] \\
\otimes \Big[ \big(\tfrac12, \tfrac12 \big) \oplus \big( -\tfrac12, -\tfrac12 \big) \oplus \big( \tfrac12, -\tfrac12 \big)  \oplus \big( -\tfrac12, \tfrac12 \big)  \Big]\;,
\end{multline}
under $SO(1,1) \times SO(2)_1 \times SO(2)_2 \times SO(2)_A \times SO(2)_B$. Note that charge conjugate representations are related by the symplectic Majorana condition, and charge conjugation preserves the $SO(1,1)$ chirality.

Let us discuss the non-flat case, $\fg_{1,2} \neq 1$, first. For generic parameters that satisfy $0 = a_1 + b_1 + \kappa_1 = a_2 + b_2 + \kappa_2$, only the complex supercharge $\big( \frac i2, \frac12, \frac12 \big) \otimes \big( \frac12, \frac12 \big)$ (and its conjugate) are covariantly constant and we preserve 2d $\cN=(0,2)$ supersymmetry. If $a_1 = a_2 = 0$ (or $b_1 = b_2 = 0$) the two supercharges $\big( \frac i2, \frac12, \frac12 \big) \otimes \big( \pm \frac12, \frac12 \big)$ are covariantly constant, the global symmetry is enhanced to $SO(3)_A \times SO(2)_B$, and we have $\cN=(0,4)$ supersymmetry. If $a_1 = b_2 = 0$ (or $b_1 = a_2 = 0$) the two supercharges $\big( \pm\frac i2, \frac12, \pm \frac12 \big) \otimes \big( \pm \frac12, \frac12 \big)$ are covariantly constant, and we have $\cN=(2,2)$ supersymmetry.

In the partially flat case, say $\fg_1 = 1$ and $\fg_2 \neq 1$, something similar to the previous section happens. The parameters satisfy $0 = a_1 + b_1 = a_2 + b_2 + \kappa_2$ and the two complex supercharges $\big( \pm \frac i2, \pm \frac12, \frac12 \big) \otimes \big( \frac12, \frac12 \big)$, together with their conjugates, are covariantly constant. The flux on $\Sigma_1$ couples to a flavor symmetry, and indeed the supercharges correspond to a gauging that preserves 4d $\cN=1$ supersymmetry on $\bR^{1,1} \times \Sigma_1$. When the flavor flux is turned on, supersymmetry requires to turn on an auxiliary field $D$ as well, which breaks supersymmetry to 2d $\cN=(0,2)$. If $a_1 = b_1 = 0$ supersymmetry is enhanced to $\cN=(2,2)$; if $a_2=0$ (or $b_2 = 0$) as well then supersymmetry is $\cN=(4,4)$. This is clear because the 6d $\cN=(2,0)$ theory on the torus with no background flux flows to 4d $\cN=4$ SYM.

Finally, in the flat case $\fg_{1,2} = 1$ the parameters satisfy $0 = a_1 + b_1 = a_2 + b_2$ and there are four covariantly constant complex supercharges, realizing a 6d $\cN=(1,0)$ supersymmetry preserved by the gauging of the flavor symmetry. When the flux is turned on together with the auxiliary fields required to cancel the off-shell gaugino variation, 2d $\cN=(0,2)$ supersymmetry is preserved. If $a_1 = b_1 = 0$ (or $a_2 = b_2 = 0$) we have $\cN=(2,2)$ supersymmetry, and if all fluxes are zero we have the $\cN=(8,8)$ gauge theory on the worldvolume of D1-branes.

\section{Five-dimensional gauged supergravity}
\label{app: 5D sugra}

To study the holographic duals to $\mathcal{N}=4$ SYM on Riemann surfaces it is sufficient to use the five-dimensional maximal gauged supergravity \cite{Gunaydin:1984qu, Pernici:1985ju, Gunaydin:1985cu}. Actually for the twists of interest we only need a subsector of the full maximal theory, consisting of  three Abelian gauge fields $A_\mu^I$ and two real scalars $\phi_i$. This is the so called STU model of five-dimensional gauged supergravity and was shown to be a consistent truncation in \cite{Cvetic:1999xp} (see also \cite{Bobev:2010de}).

The supersymmetry transformations of fermionic fields are (see \cite{Behrndt:1998ns} and Appendix A of \cite{Maldacena:2000mw} for more details):
\bea
\label{AAgravitinoD3}
\delta\psi_\mu &= \Big[ \partial_\mu + \frac14 \omega\du{\mu}{ab}\gamma_{ab} + \frac i8 X_{I}(\gamma\du{\mu}{\nu\rho} - 4 \delta_{\mu}^{\nu} \gamma^{\rho}) F^{I}_{\nu\rho} + \frac12 X^{I}V_{I} \gamma_{\mu} - \frac{3i}2 V_{I}A^{I}_{\mu} \Big] \, \epsilon\;, \\
\delta\chi_{(j)} &= \Big[ \frac38 (\partial_{\phi_j} X_{I} )F^{I}_{\mu\nu} \gamma^{\mu\nu} + \frac{3i}2 V_{I}\partial_{\phi_j} X^{I} - \frac{i}4 \delta_{jk} \partial_{\mu}\phi_k \gamma^{\mu} \Big] \, \epsilon\;, \qquad\qquad\qquad j=1,2  \;,
\eea
where we have defined
\be
\label{5d gauged sugra definitions}
X^1 = e^{-\frac{\phi_1}{\sqrt{6}}-\frac{\phi_2}{\sqrt{2}}} \;,\quad
X^2 = e^{-\frac{\phi_1}{\sqrt{6}}+\frac{\phi_2}{\sqrt{2}}} \;,\quad
X^3 = e^{\frac{2\phi_1}{\sqrt{6}}} \;,\quad
V_I = \tfrac{1}{3} \;,\quad
X_I = \tfrac{1}{3} (X^I)^{-1} \;.
\ee
The fields $X^I$ are constrained to satisfy $X^1X^2X^3=1$ and to be positive.
Since we are using an $\cN=2$ truncation of the full gauged supergravity, only a fraction of the maximal possible supersymmetry is visible. The spinors obey the following constraints (the hats denote flat indices):
\be
\gamma_{\hat r} \, \epsilon= \epsilon \;, \qquad \gamma_{\hat{x}\hat{y}} \, \epsilon= i  \epsilon \;, \qquad \partial_t \epsilon = \partial_z \epsilon = \partial_x \epsilon = \partial_y \epsilon = 0 \;.
\ee
Note that the radius of AdS$_5$ is fixed to one.

For our Ansatz the BPS equations reduce to the system%
\footnote{Note that there is a typo in the equation for $f'$ in \cite{Maldacena:2000mw}.}
in (\ref{BPS conditions 5Dsugra}).
The equation for $g$ comes from the vanishing of $\delta\psi_x$ and $\delta\psi_y$, those for $\phi_1$ and $\phi_2$ from the vanishing of $\delta\chi_{(1)}$ and $\delta\chi_{(2)}$, and the one for $f$ from the vanishing of $\delta\psi_t$. The vanishing of $\delta\psi_x$ and $\delta\psi_y$ also impose the last constraint in (\ref{BPS conditions 5Dsugra}): $a_1+a_2+a_3=-\kappa$.

\paragraph{AdS$_3$ solutions.}
We consider the Ansatz in (\ref{gravity ansatz}) with $f(r) = f_0 - \log r$ and $g$, $\phi_1$, $\phi_2$ constant. The BPS equations reduce to a system of algebraic equations:
\begin{align}
0 &= \Big( X^1 + X^2 + \frac1{X^1X^2} \Big) - e^{-2g} \Big( \frac{a_1}{X^1} + \frac{a_2}{X^2} + a_3X^1X^2 \Big)
\label{eqnD3gfp}\;, \\
0 &= \frac{e^{f_0}}3 \Big( X^1 + X^2 + \frac1{X^1X^2} \Big) + \frac{e^{f_0 - 2g}}6 \Big( \frac{a_1}{X^1} + \frac{a_2}{X^2} + a_3 X^1 X^2 \Big) - 1
\label{eqnD3ffp}\;, \\
0 &= \Big( X^1 + X^2 - \frac2{X^1X^2} \Big) +  \frac{e^{-2g}}2 \Big( \frac{a_1}{X^1} + \frac{a_2}{X^2} - 2a_3 X^1X^2 \Big)
\label{eqnD3phi1fp}\;, \\
0 &= \big( X^1 - X^2 \big) + \frac{e^{-2g}}2 \Big( \frac{a_1}{X^1} - \frac{a_2}{X^2} \Big)\;,
\label{eqnD3phi2fp}
\end{align}
where we have eliminated $X^3$ with the constraint. The difference between \eqref{eqnD3gfp} and \eqref{eqnD3phi1fp} gives
$\big( 2 - e^{-2g} (a_1X^2 + a_2X^1) \big)/X^1X^2 = 0$ which implies $a_1X^2 + a_2X^1 \neq 0$. Then
\be
e^{-2g} = \frac{2}{a_1X^2 + a_2X^1} \;.
\ee
Plugging \eqref{eqnD3gfp} into \eqref{eqnD3ffp} one finds
\be
\label{ef0sol}
e^{f_0} = \frac2{X^1+X^2+X^3} = \frac{2X^1X^2}{1+(X^1)^2X^2+X^1(X^2)^2} \;.
\ee
We have thus solved for the metric functions in terms of the $X^I$'s. We now need to solve \eqref{eqnD3gfp} and \eqref{eqnD3phi2fp} for $X^1$ and $X^2$. Let us define the positive variables
\begin{equation}
Y \equiv (X^1)^2X^2~, \qquad\qquad Z \equiv X^1(X^2)^2 \;.
\end{equation}
The sum and difference of \eqref{eqnD3gfp} and \eqref{eqnD3phi2fp} simplify to $(a_2-a_3)Y+a_1(Z-1)=0$ and $(a_1-a_3) Z + a_2(Y-1)=0$, which generically are solved by
\be
\label{YZdef}
Y = \frac{a_1\, (-a_1 + a_2 + a_3)}{a_3 \, (a_1+a_2-a_3)} \;,\qquad\qquad
Z = \frac{a_2 \, (a_1 - a_2 + a_3)}{a_3 \, (a_1+a_2-a_3)} \;.
\ee
This produces the expressions in (\ref{final sugra solution}).
Notice also that
\be
e^{2g + f_0} = \frac{a_1 Z + a_2 Y}{1+Y+Z} \;.
\ee
There are three values of $a_I$'s such that the system before (\ref{YZdef}) is degenerate: $a_1 = a_2 = \frac12$, $a_3=0$ and permutations thereof, which correspond to $\cN=(2,2)$ supersymmetry and exist for $\fg>1$. In these cases one finds a one-dimensional branch of marginal deformations instead of a single solution \cite{Maldacena:2000mw}:
\be
X^1 + X^2 = \frac1{X^1X^2}\;,
\ee
and permutations of $X^1,X^2,X^3$ respectively. For $X^1 = X^2 = 2^{-1/3}$ (and permutations thereof) the solution has enhanced $SU(2)$ isometry.

\paragraph{Analysis of positivity.}
Necessary and sufficient conditions to have a good supersymmetric AdS$_3$ solution are $X^1 > 0$, $X^2 > 0$, $e^{2g} > 0$ and $e^{f_0} > 0$. Since $Y,Z > 0$ if and only if $X^1, X^2, e^{f_0}>0$, we can equivalently require that
\be
\label{positivity}
Y>0 \;,\qquad Z>0 \;,\qquad e^{2g+f_0}>0 \;.
\ee
This implies that only in some regions of the parameter space $(a_1,a_2,a_3)$ the compactified 4d $\cN=4$ SYM theory flows to an IR fixed point with normalizable vacuum. Let us analyze these regions for different genera (we present plots in Figure \ref{fig: good SUGRA}).

For the sphere $\fg=0$, \ie{} $a_1 + a_2 + a_3 = -1$, one finds that (\ref{positivity}) are obeyed in three open regions:
$$
\fg = 0: \qquad \{ a_1>0,\, a_2 > 0\} \;\cup\: \{ a_1>0 ,\, a_1 + a_2 < -1\} \;\cup\; \{ a_2 > 0 ,\, a_1+a_2 < -1 \} \;.
$$
They correspond to requiring that two $a_I$'s are positive (then one is necessarily negative).
For the torus $\fg = 1$, \ie{} $a_1 + a_2 + a_3 = 0$, one also finds three open regions:
$$
\fg = 1:\qquad \{ a_1>0,\, a_2>0 \} \;\cup\; \{a_1>0,\, a_1 + a_2 < 0\} \;\cup\; \{ a_2>0,\, a_1 + a_2 < 0\} \;.
$$
They correspond to requiring that two $a_I$'s are positive (then one is necessarily negative).
For higher genus Riemann surfaces $\fg > 1$, \ie{} $a_1 + a_2 + a_3 = 1$, one finds four open regions and the three points where they touch:
\bea\nn
\fg>1: \qquad &\{ a_1>\tfrac12,\, a_2 > \tfrac12 \} \;\cup\; \{ a_1 > \tfrac12 ,\, a_1+a_2 < \tfrac12 \} \;\cup\; \{ a_2 > \tfrac12 ,\, a_1+a_2 < \tfrac12 \} \\
\cup\; &\{ a_1 < \tfrac12,\, a_2<\tfrac12 ,\, a_1+a_2 > \tfrac12\} \;\cup\; \{ a_1=a_2=\tfrac12,\, a_3 = 0 \text{ and permutations} \} \;.
\eea

Let us finally study the signs of the eigenvalues of the Hessian matrix (\ref{Hessian}) of second derivatives of $c_R^\text{tr}(\epsilon_1,\epsilon_2)$ at the critical point, in the allowed regions of the $(a_1,a_2)$-plane. This will tell us whether the function is maximized or minimized, and therefore what is the chirality of the flavor currents. One finds that for $\fg=0,1$ in all allowed regions the Hessian has one positive and one negative eigenvalues. For $\fg>1$ the eigenvalues have opposite sign in the three infinite regions (see Figure \ref{fig: good SUGRA}), they are both negative in the finite region in the middle, and there is a flat direction at the three special points.

\paragraph{Chern-Simons levels and anomalies.}
The five-dimensional supergravity action contains a Chern-Simons term:%
\footnote{We will use the conventions in Section 7.1 of \cite{Maldacena:2000mw}, see in particular equation (46) of that paper.}
\begin{equation}
S_{5d} \;\supset\; \frac1{4G_N^{(5)}} \int d^5x \, \epsilon^{\mu\nu\alpha\beta\rho} F^{(1)}_{\mu\nu} F^{(2)}_{\alpha\beta} A^{(3)}_{\rho} = \frac1{G_N^{(5)}} \int F^{(1)} \wedge F^{(2)} \wedge A^{(3)} \;.
\end{equation}
Expanding the action on the flux background, integrating on the Riemann surface recalling that $\frac1{2\pi} \int_\Sigma F^{(I)} = - a_I \eta_\Sigma$, and using that $G_N^{(5)} = \pi/2N^2$ in our conventions, we obtain the three-dimensional effective action:
\be
S_{3d} \;\supset\; - 4\eta_\Sigma N^2 \int \Big( a_1 A^{(2)}\wedge F^{(3)} + a_2 A^{(3)}\wedge F^{(1)} + a_3 A^{(1)} \wedge F^{(2)} \Big) \;.
\ee
More compactly we have
\be
S_{3d}^\text{CS} \;\sim\; k^{IJ} \int A^{(I)} \wedge F^{(J)}\;,
\ee
where the Chern-Simons matrix $k^{IJ}$ coincides with the matrix of 't~Hooft anomalies (\ref{anomaly matrix D3-branes}). As we describe in Section \ref{sec: vectors}, there is indeed a relation between CS couplings in an asymptotically AdS$_3$ supergravity background and 't~Hooft anomalies of the currents in the dual CFT$_2$.

The signature of the 't~Hooft anomaly matrix (equivalently of the CS matrix in AdS$_3$) determines the chirality of the currents at the IR fixed point: positive (negative) eigenvalues correspond to right-moving (left-moving) currents. On the other hand, the signature of the Hessian of $c_R^\text{tr}(\epsilon_i)$ determines the chirality of the flavor currents only. One can check that, indeed, in the regions of parameter space $(a_1,a_2,a_3)$ where there are good supergravity solutions:
\begin{itemize}
\item where $\text{Hess}(c_R^\text{tr})$ has signature $(1,1)$ (black regions in Figure \ref{fig: good SUGRA}), $k^{IJ}$ has signature $(2,1)$;
\item where $\text{Hess}(c_R^\text{tr})$ has signature $(0,2)$ (grey region in Figure \ref{fig: good SUGRA}), $k^{IJ}$ has signature $(1,2)$.
\end{itemize}
This is consistent with the fact that the R-symmetry is right-moving.

\section{Seven-dimensional gauged supergravity}
\label{app: 7D sugra}

To study the gravity duals to the twisted compactifications of the 6d $\cN=(2,0)$ theory we have used the seven-dimensional maximal gauged supergravity \cite{Pernici:1984xx}. As shown in \cite{Nastase:1999cb,Nastase:1999kf} this theory is a consistent truncation of the eleven-dimensional supergravity on $S^4$. The Lagrangian of the theory is
\begin{multline}
\cL = \frac{\sqrt{-g}}2 \bigg( R + \frac m2 (T^2-2T_{ij}T^{ij}) - P_{\mu ij}P^{\mu ij} - \frac12 (\Pi_A^i \Pi_B^j F_{\mu\nu}^{AB})^2 - m^2 \big( (\Pi^{-1})^A_i S_{A\, \mu\nu\rho} \big)^2 \bigg) \\
- 3m \delta^{AB} S_{A}\wedge F_{B} + \frac{\sqrt{3}}{2} \epsilon_{ABCDE} \delta^{AG}S_{G} \wedge F^{BC} \wedge F^{DE}  + \frac1{16 m} (2\Omega_{5}[A]-\Omega_{3}[A]) \;.
\end{multline}
The indices $A,B = 1, \ldots ,5$ are in the fundamental of the $SO(5)_g$ gauge group and the indices $i,j=1,\ldots, 5$ are of the $SO(5)_c$ local composite gauge group. There are 14 scalars parametrizing the coset $SL(5,\bR)/SO(5)_c$, which are described by the matrix $\Pi_A^i$ in the \rep{5} of both $SO(5)_g$ and $SO(5)_c$. The gauge group $SO(5)_g$ has vectors $A_\mu^{AB}$ and field strength $F^{AB}$. The gauge coupling constant is $g=2m$. There are 3-form potentials $S_A$, whose 4-form field strengths are
$F_A = dS_A + 2m A\du{A}{B} S_{B}$.  The 7-forms $\Omega_{3}[A]$ and $\Omega_{5}[A]$ are Chern-Simons forms for the gauge field $A\du{\mu A}{B}$ whose explicit form is given in \cite{Pernici:1984xx}.

To construct the solutions of interest we will turn the gauge field along the Cartan of $SO(5)$ as given in \eqref{7DsugraAnsatz}. The matrix of scalars is taken to be
\be
\Pi\du{A}{i} = \text{diag}(e^{\lambda_1},e^{\lambda_1},e^{\lambda_2},e^{\lambda_2},e^{-2\lambda_1-2\lambda_2}) \;,
\ee
where $A,i=1,\dots,5$. The other quantities are defined as follows. First on defines the matrix
\be
M_{\mu\, ij} = (\Pi^{-1})^A_i \, \big( \delta_A^B \partial_\mu + 2m A\du{\mu A}{B} \big) \Pi_B^k \delta_{kj} \;,
\ee
then we have
\be
P_{\mu\, ij} = \frac12 (M_{\mu\, ij} + M_{\mu\, ji}) \;,\qquad\qquad  Q_{\mu\, ij} = \frac12 (M_{\mu\, ij} - M_{\mu\, ji}) \;.
\ee
The potential is constructed from
\be
T_{ij} = \delta_{AB}(\Pi^{-1})^{A}_{i}(\Pi^{-1})^{B}_{j} \;,\qquad\qquad T = \delta^{ij}T_{ij} \;.
\ee
This is the same truncation of the maximal seven-dimensional theory as in \cite{Maldacena:2000mw,Liu:1999ai}.

The supersymmetry variations of the fermionic fields
are
\bea
\delta\psi_\mu &= \Big[ \partial_\mu + \tfrac14 \omega_\mu^{ab} \gamma_{ab} + \tfrac14 Q_{\mu\, ij} \Gamma^{ij} + \tfrac m{20}  T \gamma_\mu  - \tfrac1{40} \big( \gamma\du{\mu}{\nu\rho} - 8 \delta_{\mu}^{\nu}\gamma^{\rho} \big) \Gamma_{ij} \, \Pi_{A}^{i}\Pi_{B}^{j}F_{\nu\rho}^{AB} \\
&\qquad\qquad\qquad\qquad\qquad\qquad\qquad\qquad + \tfrac m{10\sqrt3} \big( \gamma\du{\mu}{\nu\rho\sigma} - \tfrac92 \delta_\mu^\nu \gamma^{\rho\sigma} \big) \Gamma^i \, (\Pi^{-1})^A_i S_{A\, \nu\rho\sigma} \Big] \epsilon \;,\\
\delta\chi_i &= \Big[ \tfrac12 P_{\mu\, ij} \gamma^\mu \Gamma^j + \tfrac m2 \big( T_{ij} - \tfrac15 \delta_{ij} T \big) \Gamma^j + \tfrac1{16} \big( \Gamma_{kl} \Gamma_i - \tfrac15 \Gamma_i\Gamma_{kl}\big) \gamma^{\mu\nu} \, \Pi_A^k \Pi_B^l F_{\mu\nu}^{AB} \\
&\qquad\qquad\qquad\qquad\qquad\qquad\qquad\qquad\qquad + \tfrac m{20\sqrt3} \gamma^{\mu\nu\rho} \big( \Gamma\du{i}{j} - 4\delta_i^j\big) (\Pi^{-1})^A_j S_{A\, \mu\nu\rho} \Big] \epsilon \;.
\eea
Here $\gamma^a$ are gamma matrices for the seven-dimensional spacetime and $\Gamma^i$ are $SO(5)_c$ gamma matrices. Notice also that $\Gamma^i \chi_i = 0$. The three-form potential is subject to the equation of motion:
\be
m^2 \delta_{AC} (\Pi^{-1})^C_i (\Pi^{-1})^B_i S_{B} = -m \star_7 (dS_A + 2m A\du{A}{B} S_B ) + \tfrac1{4\sqrt{3}}\epsilon_{ABCDE} \star_7(F^{BC}\wedge F^{DE}) \;.
\ee
In the following we will fix the constant $m=2$.

\paragraph{AdS$_3$ solutions.} For the Ansatz of interest the BPS equations reduce to the system of equations in (\ref{7D BPS eqns}). To find AdS$_3$ solutions we take $e^f(r) = e^{f_0}/r$ and $g_{1,2}$, $X_1 \equiv e^{2\lambda_1}$ and $X_2 \equiv e^{2\lambda_2}$ to be constant. We obtain the following algebraic system:
\begin{align}
e^{-f_0} &= \frac15 \Big( \frac2{X_1} + \frac2{X_2}+ X_1^2 X_2^2 \Big) + \frac3{80}(a_1b_2+a_2b_1) \,  \frac{e^{-2g_1-2g_2}}{X_1X_2} \nn\\
&\qquad\qquad\qquad\qquad + \frac1{20} \Big( X_1(a_1 e^{-2g_1}+a_2 e^{-2g_2}) + X_2(b_1 e^{-2g_1}+b_2 e^{-2g_2}) \Big) \;, \label{Eqf} \\
0 &= \frac15 \Big( \frac2{X_1} + \frac2{X_2} + X_1^2 X_2^2 \Big) - \frac1{40} (a_1b_2+a_2b_1) \, \frac{e^{-2g_1-2g_2}}{X_1X_2} \nn\\
&\qquad\qquad\qquad\qquad - \frac1{20} \Big( X_1(4a_1 e^{-2g_1}-a_2 e^{-2g_2}) + X_2(4b_1 e^{-2g_1}-b_2 e^{-2g_2}) \Big) \;, \label{Eqg1}\\
0 &= \frac15 \Big( \frac2{X_1} + \frac2{X_2} + X_1^2 X_2^2 \Big) - \frac1{40} (a_1b_2+a_2b_1) \, \frac{e^{-2g_1-2g_2}}{X_1X_2} \nn\\
&\qquad\qquad\qquad\qquad - \frac1{20} \Big( X_1(4a_2 e^{-2g_2}-a_1 e^{-2g_1}) + X_2(4b_2 e^{-2g_2}-b_1 e^{-2g_1}) \Big) \;, \label{Eqg2}\\
0&= \frac25 \Big( \frac3{X_1} - \frac2{X_2} - X_1^2 X_2^2 \Big) - \frac1{80} (a_1b_2+a_2b_1) \, \frac{e^{-2g_1-2g_2}}{X_1X_2} \nn\\
&\qquad\qquad\qquad\qquad + \frac1{20} \Big( 3X_1(a_1 e^{-2g_1}+a_2 e^{-2g_2}) - 2X_2(b_1 e^{-2g_1}+b_2 e^{-2g_2}) \Big) \;, \label{Eqlambda1}\\
0&= \frac25 \Big( \frac3{X_2} - \frac2{X_1} - X_1^2 X_2^2 \Big) - \frac1{80} (a_1b_2+a_2b_1) \, \frac{e^{-2g_1-2g_2}}{X_1X_2} \nn\\
&\qquad\qquad\qquad\qquad + \frac1{20} \Big( 3X_2(b_1 e^{-2g_1}+b_2 e^{-2g_2}) - 2X_1(a_1 e^{-2g_1}+a_2 e^{-2g_2}) \Big) \;. \label{Eqlambda2}
\end{align}
To solve it, we take the linear combinations $(\ref{Eqg1})-(\ref{Eqlambda1})-(\ref{Eqlambda2})$ and $(\ref{Eqg2})-(\ref{Eqlambda1})-(\ref{Eqlambda2})$ that reduce to
\be
e^{2g_1} = \frac{a_1 X_1 + b_1 X_2}{4X_1^2 X_2^2} \;,\qquad\qquad e^{2g_2} = \frac{a_2 X_1 + b_2 X_2}{4X_1^2X_2^2} \;.
\ee
In particular $a_1X_1 + b_1X_2\neq 0$ and $a_2X_1 + b_2X_2 \neq 0$. Then we define the non-vanishing combinations\footnote{We use the same letters as in Appendix \ref{app: 5D sugra} and hope this will not cause confusion.}
\be
Y \equiv X_1^3 X_2^2 \;,\qquad\qquad Z \equiv X_1^2 X_2^3\;,
\ee
in terms of which (\ref{Eqlambda1}) and (\ref{Eqlambda2}) can be reduced to a linear system.%
\footnote{To do this rewrite (\ref{Eqlambda1}) and (\ref{Eqlambda2}) in terms of $Y,Z$. Then $0 = (3Y-2Z)(\ref{Eqlambda1}) + (2Y-3Z)(\ref{Eqlambda2})$ is a linear homogeneous equation in $Y,Z$. Substituting back into $0 = 3(\ref{Eqlambda1}) + 2(\ref{Eqlambda2})$ gives a linear equation for $Y$. The system degenerates for $a_1 = b_1 = 0$ or $a_2 = b_2 = 0$ but these are not acceptable fluxes.}
The solution is
\be
Y = \frac{(a_1^2b_2 + a_2b_1^2)(a_2^2b_1 + a_1b_2^2)}{(a_1^2b_2^2+a_2^2b_1^2+a_1a_2b_1b_2) (a_1b_2+a_2b_1-a_1a_2)} \qquad Z = \dfrac{a_1b_2 + a_2b_1 - a_1a_2}{a_1b_2 + a_2b_1 - b_1b_2}\, Y \;.
\ee
We then obtain
\be
X_1^5 = e^{10\lambda_1} = \frac{Y^3}{Z^2} \;,\qquad\qquad X_2^5 = e^{10\lambda_2} = \frac{Z^3}{Y^2} \;.
\ee
Finally we solve (\ref{Eqf}):
\be
e^{f_0} = \frac{b_1b_2 - a_1b_2 - a_2b_1}{a_1a_2 + b_1b_2 - 2a_1b_2 - 2a_2b_1} \, X_2 \;.
\ee
The combination that appears in the computation of the central charge is
\be
e^{f_0 + 2g_1 + 2g_2} =\frac{a_1^2b_2^2 + a_2^2b_1^2 + a_1a_2b_1b_2}{16(2a_1b_2 + 2a_2b_1 - a_1a_2 - b_1b_2)} \;.
\ee

\begin{figure}[t]
\begin{center}
\includegraphics[width=.32\textwidth]{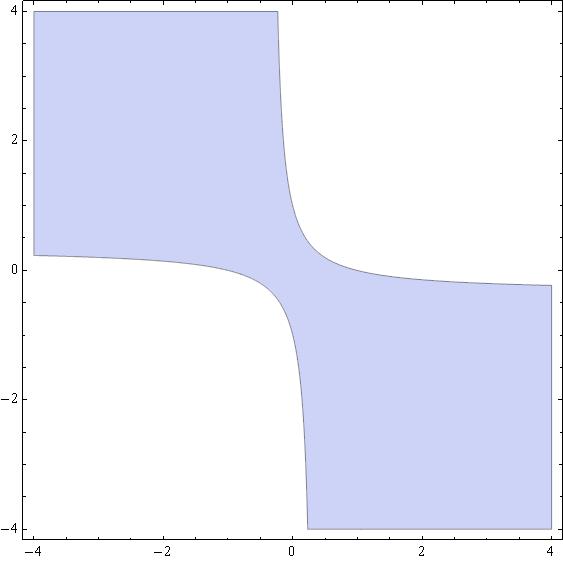}
\hfill
\includegraphics[width=.32\textwidth]{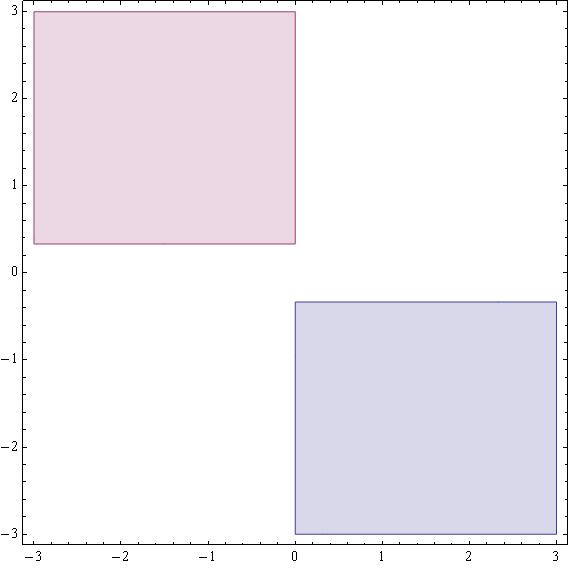}
\hfill
\includegraphics[width=.32\textwidth]{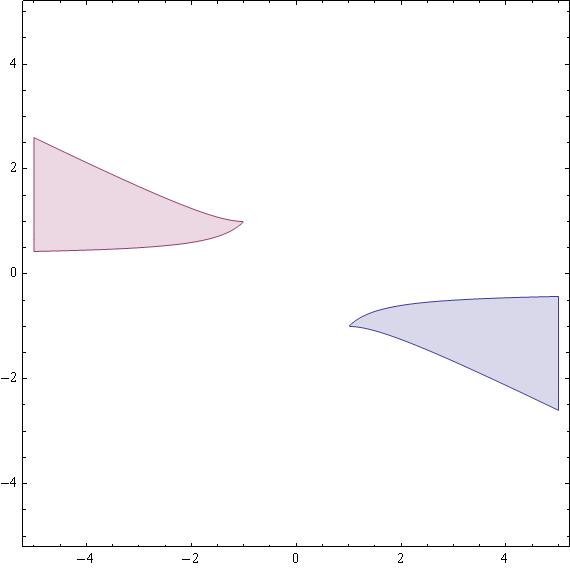}

\caption{Regions of the parameter space $(z_1,z_2)$ where there exist good AdS$_3$ vacua. Boundaries are excluded. We have plotted the three cases $(\fg_1>1,\fg_2>1)$, $(\fg_1=1,\fg_2>1)$ and $(\fg_1=0,\fg_2>1)$ from left to right.
\label{fig: good SUGRA M5}}
\end{center}
\end{figure}

\paragraph{Analysis of positivity.} The necessary and sufficient conditions to have good supersymmetric AdS$_3$ solutions are $Y>0$, $Z>0$, $a_1/Z + b_1/Y>0$, $a_2/Z + b_2/Y>0$ and $e^{f_0}>0$. Recall that the parameters are constrained as $a_\sigma + b_\sigma = -\kappa_\sigma$. After analyzing all cases in turn, we found that there are good solutions if at least one of the two genera $\fg_{1,2}$ is larger than 1. We use the parametrization in terms of $z_\sigma$ and plot the regions of parameter space where good supergravity solutions exist in Figure \ref{fig: good SUGRA M5}. For $\fg_1>1$ and $\fg_2>1$ we get one connected region:
\be
\{ 1 - 3z_1z_2 > |z_1 + z_2| \} \;.
\ee
For $\fg_1=1$ and $\fg_2>1$ we get two regions:
\be
\{z_1<0 \,,\, z_2 > \frac13 \} \,\cup\, \{ z_1>0 \,,\, z_2 < - \frac13 \} \;.
\ee
The case $\fg_1>1$, $\fg_2=1$ is obtained by exchanging $z_1$ and $z_2$. For $\fg_1 = 0$ and $\fg_2>1$ we get two regions:
\be
\Big\{ - \frac{1+z_1^2}{2z_1} < z_2 < \frac{z_1+1}{1-3z_1} \,,\, z_1>1 \Big\} \,\cup\, \Big\{ \frac{z_1-1}{1+3z_1} < z_2 < - \frac{1+z_1^2}{2z_1} \,,\, z_1<-1 \Big\} \;.
\ee
The case $\fg_1>1$, $\fg_2=0$ is again obtained by exchanging $z_1$ and $z_2$.

\paragraph{Chern-Simons levels and anomalies.}
The CS part of the seven-dimensional supergravity action simplifies when the gauge field is Abelian and reads \cite{Pernici:1984xx}
\begin{multline}
S_{7D} \supset \dfrac{1}{16m}\int (2\Omega_{5}[A]-\Omega_{3}[A]) \\
= \dfrac{1}{16m}\int d^7x \sqrt{-g}\epsilon^{\alpha\beta\gamma\delta\epsilon\eta\zeta}(2\text{Tr}(A_{\alpha}F_{\beta\gamma}F_{\delta\epsilon}F_{\eta\zeta})-\text{Tr}(A_{\alpha}F_{\beta\gamma})\text{Tr}(F_{\delta\epsilon}F_{\eta\zeta})) \\
= - \dfrac{1}{4m} \int d^7x \sqrt{-g}\epsilon^{\alpha\beta\gamma\delta\epsilon\eta\zeta} (A^{A}_{\alpha} F^{A}_{\beta\gamma}F^{B}_{\delta\epsilon}F^{B}_{\eta\zeta}+A^{B}_{\alpha} F^{A}_{\beta\gamma}F^{A}_{\delta\epsilon}F^{B}_{\eta\zeta})\\
= - \dfrac{1}{2m} \int [ b_1b_2 A^{A}\wedge F^{A}+a_1a_2 A^{B}\wedge F^{B}+2(a_1b_2+a_2b_1) A^{A}\wedge F^{B}]\,.
\end{multline}
The trace in the second line is over the $SO(5)$ indices of the gauge field $A$. From here we read off the matrix of Chern-Simons levels in the effective three-dimensional supergravity in AdS$_3$.

On the other hand in field theory we can compute the matrix $k^{IJ}$ of 't~Hooft anomalies. It is computed using the anomaly polynomial of the 6d $\cN=(2,0)$ theory and integrating it over $\Sigma_1 \times \Sigma_2$, as was done in Section \ref{sec: M5 field theory} for the two-dimensional central charges. The result is:
\be
k^{IJ} = \frac1{48} \eta_1\eta_2 \mat{ b_1b_2(4d_Gh_G+3r_G)+3a_1a_2r_G & (a_1b_2+a_2b_1)(4d_Gh_G+3r_G) \\ (a_1b_2+a_2b_1)(4d_Gh_G+3r_G) & a_1a_2(4d_Gh_G+3r_G)+3b_1b_2r_G } \;.
\ee
The matrix $k^{IJ}$ has one positive and one negative eigenvalue for all values of $(a_{\sigma},b_{\sigma})$ that lead to a good AdS$_3$ solution. This means that the flavor current is always left-moving for the SCFTs in question. This result is in harmony with the fact that the second derivative of the trial central charge evaluated at the extremum is negative in the allowed regions for $(a_{\sigma},b_{\sigma})$, indicating again that the flavor current is left-moving.

For the $A_{N}$ theory at large $N$ the matrix of 't~Hooft anomalies simplifies to
\be
k^{IJ} \;\simeq\; \frac1{12} \eta_1\eta_2 N^3 \mat{ b_1b_2 & a_1b_2+a_2b_1 \\ a_1b_2+a_2b_1 & a_1a_2 } \;,
\ee
exactly proportional to the CS matrix in supergravity.

\section{Some geometry of four-manifolds}
\label{app: some geometry}

Let us collect here some known facts---mainly taken from \cite{Anderson}---about the geometry of four-manifolds, that are relevant to Section \ref{sec: other 4manifolds}. Throughout this section $M$ will be a compact \emph{oriented} four-manifold.

\paragraph{Topology.} The most basic topological data are $\pi_1(M)$ and the intersection pairing
\be
I:\, H_2(M,\bZ) \otimes H_2(M,\bZ) \;\to\; \bZ \;.
\ee
When $M$ is simply connected (so that $H_2(M,\bZ)$ is torsion-free), $I$ is a symmetric non-degenerate bilinear form. Let $(n_+,n_-)$ be the signature of the pairing.
Then we define the Betti numbers $b_2^+(M) = n_+$, $b_2^-(M) = n_-$ and the index (or signature) $\tau(M) = n_+ - n_-$. A simply connected four-manifold $M$ is spin if and only if $I$ is even (\ie{} $I(a,a)=0 \mod{2}$ $\forall\, a$).

\paragraph{K\"ahler-Einstein metrics.}
Let $(M,J)$ be a compact complex four-manifold. Let $J\ud{\mu}{\nu}$ be the complex structure, such that $J\ud{\mu}{\rho} J\ud{\rho}{\nu} = - \delta^\mu_\nu$ and integrable. Let $(M,J)$ admit a K\"ahler metric $g$: the metric is Hermitian, $g_{\rho\sigma} J\ud{\rho}{\mu} J\ud{\sigma}{\nu} = g_{\mu\nu}$, which allows to construct a K\"ahler form $\omega$:
\be
\omega = g(J \cdot, \cdot) \qquad\text{ or }\qquad \omega_{\mu\nu} = g_{\rho\nu} J\ud{\rho}{\mu}
\ee
and $d\omega = 0$. Similarly, out of the Ricci tensor $\text{Ric}$ (or $R_{\mu\nu}$) we can construct a Ricci form $\cR$:
\be
\cR = \text{Ric}(J \cdot, \cdot) \qquad\text{ or }\qquad \cR_{\mu\nu} = R_{\rho\nu} J\ud{\rho}{\mu} \;.
\ee
The Ricci form is given---up to a factor of $i$---by the curvature form of the canonical bundle, therefore it only depends on the complex structure $J$ and the volume form $\mu_g$. One also finds
\be
\frac1{2\pi} \, [\cR] = [c_1(M)] \qquad \in \; H^2(M,\bC) \;.
\ee
The Einstein condition is $R_{\mu\nu} = \lambda\, g_{\mu\nu}$ with $\lambda \in \bR$ constant; if $M$ is K\"ahler we can equivalently state it as
\be
c_1(M) = \frac1{2\pi} \, \cR = \frac{\lambda}{2\pi} \,  \omega \;.
\ee
This gives strong restrictions on the existence of KE metrics on $M$.

\paragraph{Obstructions to Einstein metrics and inequalities.} The Chern-Gauss-Bonnet theorem and the Hirzebruch signature formula in four dimensions are:
\bea
\label{curvature formulae}
\chi(M) &= \frac1{32\pi^2} \int_M \Big[ |\text{Rm}|^2 - 4|z|^2 \Big] = \frac1{32\pi^2} \int_M \Big[ |W_+|^2 + |W_-|^2 - 2 |z|^2 + \frac16 R^2 \Big]\;, \\
\tau(M) &= \frac1{48\pi^2} \int_M \Big[ |W_+|^2 - |W_-|^2 \Big]\;,
\eea
where $\text{Rm}$ is the Riemann tensor, $z = \text{Ric} - \frac 14 g R$ is the traceless Ricci tensor, $R$ is the scalar curvature and $W_\pm$ are the (anti)self-dual parts of the Weyl tensor. Squares are taken by simple contraction of all indices, and integrals are taken with the usual measure $d^4x\, \sqrt g$.
An Einstein manifold is characterized by $z=0$ and $R = 4\lambda$. Moreover the Hirzebruch $L$-genus, given by
\be
L(x) = \prod_{j=1}^k \frac{x_j}{\tanh x_j} = 1 + \frac{p_1}3 + \frac{7p_2 - p_1^2}{45} + \frac{62p_3 - 12 p_1p_2 + 2p_1^3}{945} + \dots \;,
\ee
is such that, integrated over a closed smooth oriented manifold of dimension $4n$, computes the signature of the intersection form on the $2n$-th cohomology of $M$. In particular for a four-manifold:
\be
P_1(M) = 3\tau(M) \;.
\ee

\begin{teo} If $(M,g)$ is an Einstein metric, then
\be
\chi(M) \geq 0 \;,
\ee
with equality if and only if $(M,g)$ is flat (that is $\operatorname{Rm} = 0$).
\end{teo}
This simply follows from the Chern-Gauss-Bonnet formula. In particular products of surfaces whose curvatures have unequal sign, \eg{} $S^2 \times \Sigma_{g\geq 1}$ and $T^2 \times \Sigma_{g\geq 2}$, as well as such surface bundles over surfaces, do not admit Einstein metrics. A stronger result is:

\begin{teo} If $(M,g)$ is an Einstein metric, then
\be
\chi(M) \geq \frac32 \, |\tau(M)|\;,
\ee
with equality if and only if $(M,g)$ is flat or it is a Ricci-flat K\"ahler metric on K3 (or orbifold thereof).
\end{teo}
The inequality follows from (\ref{curvature formulae}), while the equality requires a little bit more thought \cite{Anderson}.
For complex surfaces with $c_1<0$ we have
\be
c_2 = \chi \;,\qquad\qquad c_1^2 = 2\chi  + 3\tau \;.
\ee
Notice also that every complex manifold is orientable. This leads to an even stronger inequality:

\begin{teo} If $(M,J)$ admits a K\"ahler-Einstein metric and $c_1<0$, then
\be
\chi(M) - 3\tau(M) = 3c_2 - c_1^2 \geq 0\;,
\ee
with equality if and only if $(M,J)$ is biholomorphic to a complex hyperbolic space-form $\bC\bH^2/\Gamma$.
\end{teo}

\paragraph{Volumes of four-manifolds.}
Let us apply the formul\ae{} in (\ref{curvature formulae}) and the theorems above to obtain the volume of various four-manifolds, of interest for the discussion in Section \ref{sec: other 4manifolds}, in terms of topological data.
For a negatively-curved K\"ahler-Einstein four-manifold $KE_4$ we have:
\be
\label{volume KE4}
\vol(KE_4) = \int_{KE_4} \frac{\omega^2}2 = \frac{2\pi^2}{\lambda^2} \, c_1^2 = \frac{2\pi^2}{\lambda^2}\, (2\chi + 3\tau) = \frac{2\pi^2}{\lambda^2} \, (2\chi + P_1) \;.
\ee
For the special case of a complex hyperbolic space-form $\bC\bH^2/\Gamma$ with K\"ahler-Einstein metric, the same volume formula is valid but we also have
\be
\label{volume extra condition CH2}
\chi(M) = 3\tau(M) = P_1 \;.
\ee
Instead if the four-manifold $M$ has an Einstein metric with self-dual or anti-self-dual Weyl tensor, by direct application of (\ref{curvature formulae}) one finds:
\be
\label{volume HFE4}
\vol(M) = \frac{6\pi^2}{\lambda^2} \, \big( 2\chi - 3 |\tau| \big) = \frac{6\pi^2}{\lambda^2} \, \big( 2\chi - |P_1| \big) \;.
\ee
In particular $3\tau = P_1 > 0$ for $W_- = 0$, and $3\tau = P_1 < 0$ for $W_+ = 0$.

{\small
\bibliographystyle{utphys}
\bibliography{c-ext}
}

\end{document}